\documentclass[traditabstract]{aa} % for the abstract without structuration 
\usepackage{graphicx}
\usepackage{lscape}
\usepackage{longtable}
\def\nodata{{\ldots}}
\usepackage{txfonts}
\usepackage{txfonts}
\usepackage{natbib}
\bibpunct{(}{)}{;}{a}{}{,} % to follow the A&A style
\def\vsini{$v \sin i$}

\def\lam{$\lambda$}
\def\ll{$\lambda\lambda$}

\def\hb{H\,{\sc ii}}
\def\hea{He\,{\sc i}}
\def\heb{He\,{\sc ii}}
\def\cc{C\,{\sc iii}}
\def\cd{C\,{\sc iv}}
\def\nb{N\,{\sc ii}}
\def\nc{N\,{\sc iii}}
\def\nd{N\,{\sc iv}}
\def\ne{N\,{\sc v}}

\def\sic{Si\,{\sc iii}}
\def\sid{Si\,{\sc iv}}
\def\naa{Na\,{\sc i}}
\def\feb{Fe\,{\sc ii}}
 
\begin{document}
   \title{The VLT-FLAMES Tarantula Survey}

   \subtitle{XIV. The O-Type Stellar Content of 30~Doradus}

\author{
N.~R.~Walborn\inst{1},
H.~Sana\inst{1,2}, 
S.~Sim\'{o}n-D\'{i}az\inst{3,4}, 
J.~Ma\'{i}z Apell\'aniz\inst{5},
W.~D.~Taylor\inst{6,7},
C.~J.~Evans\inst{7},
N.~Markova\inst{8}, 
D.~J.~Lennon\inst{9}, and
A.~de~Koter\inst{2,10}
}

\authorrunning{N.~R.~Walborn et al.}
   
\titlerunning{O-Type Stellar Content of 30~Dor}

\institute{
           Space Telescope Science Institute,
           3700 San Martin Drive,
           Baltimore, MD 21218, USA
             \and
	   Astronomical Institute Anton Pannekoek, 
           University of Amsterdam, 
           Kruislaan 403, 1098 SJ, Amsterdam, The Netherlands
             \and
           Instituto de Astrof\'{i}sica de Canarias, 
           E-38200 La Laguna, Tenerife, Spain 
             \and 
	   Departamento de Astrof\'{i}sica, 
	   Universidad de La Laguna,
           E-38205 La Laguna, Tenerife, Spain
             \and
	   Instituto de Astrof\'{i}sica de Andaluc\'{i}a-CSIC, 
	   Glorieta de la Astronom\'{i}a s/n, 
	   E-18008 Granada, Spain   
	     \and
	   Scottish Universities Physics Alliance, 
	   Institute for Astronomy, University of Edinburgh, 
	   Royal Observatory Edinburgh, Blackford Hill, 
	   Edinburgh, EH9 3HJ, UK  
	     \and
	   UK Astronomy Technology Centre, 
	   Royal Observatory Edinburgh, Blackford Hill, 
	   Edinburgh EH9 3HJ, UK   
	     \and    
           Institute of Astronomy, 
           National Astronomical Observatory, 
           Bulgarian Academy of Sciences, 
           PO Box 136, 4700 Smoljan, Bulgaria
             \and
	   European Space Agency, European Space Astronomy Centre, 
	   Camino Bajo del Castillo s/n, Urbanizaci\'on Villafranca 
	   del Castillo, E-28691 Villanueva de la Ca\~nada, Madrid, Spain  
	     \and
           Instituut voor Sterrenkunde, KU Leuven, 
	   Celestijnenlaan 200D, 
	   3001 Leuven, Belgium
}

\date{}
 
 \abstract
{Detailed spectral classifications are presented for 352 O--B0 stars in the 
VLT-FLAMES Tarantula Survey ESO Large Programme, of which 213 O-type 
are judged of sufficiently high quality for further morphological analysis.  
Among them, six subcategories of special interest are distinguished.  
(1) Several new examples of the earliest spectral types O2--O3 have been 
found, while a previously known example has been determined to belong to 
the nitrogen-rich ON2 class.  (2) A group of extremely rapidly rotating main-sequence objects has been isolated, including the largest \vsini\ 
values known, the spatial and radial-velocity distributions of which suggest ejection from the two principal ionizing clusters NGC~2070 and NGC~2060.  
(3) Several new examples of the evolved, rapidly rotating Onfp class show similar evidence, although at least some of them are spectroscopic binaries.    
(4) No fewer than 48 members of the Vz category, hypothesized to be on or 
near the zero-age main sequence, are found in this sample; in 
contrast to the rapid rotators, they are strongly concentrated to the 
ionizing clusters and a newly recognized region of current and recent star 
formation to the north, supporting their interpretation as very young objects, 
as do their relatively faint absolute magnitudes.  (5) A surprisingly large 
fraction of the main-sequence spectra belong to the recently recognized V((fc)) 
class, with \cc\ emission lines of similar strength to the usual \nc\ in V((f)) 
spectra, although a comparable number of the latter are also present, as well 
as six objects with very high-quality data but no trace of either emission 
feature, presenting new challenges to physical interpretations.  
(6) Two mid-O~Vz and three late-O giant/supergiant spectra with morphologically enhanced nitrogen lines have been detected.  Absolute visual magnitudes 
have been derived for each star with individual extinction laws, and composite 
Hertzsprung-Russell Diagrams provide evidence of the multiple generations present in this field.  Spectroscopic binaries, resolved visual multiples, and  
possible associations with X-ray sources are noted.  Astrophysical and dynamical
analyses of this unique dataset underway will provide new insights into the 
evolution of massive stars and starburst clusters.}

% \abstract{}{}{}{}{} 
% 5 {} token are mandatory

   \keywords{Galaxies: star clusters: individual: 30~Doradus -- Magellanic
Clouds -- Stars: early-type -- Stars: fundamental parameters -- Stars: massive
-- Stars: spectral classification}

   \maketitle

\section{Introduction} \label{sect: intro}

30~Doradus in the Large Magellanic Cloud (LMC) comprises the most massive starburst cluster and giant \hb\ region in the Local Group, and it contains the most massive stars known (Crowther et al. 2010; Bestenlehner et al. 2011).  It is the paradigm for understanding early massive stellar and cluster evolution, and for interpreting more distant starbursts.  Thus it is appropriate that 30~Dor is the subject of the unprecedented spectroscopic dataset obtained by the VLT-FLAMES Tarantula Survey (VFTS; ESO Large Programme 182.D-0222; Evans et al. 2011), which contains high-resolution ($R \sim 10^4$) observations of about 800 OB stars in this field.  Of these, 352 from the Medusa-Giraffe multi-object spectrograph configuration have been determined to be of spectral type O (or B0 in 10 cases included here) and are the subject of this morphological investigation.

The prior state of the art in digital spectral classification of the O stars 
is represented by the Galactic O-Star Spectroscopic Survey 
(GOSSS; Ma\'{i}z Apell\'aniz et al. 2011; Sota et al. 2011), 
in which the system and procedures have been further developed and refined.  
An expanded list of standard stars and an extensive new classification
atlas from those high-S/N data are presented there; many of the GOSSS
developments have been applied in the present work.  However, the resolution of
the VFTS data is substantially higher, which has required additional
developments and another new atlas of primarily Galactic spectra, from the IACOB survey (Sim\'on-D\'{\i}az et al. 2011) in the northern hemisphere and the ESO archives in the southern, as discussed in detail and presented by Sana et al. (2014a, in prep.).  Most of that discussion will not be repeated here, but it was determined that $R \sim 4000$ is the practical upper limit for reliable visual spectral classification of O-type spectra, because of the effects of line-profile differences and numerous weak lines at higher resolving powers, so  that value has been adopted for the new atlas and for this work.

Section~2 provides some further details about the data and classification
procedures used here, including the specification of the high-quality subset of 213 objects isolated for further analysis, based on not only data characteristics but also multiplicity considerations.  Experience has shown that substantial advances in data quality and/or sample size generally produce both foreseeable and unexpected new scientific results; the present study is no exception, and Section~3 describes six spectroscopic categories of special interest encountered among this subsample, some previously recognized (with new members added here) but others newly distinguished. Section~4 references the essential, tailored extinction corrections required for astrophysical analysis in regions of this kind and presents empirical and theoretical  Hertzsprung-Russell Diagrams (HRDs) for the high-quality subsample.  Possible associations with {\it Chandra} X-ray sources (Townsley et al. 2006) are noted in Section~5.  Section~6 provides a summary, conclusions, and outlook for further analyses of these data by the VFTS Team.

\section{Data and Analysis}

Full details of the instrumental parameters and data reductions were
provided by Evans et al. (2011).  In brief, the VFTS data discussed here
were obtained with the Medusa--Giraffe mode of the Fibre Large Array
Multi-Element Spectrograph (FLAMES) instrument (Pasquini et al. 2002) at
the Very Large Telescope (VLT) on Cerro Paranal, Chile.  Each target was observed with the standard LR02 and LR03 settings of the Giraffe spectrograph (providing coverage of 3960--5071~\AA\ at $R$ of 7000--8500).  At least six observations were obtained with the LR02 setting to support the
investigation of spectroscopic binaries (Sana et al. 2013).  Most of the observations were done during 2008 October through 2009 February, with a final epoch in 2009 October to extend the binary period sensitivity.

The LR02 and LR03 observations were coadded to produce the spectrograms
studied here, so data below and above $\sim$4500~\AA\ correspond to
different observations/epochs.  Thus, in a rare, unlucky case of spectral
variability, it is possible that an unreal spectrum has been synthesized.
However, the individual observations were each inspected, and those of
lower quality or subject to apparent artifacts were omitted from the sums.
Most real variations correspond to either SB2 or SB1 systems; the former
have been treated separately as further discussed below, while the latter
were shifted for coaddition.  Further processing steps undertaken for this
work were rectification and rebinning to $R \sim 4000$, prior to plotting
each spectrogram with a large scale matching that of the Sana et al. (2014a,in
prep.) atlas for visual classification.

\begin{figure*}
  \centering
  \includegraphics[angle=90,width=\textwidth]{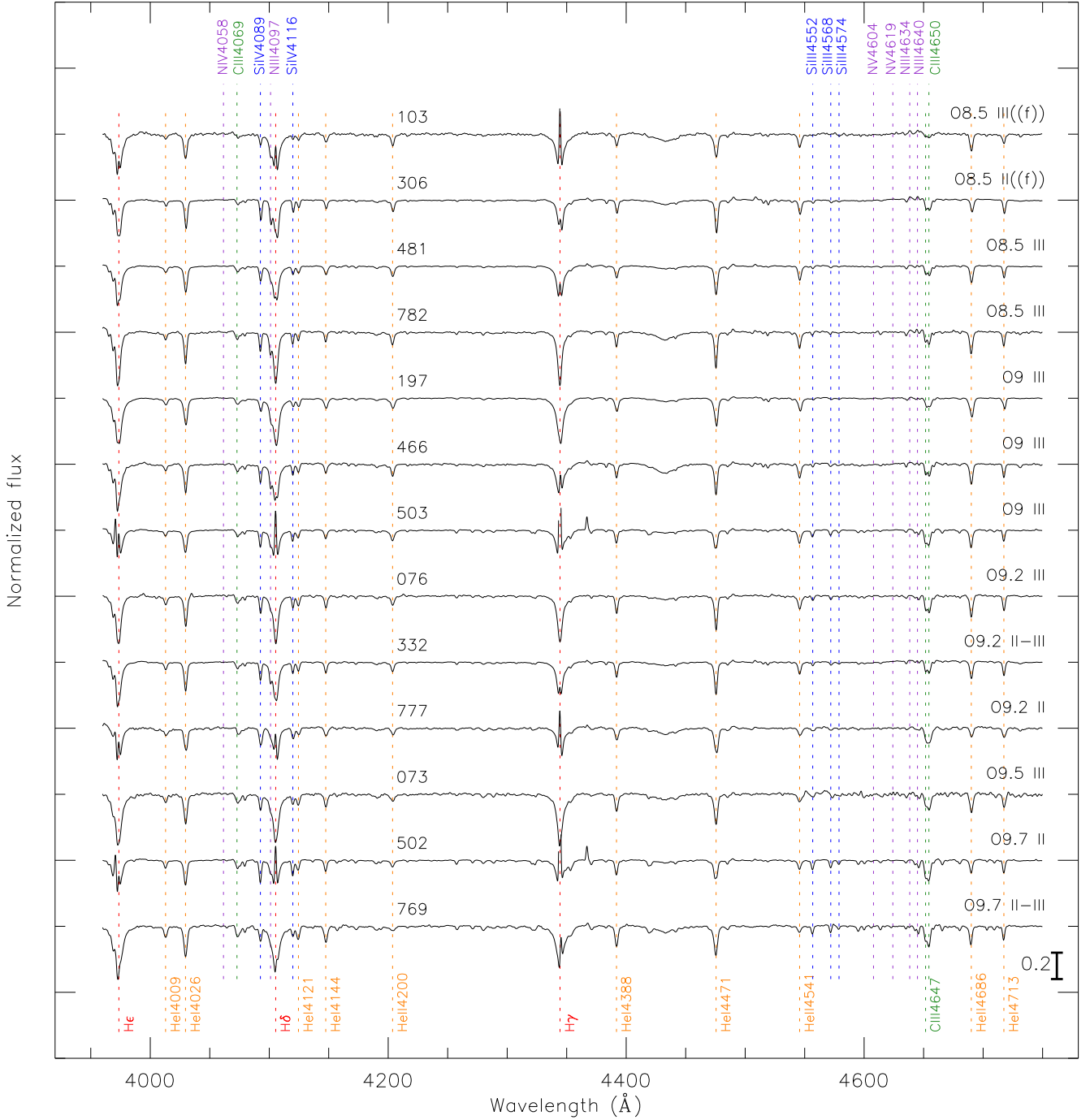}
  \caption{A sequence of O8.5 through O9.7 spectra from the VFTS sample, to illustrate the new O9.2 type in terms of the \heb~$\lambda\lambda4541$,4200/\hea~$\lambda\lambda4387$,4144 and 
\heb~$\lambda4541$/\sic~$\lambda4552$\ criteria. $\lambda4026$ is a blend of \hea\ and \heb, with the former predominating at late-O types. The ordinate scales in all spectroscopic figures are given in continuum units by the bar at lower right.  These figures should be viewed online with at least 200\% enlargement to fully appreciate their information content.}
  \label{fig: 1}
\end{figure*}

Among other refinements in GOSSS (Sota et al. 2011), the late-O/early-B spectral
types and standards were redefined to maintain constant He line ionization
ratios at all luminosity classes, in order to improve the consistency and reliability of the results.  To this end, the O9.7 spectral subclass was introduced at the lower luminosity classes (V through III) for the first time, and some previous standards were shifted to the adjacent earlier subclasses.
As also discussed by Sana et al. (2014a, in prep.), during the present work a further development along these lines was found to be warranted and useful, namely the introduction of the new subclass O9.2 to describe spectra with the \heb\ lines (\ll4541, 4200) in the relevant ratio criteria just slightly weaker than the respective \hea\ lines (\ll4387, 4144).  The range in these ratios at O9.5 appeared excessive and the number of VFTS objects placed in that subclass is larger than in any other (Table~1).  Thus, O9.2 is symmetrical with respect to O8.5, in which the same \heb\ lines are slightly stronger than the \hea, i.e., these ionization ratios are inverted between those two subclasses, on opposite sides of the unit ratios at O9.  A sequence of these spectral types selected from the VFTS data is shown in Figure~1\footnote{It must be noted that there is a \nc\ feature blended with \heb\ at 4200~\AA\ with the present resolution, which may affect the appearance of the relevant He ionization ratio when the \heb\ is weak, i.e., at late-O types.  Such an effect would be most pronounced in supergiant spectra, in which the \nc\ lines are normally enhanced, or in N-enhanced spectra at any luminosity class.  Indeed, there are some spectra in which these two He ionization ratios appear somewhat discrepant, frequently in the sense that \lam4200/\lam4144 is greater than \lam4541/\lam4387, which may well be due to the \nc\ contribution at \lam4200.  Clarification of this issue will require higher resolution and/or spectral synthesis.}.

\begin{table}
\caption{Spectral-Type Distribution of AAA-Rated Classifications for VFTS O Stars}
\label{table:1}
\smallskip
\begin{tabular}{lr}
\hline
SpT &No.\\ 
\hline
O2   &7\\
O3   &11\\
O4   &12\\
O5   &11\\
O6   &27\\
O7   &24\\
O8   &14\\
O8.5 &17\\
O9   &21\\
O9.2 &8\\
O9.5 &42\\
O9.7 &19\\
\\
\hline
\end{tabular}
\end{table}

\begin{figure*}
  \centering
  \includegraphics[angle=0,width=\textwidth]{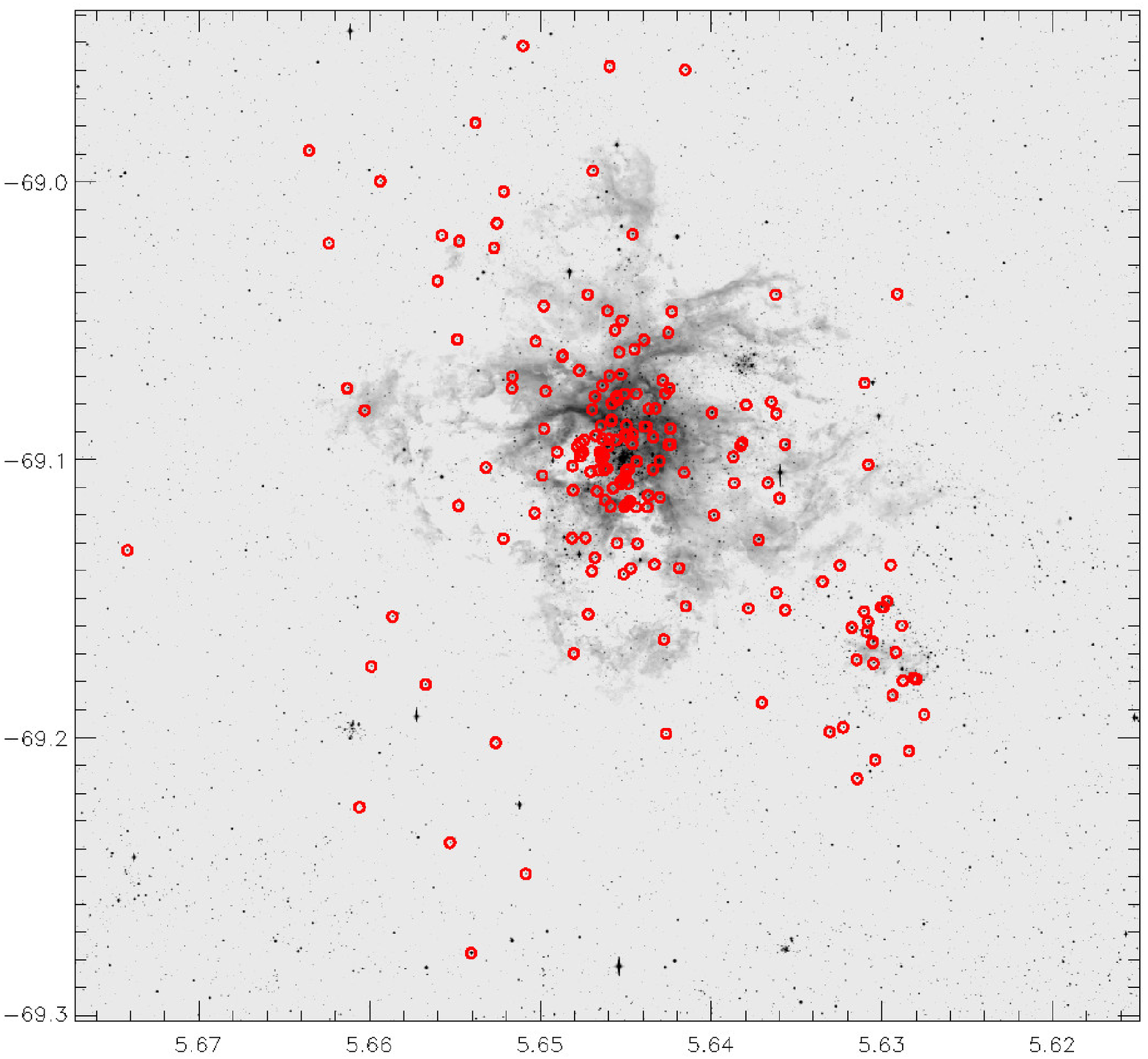}
  \caption{Locations of the 213 AAA O stars in the 30~Dor field, represented by a visual image from the ESO 2.2~meter Wide Field Imager.  The axes are labeled with equatorial coordinates in hours of right ascension and degrees of declination in this and all subsequent direct images.  NGC~2070 (including R136, the bright object above center) is the ionizing cluster of the Tarantula Nebula, while NGC~2060 is a somewhat older association 6\arcmin\ (90~pc in projection) to the SW.}
  \label{fig: 2}
\end{figure*}

The present spectral classifications of the 213 most highly rated (AAA) VFTS O-type data are listed in Table~2, and their spatial distribution in the 30~Doradus field is shown in Figure~2.  Their positions are given by VFTS identification number in Evans et al. (2011) and are not reproduced here, although their $BV$ photometry is.  Absolute visual magnitudes, effective temperatures, and bolometric luminosities derived here as referenced later are also listed.  Spectroscopic and visual multiplicity information is given, as are cross-identifications with previous work to facilitate comparisons and recognition of new discoveries of special interest here.  The remaining 139 (BBB) VFTS O--B0 classifications are listed in Appendix~A, along with the various reasons for their lower ratings and discussion of some of them.
 
The visual spectral classifications presented here for the apparently constant-velocity and (shifted) SB1 objects have been performed on large-scale paper plots.  The GOSSS classifications are being done with an electronic standard-spectra comparator (MGB; Ma\'{\i}z Apell\'aniz et al. 2012).  That program also supports the simultaneous classification of double-lined spectra, which has been carried out for the SB2 in the present sample by JMA; these results are included in the BBB list here, pending more detailed analysis of the separate components.  

\section{Special Spectroscopic Categories}

The VFTS ID numbers of several special categories of O-type spectra in 30~Doradus are listed in Table~3.  The spectra and spatial distributions of most of these categories are presented and discussed in turn in this section; note that many objects are in more than one.  The SB? category consists of spectra that exhibit radial-velocity displacements between the stellar absorption and nebular emission lines, but in which significant radial-velocity variations have not been detected.  The Galactic SB 9~Sagittarii has a period of 8.6~yrs (Rauw et al. 2012) and would likely display little motion during the interval of the initial VFTS observations.  While most of the SB? are likely SB, the alternative explanation of anomalous stellar or nebular motions cannot be excluded a priori.

%\newpage

\setcounter{table}{+2}
\begin{table}
\caption{Special Categories of VFTS O Stars with AAA-Rated Spectral Classifications}
\label{table:3}
\smallskip
\begin{tabular}{l}
\hline\hline\\
O2  (7)\\
016, 072, 169, 468, 506, 512, 621\\[9pt]
O3  (11)\\
094, 143, 180, 267, 404, 518, 532, 566, 599, 755, 797\\[9pt]
Vn/nn/nnn  (18)\\
074, 138, 184, 190, 249, 285, 406, 592, 654, 660, 706, 722, 724, 746,\\ 
751, 755, 768, 770\\[9pt]
V((n))/(n) high inclination?  (10)\\
065, 072, 243, 250, 355, 356, 404, 761, 797, 830\\[9pt]
IIIn/nn  (8)\\
012, 091, 399, 530, 531, 574, 615, 843\\[9pt]
Onfp  (7)\\
094(= ST1-28), 177, 190, 208(= ST1-93), 526, 626, 656\\[9pt]
Vz  (48)\\
014, 067, 089, 096, 110, 117, 123, 132, 140, 168, 184, 252, 256, 266,\\ 
355, 356, 380, 382, 390, 392, 398, 409, 418, 470, 472, 479, 488, 511,\\ 
532, 536, 537, 549, 550, 555, 577, 586, 601, 613, 621, 638, 651, 706,\\ 
722, 724, 751, 761, 802, 849\\[9pt]
O3.5-7 V((fc))  (19)\\
096, 143, 216, 355, 382, 385, 390, 404, 418, 479, 491, 511, 537, 550,\\ 
577, 581, 586, 797, 812\\[9pt]
O3.5-7 V((f))  (14)\\
089, 190, 243, 266, 392, 398, 409, 432, 470, 488, 564, 667, 761, 830\\[9pt]
O3.5-7 V high-S/N with no trace of N III, C III emission lines  (6)\\
110, 117, 472, 484, 601, 849\\[9pt]
ON/N strong/N weak  (7)\\
045, 089, 506, 761, 764, 807, 819\\[9pt]
SB? from stellar/nebular shift  (36)\\
021, 046, 067, 117, 123, 168, 169, 190, 216, 235, 251, 290, 306, 356,\\ 
418, 436, 491, 536, 549, 550, 586, 609, 620, 626, 639, 663, 664, 667,\\ 
679, 704, 710, 717, 722, 775, 777, 797\\
\\
\hline
\end{tabular}
\end{table}

\subsection{O2-O3 Stars}

\begin{figure*}
  \centering
  \includegraphics[angle=90,width=\textwidth]{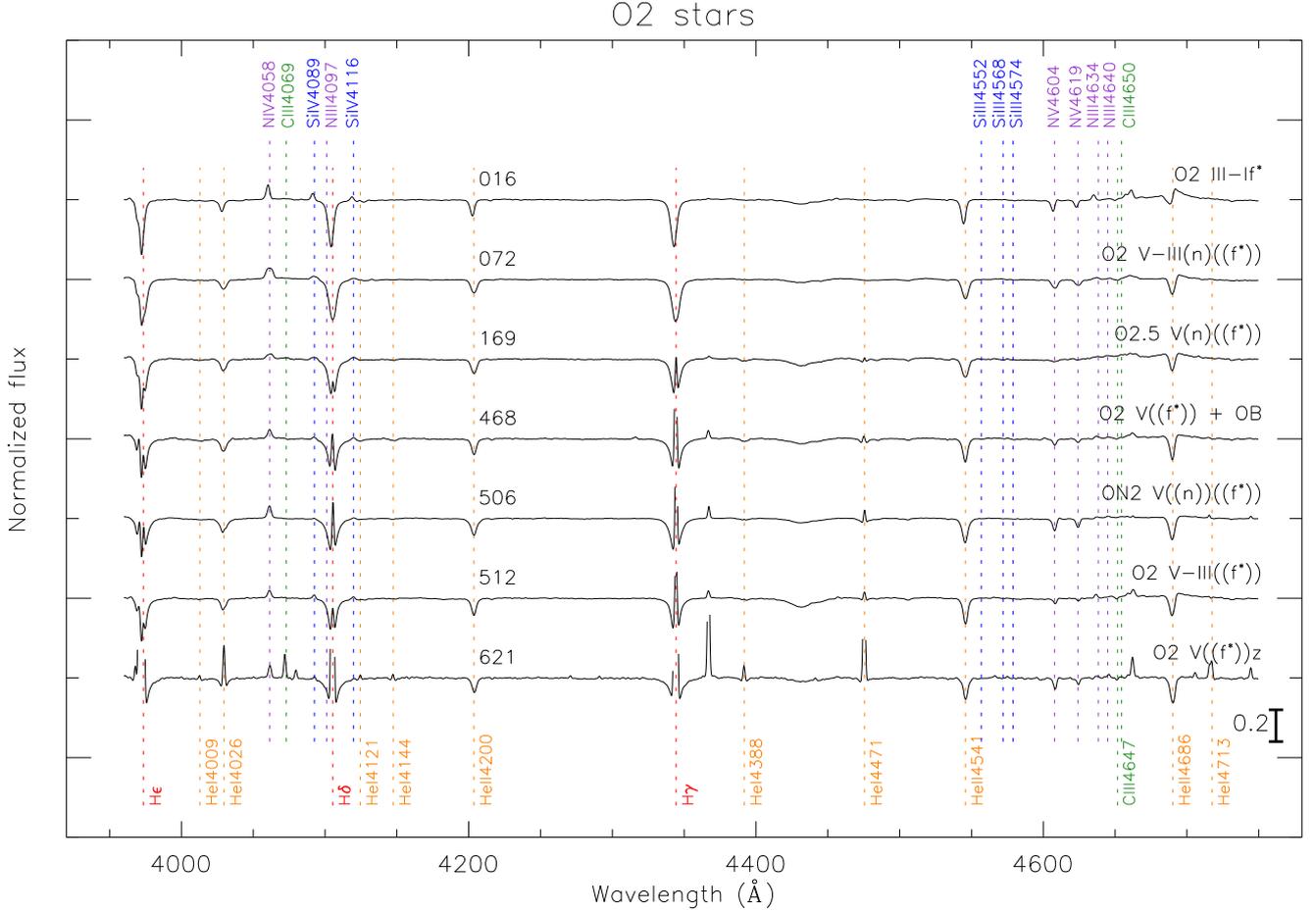}
  \caption{VFTS spectra of type O2.  $\lambda4026$ is a blend of \hea\ and \heb, with the latter predominating at early-O types.  \cc~$\lambda\lambda4069$, 4650 and \nc~$\lambda4640$ are blends, primarily with other lines from the same multiplets in O-type spectra.  Strong nebular emission lines are truncated to avoid overlapping adjacent spectra.}
  \label{fig: 3}
\end{figure*}

\begin{figure*}
  \centering
  \includegraphics[angle=90,width=\textwidth]{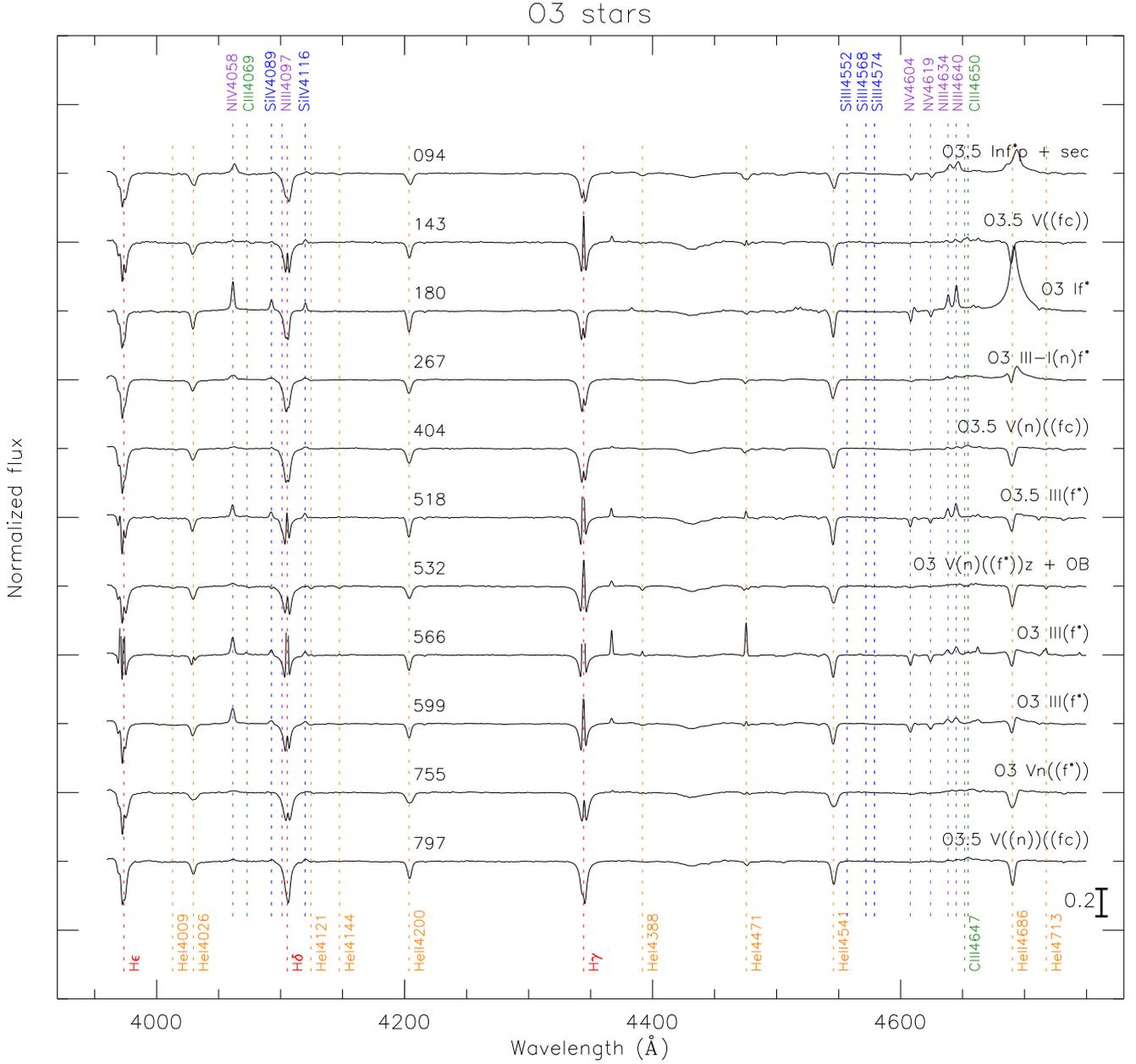}
  \caption{VFTS spectra of type O3.}
  \label{fig: 4}
\end{figure*}

\begin{figure*}
  \centering
  \includegraphics[angle=0,width=\textwidth]{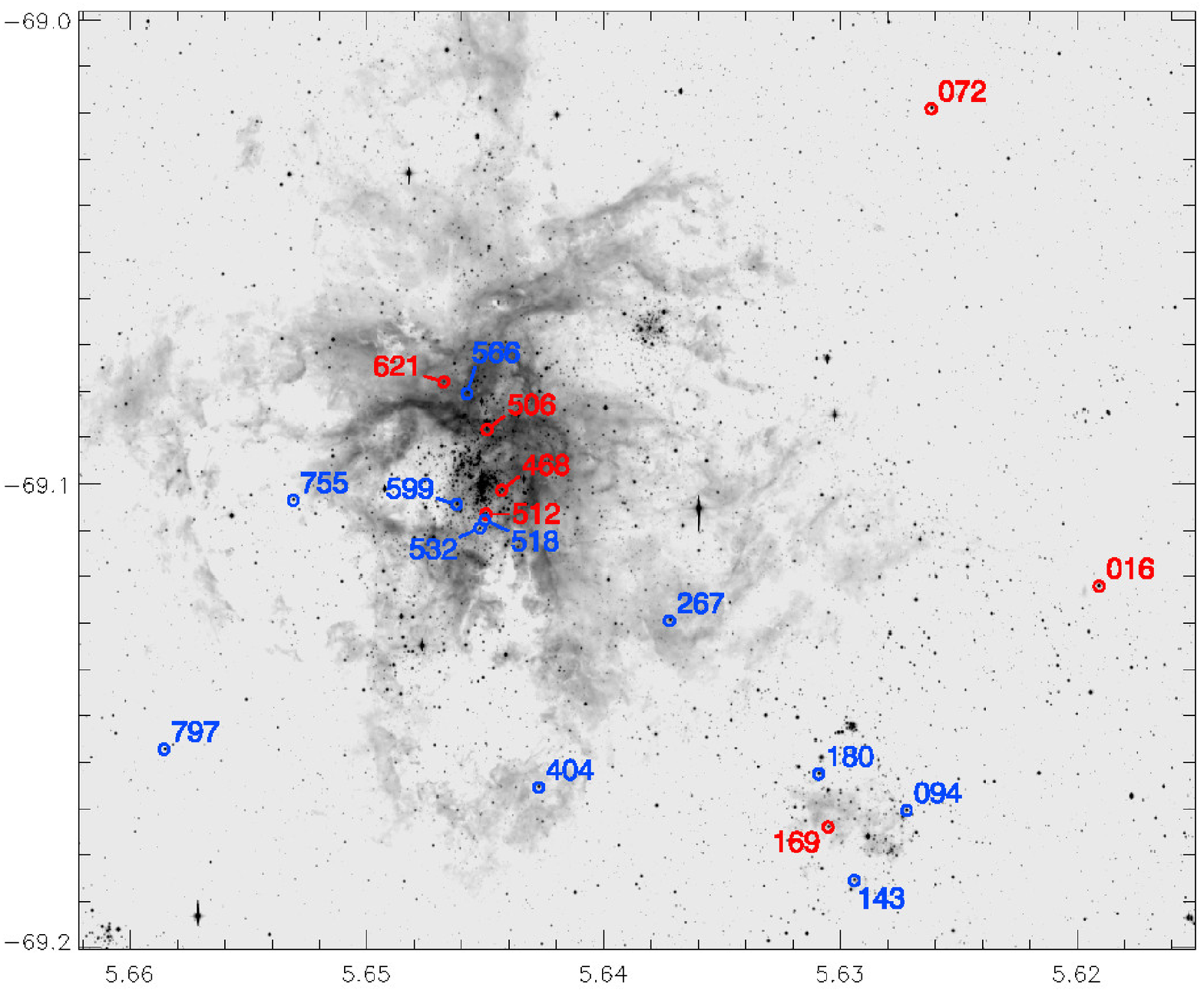}
  \caption{Spatial distributions of the O2 (red) and O3 (blue) stars.}
  \label{fig: 5}
\end{figure*}

The classification of the hottest Population I spectra was discussed by Walborn et al. (2002a, 2004) and has been recently analyzed quantitatively by Rivero Gonz\'alez et al. (2012a).  30~Doradus has been known for some time to contain the strongest resolved concentration of these extreme objects (Parker 1993; Walborn \& Blades 1997; de Koter et al. 1997, 1998; Massey \& Hunter 1998; Bosch et al. 1999).  Our sample includes 18 stars of these spectral types (not including the intermediate Of*/WN objects discussed by Crowther \& Walborn 2011), of which 9 (VFTS~16, 143, 267, 404, 512, 532, 621, 755, 797) are believed to be newly classified as such (cf. cross-identifications with earlier work in Table~2)--although all of the present classifications are of higher quality and many are not identical to previous results for that reason.  The VFTS blue-violet spectrograms of all these stars are shown in Figures~3 and 4, while their spatial distribution is in Figure~5.

Objects to be distinguished even among this exceptional group include the runaway star VFTS~16 (Evans et al. 2010).  VFTS~72~=~BI~253 is another runaway candidate based on its location (Fig.~5 here; Walborn et al. 2002a); in that reference it was adopted as a luminosity class V standard, but it is assigned V-III here, indicating either a data-quality effect or some variability.  

VFTS~506 is newly recognized as a member of the surprising ON2 subtype introduced and discussed by Walborn et al. (2004).  VFTS~169 has been classified O2.5~V(n)((f*)); the type corresponds to an interpolation of the \nd\ / \nc\ emission-line ratio between O2 and O3.  This object is not an exemplar of the type because of its weak, broad lines; published precedents are N11-26 (Evans et al. 2006) and HD~93162 (Crowther \& Walborn 2011).

VFTS~566 and 599 are the first high-quality representatives of the O3~III(f*) spectral type, for which no clear example could be identified by Walborn et al. (2002a); VFTS~518~=~P901 (Bosch et al. 1999) is a nice paradigm for O3.5~III(f*).    VFTS~180 was originally classified as an intermediate Of*/WN type by Schild \& Testor (1992) and hence appears in the LMC Wolf-Rayet (WR) catalogue (Breysacher et al. 1999); but in their detailed rediscussion of that category, Crowther \& Walborn (2011) reclassify it as a pure Of* type.  

VFTS~755 is a member of the extremely rapidly rotating dwarf category discussed in the next Section~3.2.  VFTS~94~=~ST1-28 is also a member of the Onfp category discussed in Section~3.3; it is definitely an SB but its double-lined status is uncertain, which is why it is included in the AAA list (rather than the BBB where most SB2 are found).  VFTS~532 and 621 also belong to the Vz category (Section~3.4).  VFTS~143, 404, and 797 are also in the V((fc)) category (Section~3.5); the first two are large-amplitude SB1 and the third a possible SB.     

\subsection{Extreme Rotator Runaways}

\begin{figure*}
  \centering
  \includegraphics[angle=90,width=\textwidth]{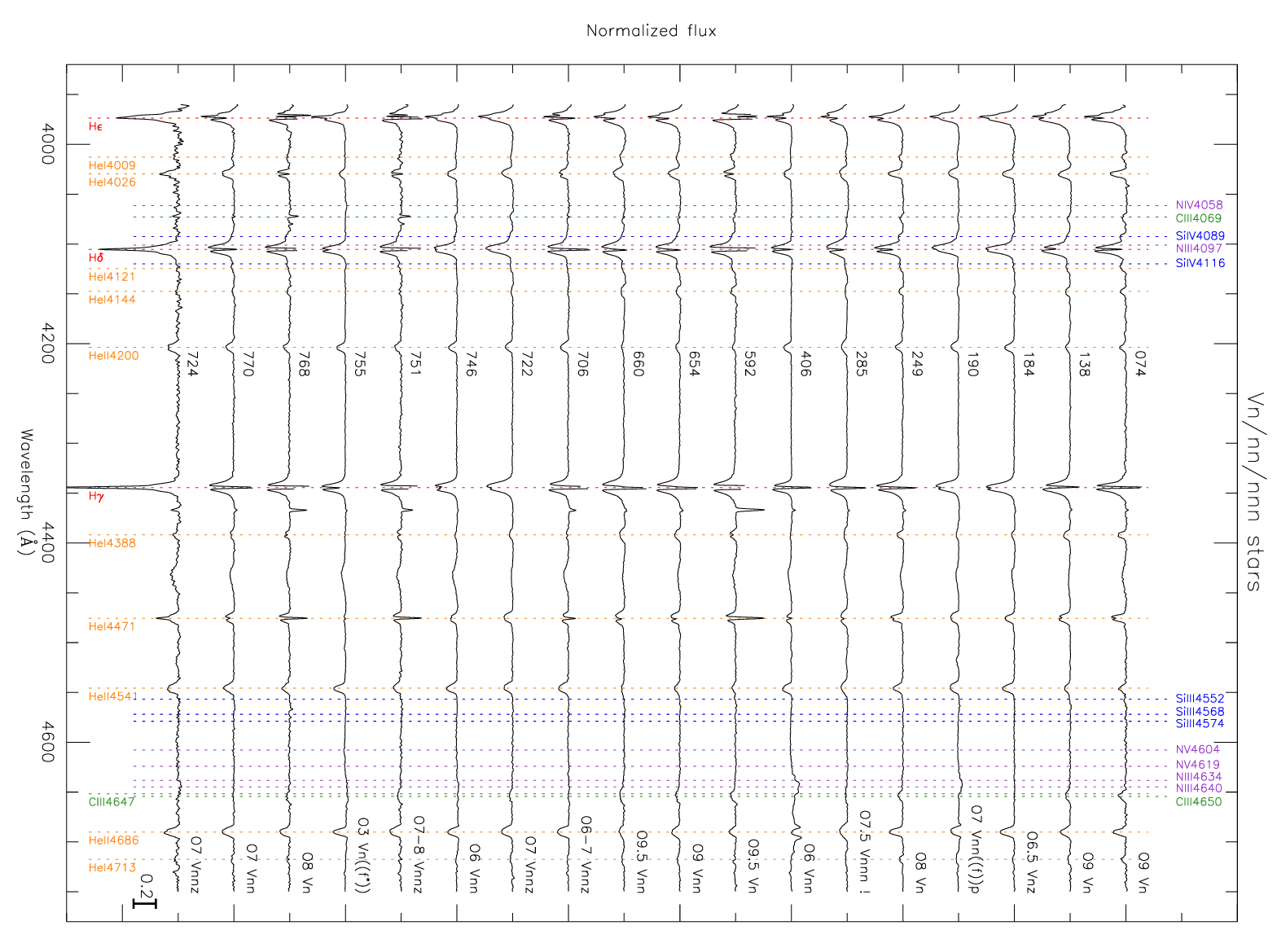}
  \caption{VFTS spectra of very rapidly rotating luminosity class V stars.  VFTS~285 is the most rapidly rotating O star known (the exclamation mark is not part of the spectral type!).  VFTS~406 is contaminated by a nearby emission-line spectrum on the detector, i.e., the apparent \nc\ and \heb\ emission features are spurious.   Note that VFTS~724 is out of the ID sequence because of the oversubtracted nebular emission lines.}
  \label{fig: 6}
\end{figure*}

\begin{figure*}
  \centering
  \includegraphics[angle=0,width=\textwidth]{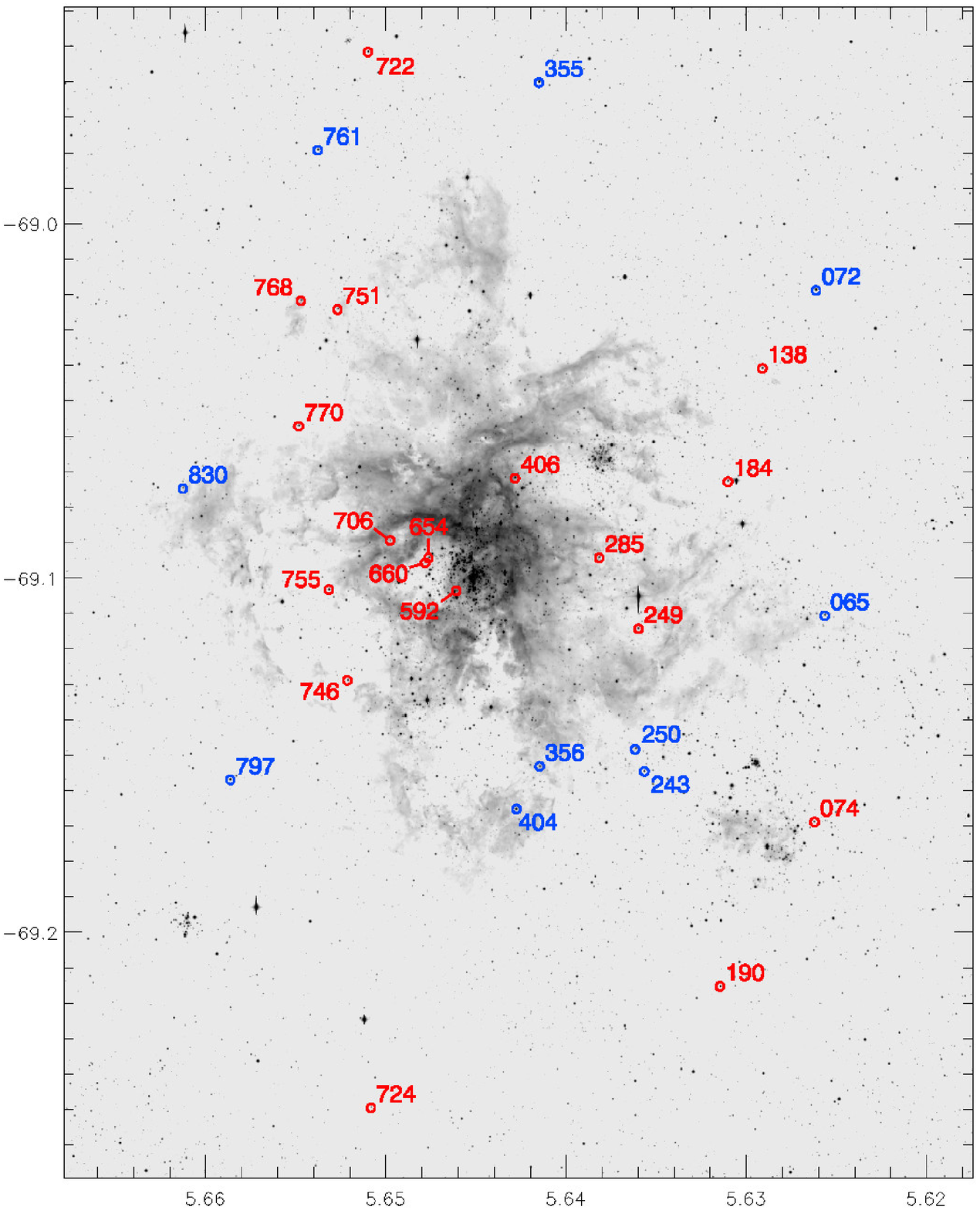}
  \caption{Spatial distribution of the VFTS luminosity class V extreme rotators (red), and a selection of peripheral moderate (projected) rotators (blue). }
  \label{fig: 7}
\end{figure*}

Perhaps the most striking and unexpected result of this investigation is described in this subsection.  It consists of the discovery of a category of 18 extremely rapid rotators (Figure~6), including the fastest known, whose peripheral space distribution with respect to the ionizing clusters (Figure~7) immediately suggests a runaway nature.  This interpretation is already supported by their peculiar radial-velocity distribution, with a high fraction of large relative values (Sana et al. 2012; Sana et al. 2014b, in prep.).  The average projected distance from R136 of the 13 of them within 81~pc is $40 \pm 7$~(error of the mean, hereafter m.e.)~pc.  For comparison with the distribution of the Vz stars as discussed in Section~3.4 below, their average separation in RA from R136 (i.e., from the NS axis of the Tarantula) is $30 \pm 5$~(m.e.)~pc.  Remarkably, the projected distances from R136 of VFTS~722 and 724, which have identical spectral types but are located at opposite NS extremes of the field, are both 132~pc; moreover, both have peculiar radial velocities on opposite sides of the cluster mean.   

\begin{figure*}
  \centering
  \includegraphics[angle=90,width=\textwidth]{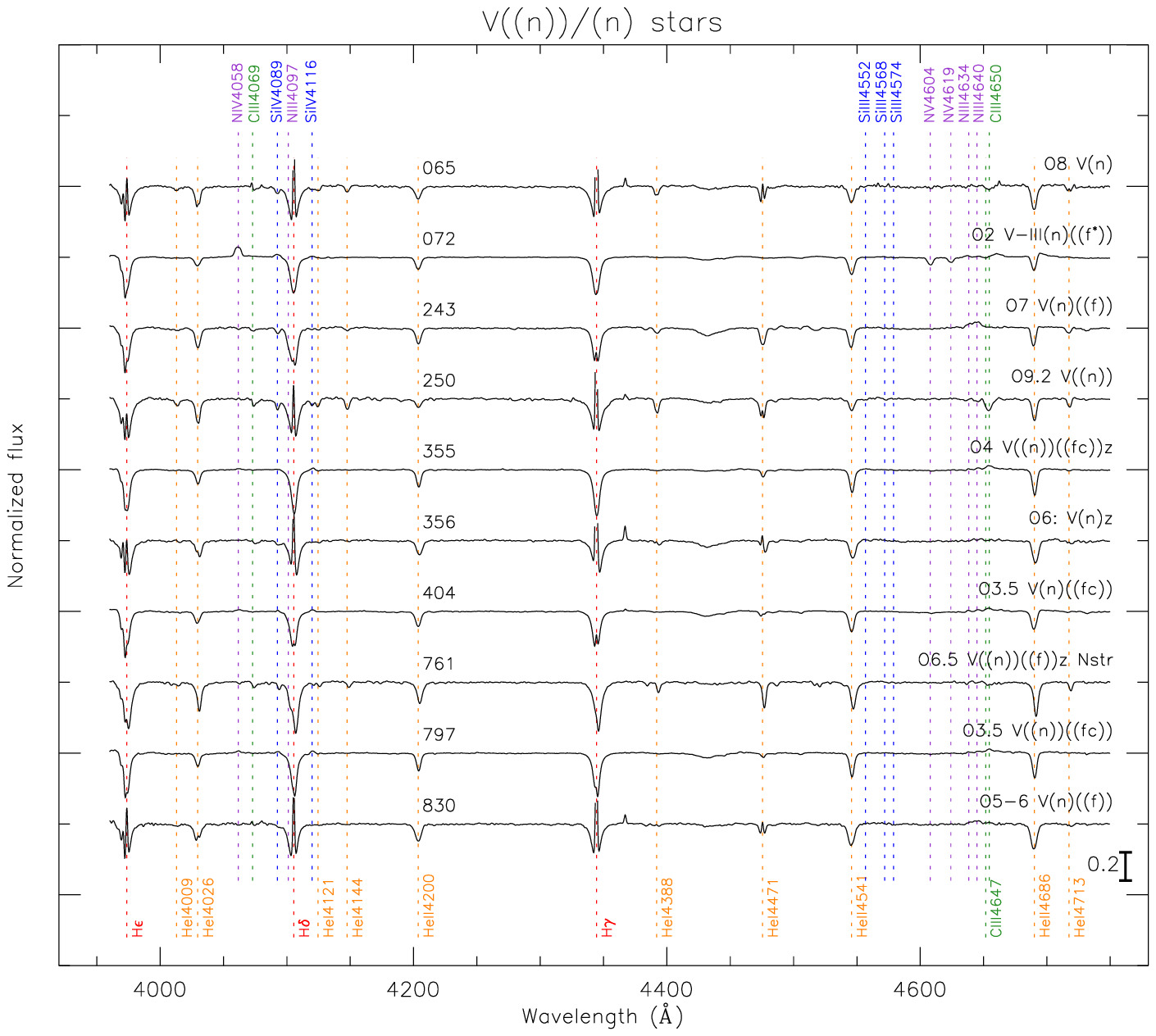}
  \caption{VFTS spectra of selected peripheral moderate (projected) rotators.}
  \label{fig: 8}
\end{figure*}

\begin{figure*}
  \centering
  \includegraphics[angle=0,width=\textwidth]{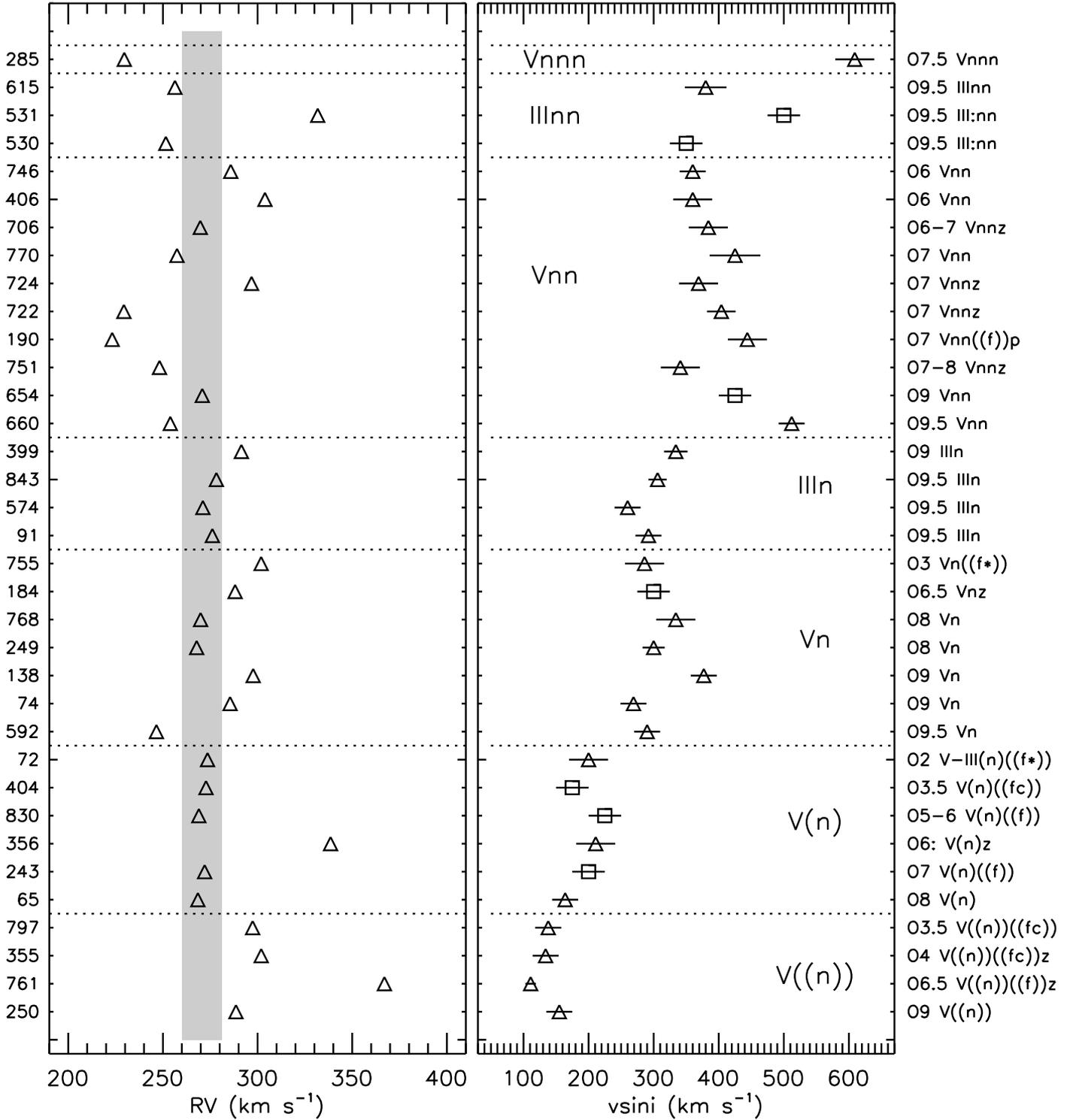}
  \caption{Moderate to extreme rotators, grouped from bottom to top first by line-broadening parameter, then by luminosity class, and finally inversely by spectral subclass within each of the former parameters.  VFTS numbers are given on the left vertical axis and the full spectral types on the right.  Left panel: radial velocities from Sana et al. (2013) relative to the cluster means (vertical bar). Right panel: projected rotational velocities determined mostly by Ram\'{\i}rez-Agudelo et al. (2013) but some by SSD for this paper with similar techniques; squares indicate SB1 not included in the former reference and bars denote the uncertainties.}
  \label{fig: 9}
\end{figure*}

\begin{figure*}
  \centering
  \includegraphics[angle=90,width=\textwidth]{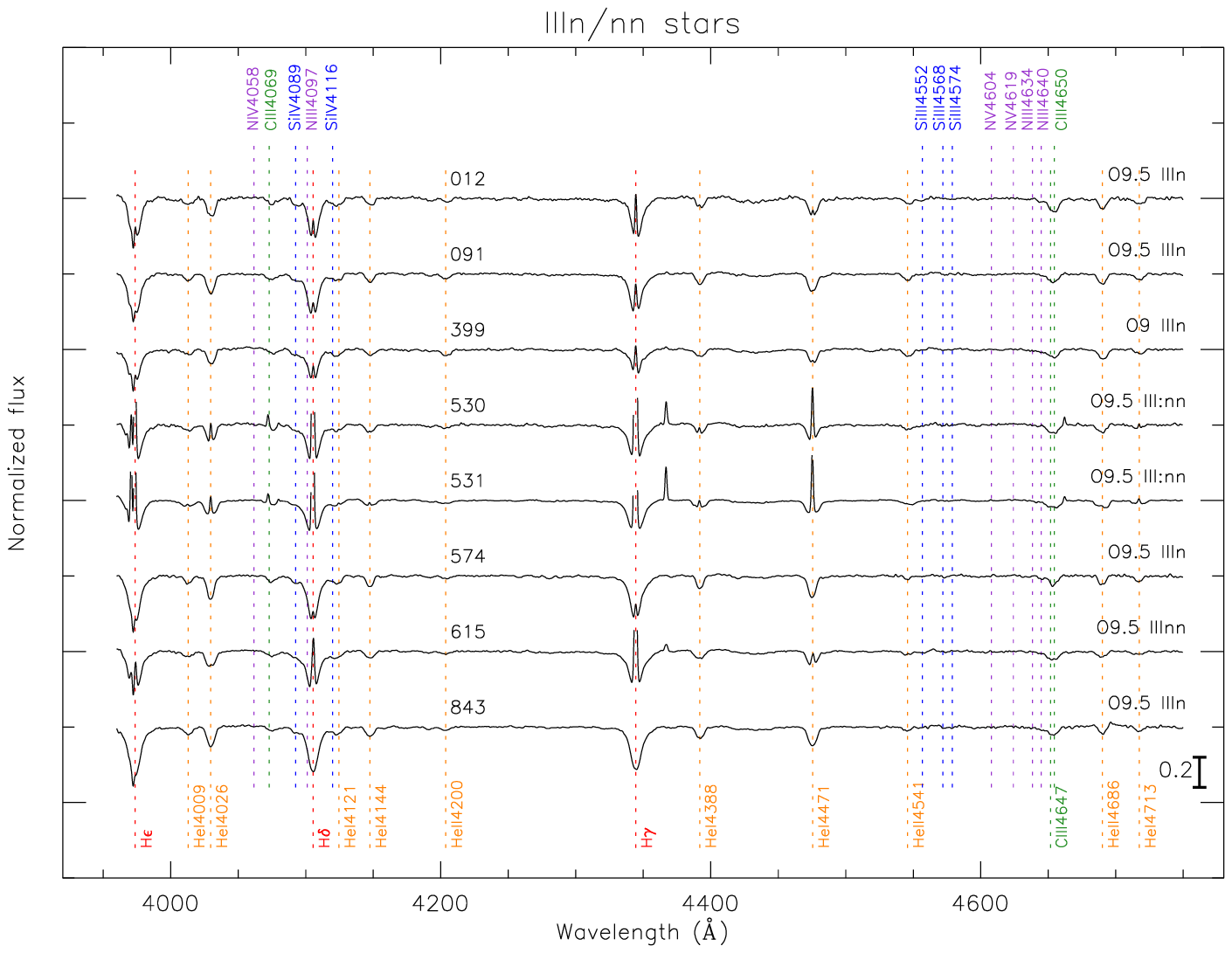}
  \caption{VFTS spectra of very rapidly rotating late-O, luminosity class III stars.}
  \label{fig: 10}
\end{figure*}

Figures~7 and 8 also display a number of moderate (projected) rotators in the same peripheral areas, for further investigation as {\it hypothetical} very rapid rotators with high axial inclinations.  Two of them, VFTS~356 and 761, do have high peculiar radial velocities consistent with a runaway nature (Sana et al. 2014b, in prep.), while three more, VFTS 250, 355, and 797, have moderately anomalous values.  

The projected equatorial rotational velocities (Ram\'{\i}rez-Agudelo et al. 2013 and this paper) together with the radial velocities (Sana et al. 2013) of both the high and these selected moderate rotators are displayed graphically in Figure~9, which also shows the progression of the line-broadening classification n-parameter and its high degree of correspondence with the \vsini\ values.  Technical details and uncertainties of the \vsini\ derivations can be found in Ram\'{\i}rez-Agudelo et al. (2013); as discussed there and by Sim\'on-D\'{\i}az \& Herrero (2014), the effect of macroturbulence is essentially negligible for \vsini\ $>120$~km~s$^{-1}$, i.e., the range of interest here.  Other potential effects, such as that of gravity darkening in very rapid rotators, will be addressed in future VFTS investigations of the rotational velocities.
 
The objects in the field image (Fig.~7) are all of luminosity class V, but Figures~9 and 10 also display several extreme rotators classified as late-O giants, for further consideration of their actual luminosities.  E.g., if the \heb\  $\lambda4686$ absorption line were filled in by emission due to some cause other than higher luminosity in these extreme objects, they could have lower actual luminosities; and the luminosity classes of two of them are flagged as uncertain.  In fact, the average $M_V$ of 4 late-O~V and 5 late-O~III rapid rotators with determinations in Table~2 are indistinguishable at $-3.85 \pm 0.07$, compared to calibration values of $-4.2$ for the former and $-5.45$ for the latter (Walborn 1973), indicating that all of them are actually dwarfs, as further discussed later.  VFTS~399 is an extreme X-ray flare object, as mentioned in Section~5.   

The previously most rapidly rotating O stars known were recently discussed by Walborn et al. (2011), with highest \vsini\ values somewhat exceeding 400~km~s$^{-1}$.  Two objects in the VFTS 30~Dor sample surpass them substantially, with values $\sim600$~km~s$^{-1}$.  The first to be discovered and assigned the nnn qualifier was VFTS~285 (Walborn et al. 2012), which has very high-S/N data (Fig.~6).  The other, VFTS~102, has lower quality data with a somewhat uncertain spectral type and is hence in the present BBB list (Appendix~A); this object and its possible association with the X-ray pulsar PSR~J0537-6910 are discussed separately by Dufton et al. (2011).  

Remarkably, none of these stars (except VFTS~102) shows evidence of disk emission lines, even at H$\alpha$ (also covered by VFTS albeit not discussed here).  VFTS~190 is a weak but definite member of the Onfp category discussed in the next Section~3.3, with weak emission wings at \heb\  $\lambda4686$ (Fig.~6), and it is a possible SB as many members of that category are.  VFTS~406 was originally thought to be a pronounced member of that category but it was subsequently realized that its apparent emission features are due to contamination on the detector from the adjacent WN spectrum of R135, so it is included here as a normal extremely broad-lined dwarf.

The moderate (projected) rotator VFTS~72 is of type O2, while 404 and 797 are O3.5.  As already noted in the previous section, the O3 type of VFTS~755 is the earliest among the extreme rotators.  VFTS~184, 355, 356, 706, 722, 724, 751, and 761 are also members of the Vz category believed to be indicative of extreme youth (Section~3.4).

A significant population of high-mass, extreme rotators ejected from a massive young cluster is a new observational phenomenon.  A current {\it HST} imaging program (PI DJL) will measure the proper motions of these stars if they are high, enabling a full kinematical analysis.  Of course, the origins and destinies of these objects are of considerable interest.  Spin-up by mass transfer in binaries (Langer et al. 2008) and ejection by dynamical processes 
in dense young clusters (Fujii \& Portegies Zwart 2011) are strong hypotheses.  There is evidence for a runaway nature of some gamma-ray burst progenitors, which should also be rapid rotators (Dale \& Davies 2006; Allison et al. 2010).  A possibly related phenomenon in the Galactic cluster Westerlund~2 has recently been discussed by Roman-Lopes et al. (2011).

\subsection{Onfp Spectra}

\begin{figure*}
  \centering
  \includegraphics[angle=90,width=\textwidth]{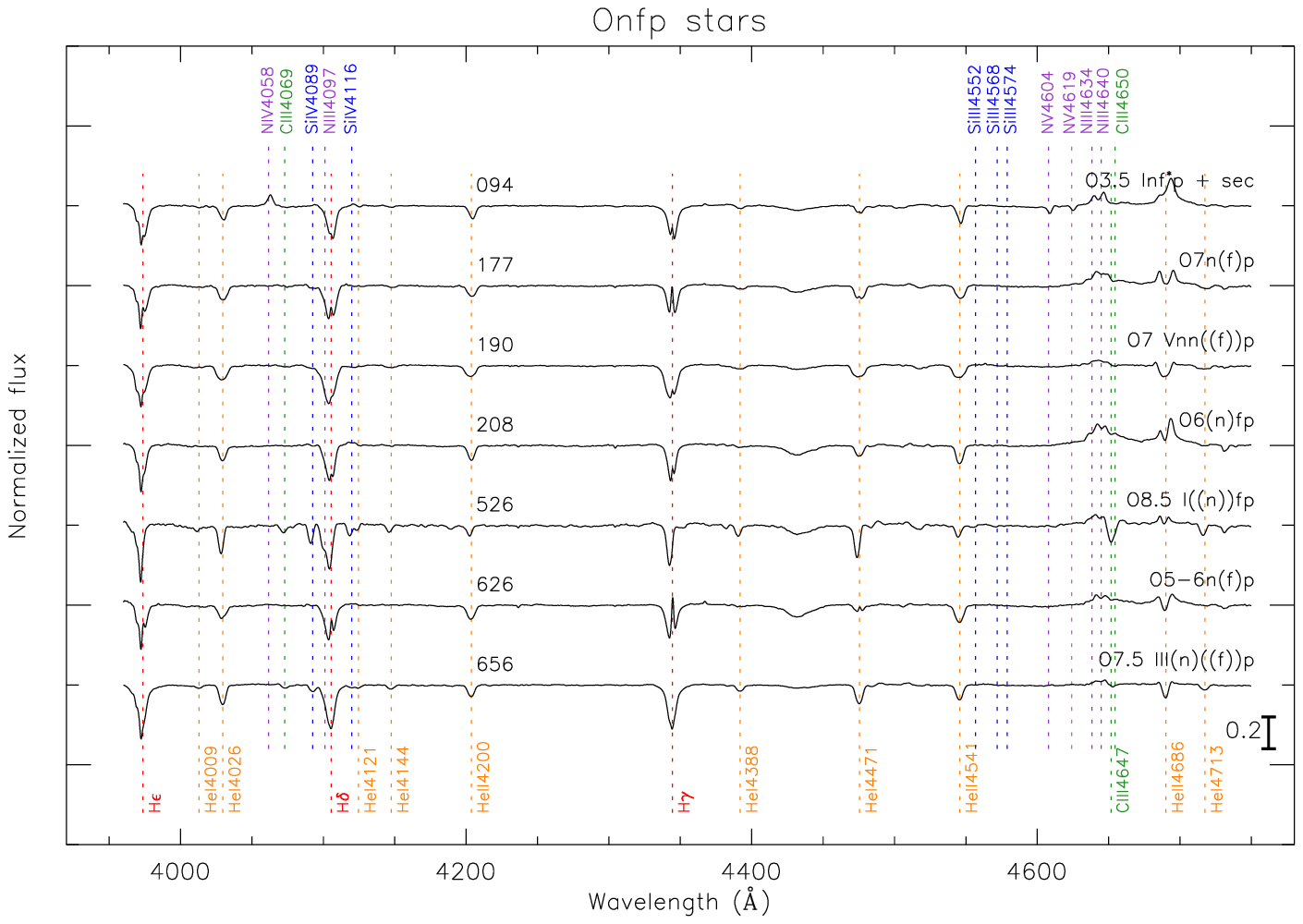}
  \caption{VFTS Onfp spectra.}
  \label{fig: 11}
\end{figure*}

\begin{figure*}
  \centering
  \includegraphics[angle=0,width=\textwidth]{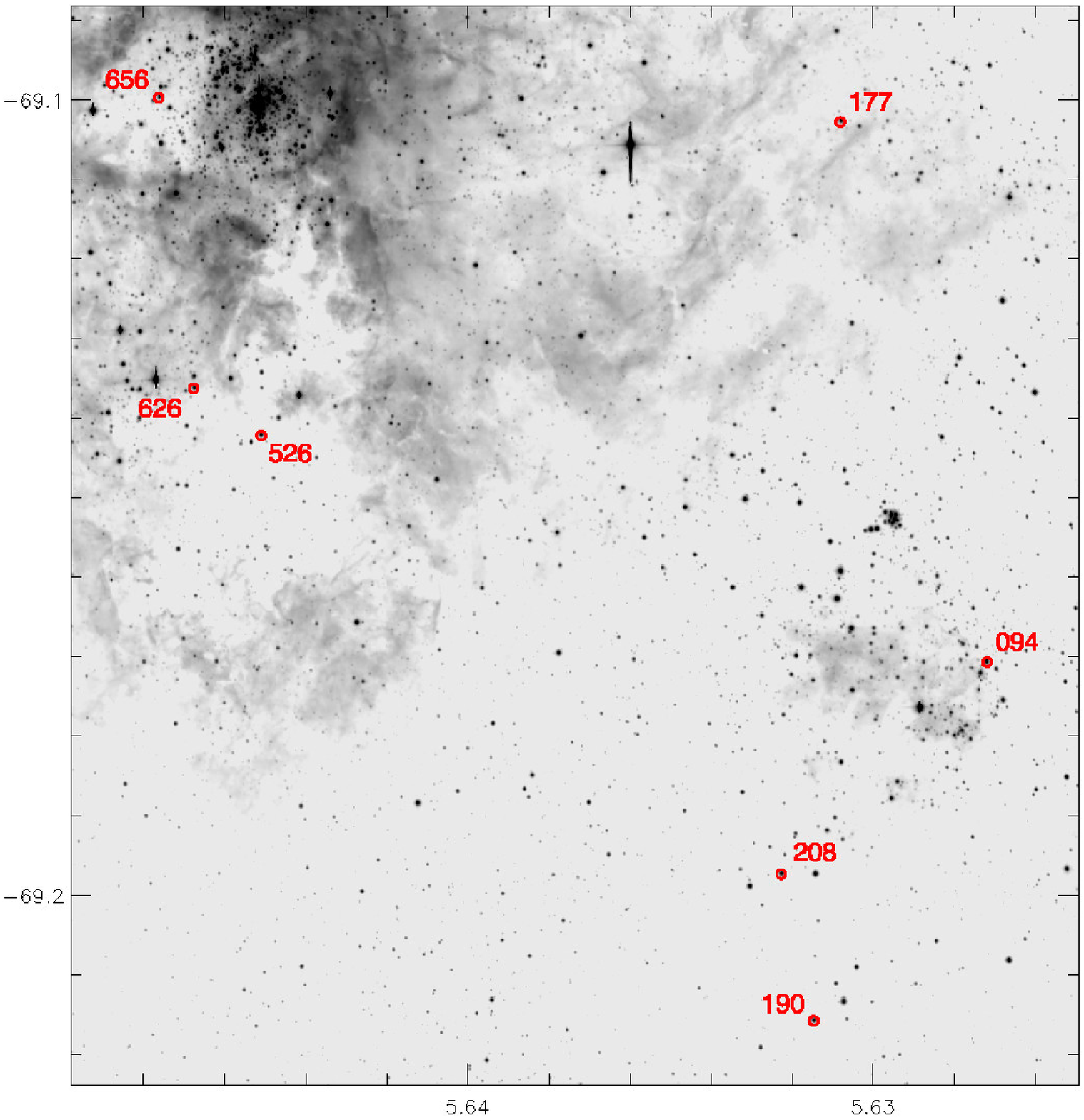}
  \caption{Spatial distribution of Onfp stars.}
  \label{fig: 12}
\end{figure*}

This peculiar category of rapid rotators with centrally reversed \heb\  $\lambda4686$ emission features was recently discussed by Walborn et al. (2010a), and its defining composite line profiles have been modeled in terms of rotating, clumped winds by Hillier et al. (2012).  The VFTS sample includes 7 members of the class (Figure~11), 5 of which are new.  Although the number is small, their spatial distribution (Figure~12) also suggests a runaway nature, which is a property of several previously known members elsewhere.  There is growing evidence for a high fraction of binaries among them, while other apparently single objects have been suggested as possible mergers; spin-up by mass transfer and merging in binary systems could explain their current rapid rotation despite strong stellar winds, which would otherwise be expected to have braked these evolved objects.

VFTS~94 = ST1-28 and VFTS~208 = ST1-93 are spectroscopic binaries and spectrum variables associated with NGC~2060, as illustrated and discussed by Walborn et al. (2010a).  The period of VFTS~94 is 2.35~d while that of VFTS~208 has not yet been determined.  VFTS~177 is far from the ionizing clusters and thus a runaway candidate (Fig.~12).  VFTS~190 is a possible SB and also a candidate runaway from NGC~2060 (Figs.~11 and 12).  VFTS~526 = P925 is a large-amplitude SB1; note the essential confirmation here of the basic spectral type derived from low-S/N data by Walborn \& Blades (1997).  VFTS~626 is another possible SB; this star and VFTS~526 are likely members of the R143 association rather than runaways (Walborn \& Blades 1997).  VFTS~656 is yet another large-amplitude SB1.  The preponderance of binaries among these objects is consistent with prior results for the class.

As noted in the previous section, VFTS~406 was initially classified as a pronounced Onfp object, but the apparent characteristics of the class in this case were subsequently determined to be due to contamination by an adjacent WN spectrum on the detector.  The resemblance of the contaminated spectrum to some Onfp types is remarkable (Fig.~6).  Clearly one must be alert to this insidious effect in multiobject data.  Also, some real Onfp spectra could arise from analogous composite effects in binary systems, possibly including colliding-wind regions as found in the case of the Galactic star HD~152248 by Sana et al. (2001).      

\subsection{ZAMS Candidates}

\begin{figure*}
  \centering
  \includegraphics[angle=90,width=\textwidth]{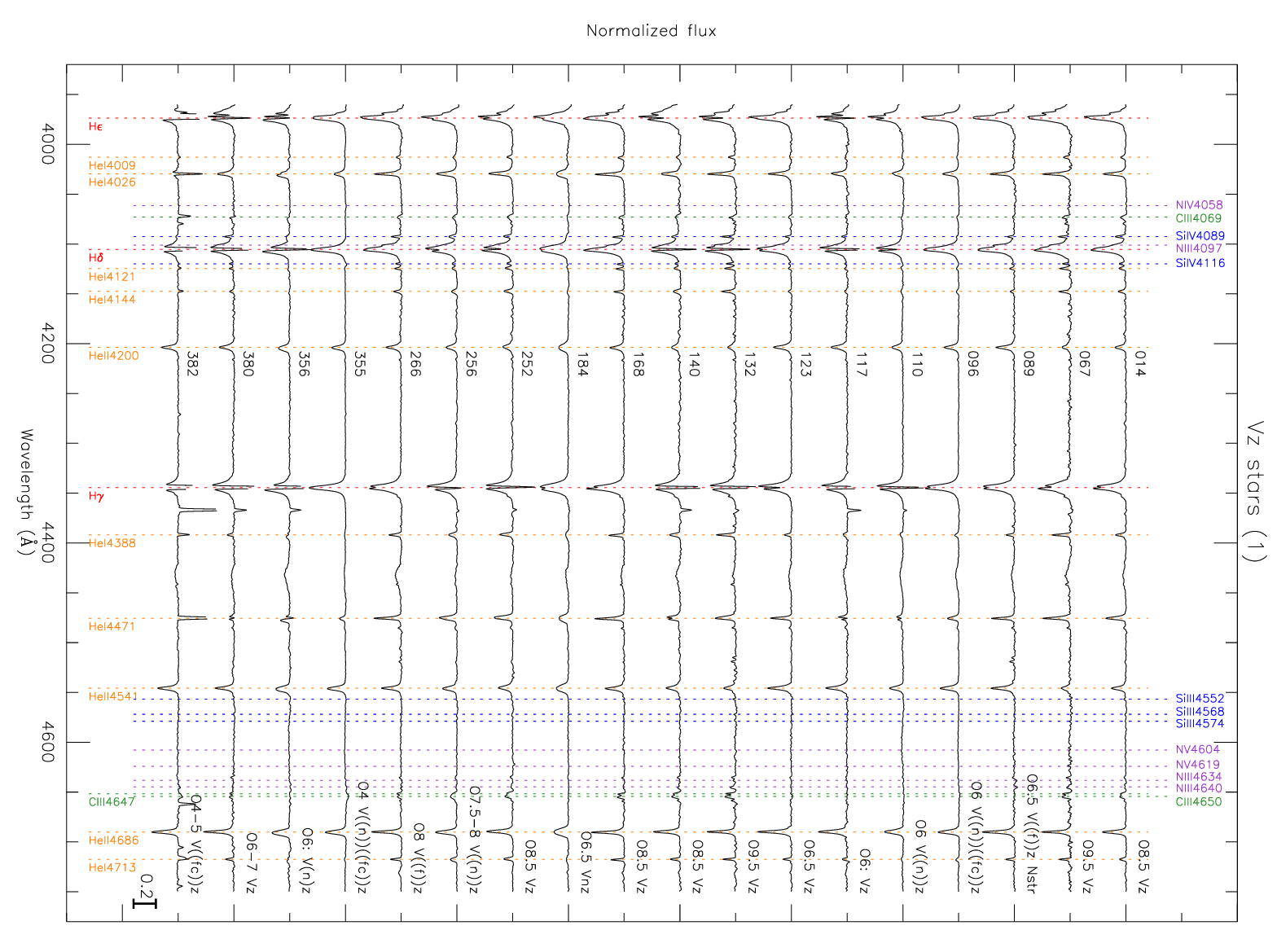}
  \caption{VFTS Vz spectra.}
  \label{fig: 13}
\end{figure*}

\begin{figure*}
  \centering
  \includegraphics[angle=90,width=\textwidth]{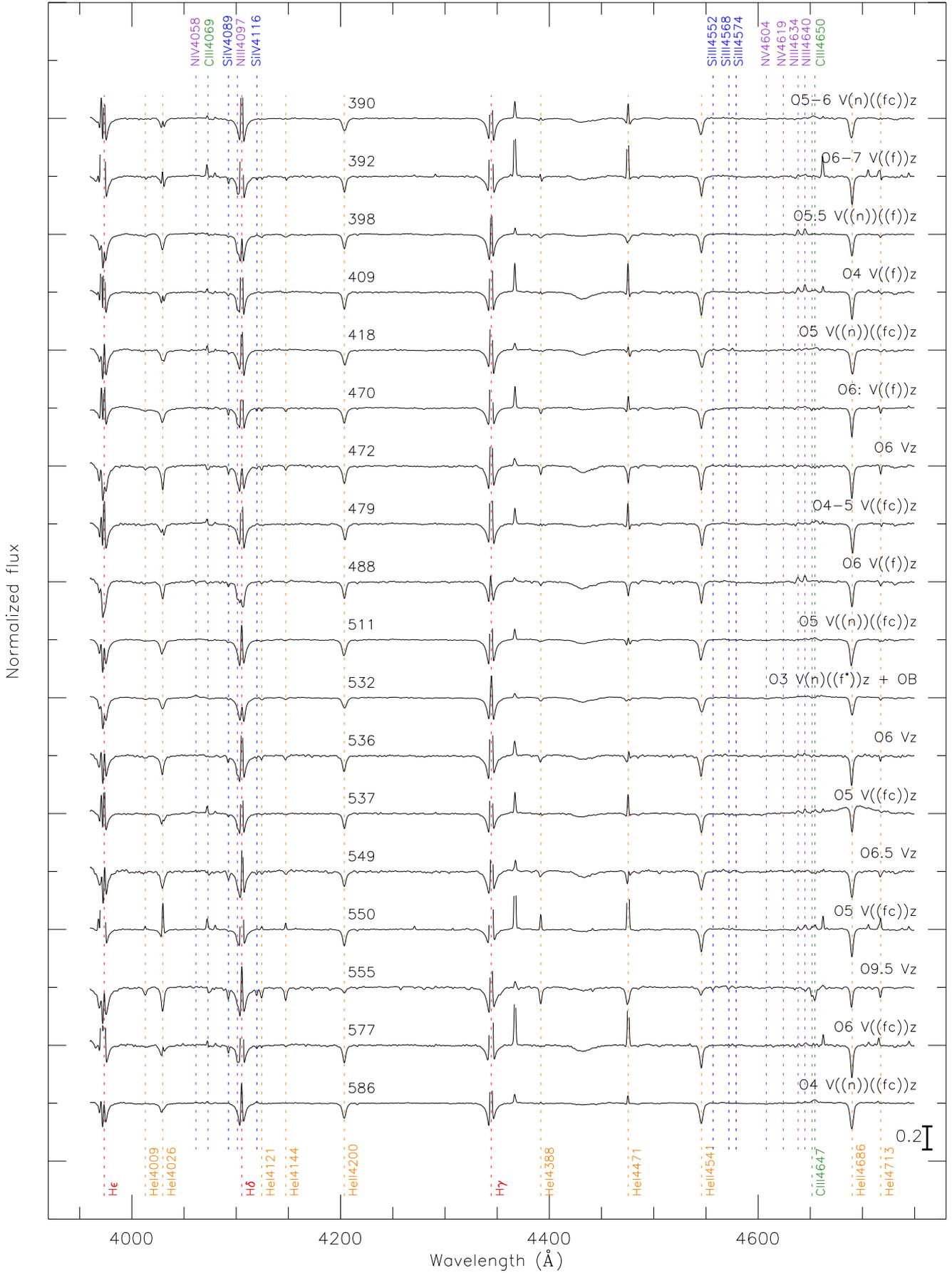}
  \caption{Vz sample continued.  VFTS~537 is contaminated by a nearby emission-line spectrum on the detector.}
  \label{fig: 14}
\end{figure*}

\begin{figure*}
  \centering
  \includegraphics[angle=90,width=\textwidth]{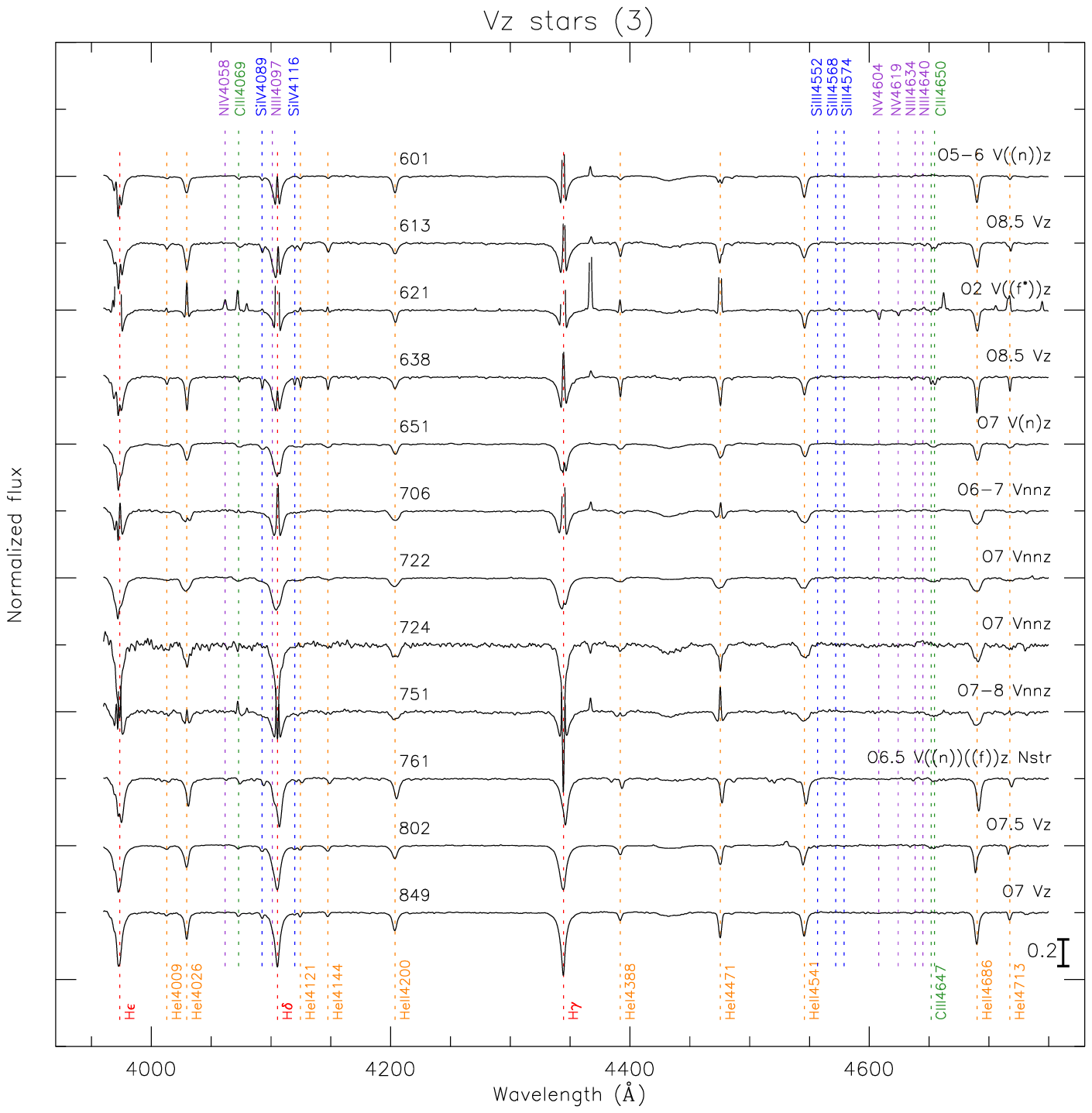}
  \caption{Vz sample continued.  VFTS~724 has oversubtracted nebular emission lines.}
  \label{fig: 15}
\end{figure*}

The defining characteristic of the Vz class, reviewed by Walborn (2009), is 
an \heb~$\lambda4686$ absorption feature stronger than any other He line in the blue-violet region.  Since emission filling in that line, namely the Of effect, has been established as a luminosity indicator in normal O-type spectra, the Vz characteristic has been hypothesized as the inverse effect, i.e., less emission than in even normal class V spectra, corresponding to lower luminosity and extreme youth.  Consistently with that hypothesis, Vz spectra are typically found in very young regions.  Nevertheless, we were surprised to find no fewer than 48 such spectra among our 30~Dor AAA sample of 213 O stars  (Figures~13--15).  

\begin{figure*}
  \centering
  \includegraphics[angle=0,width=\textwidth]{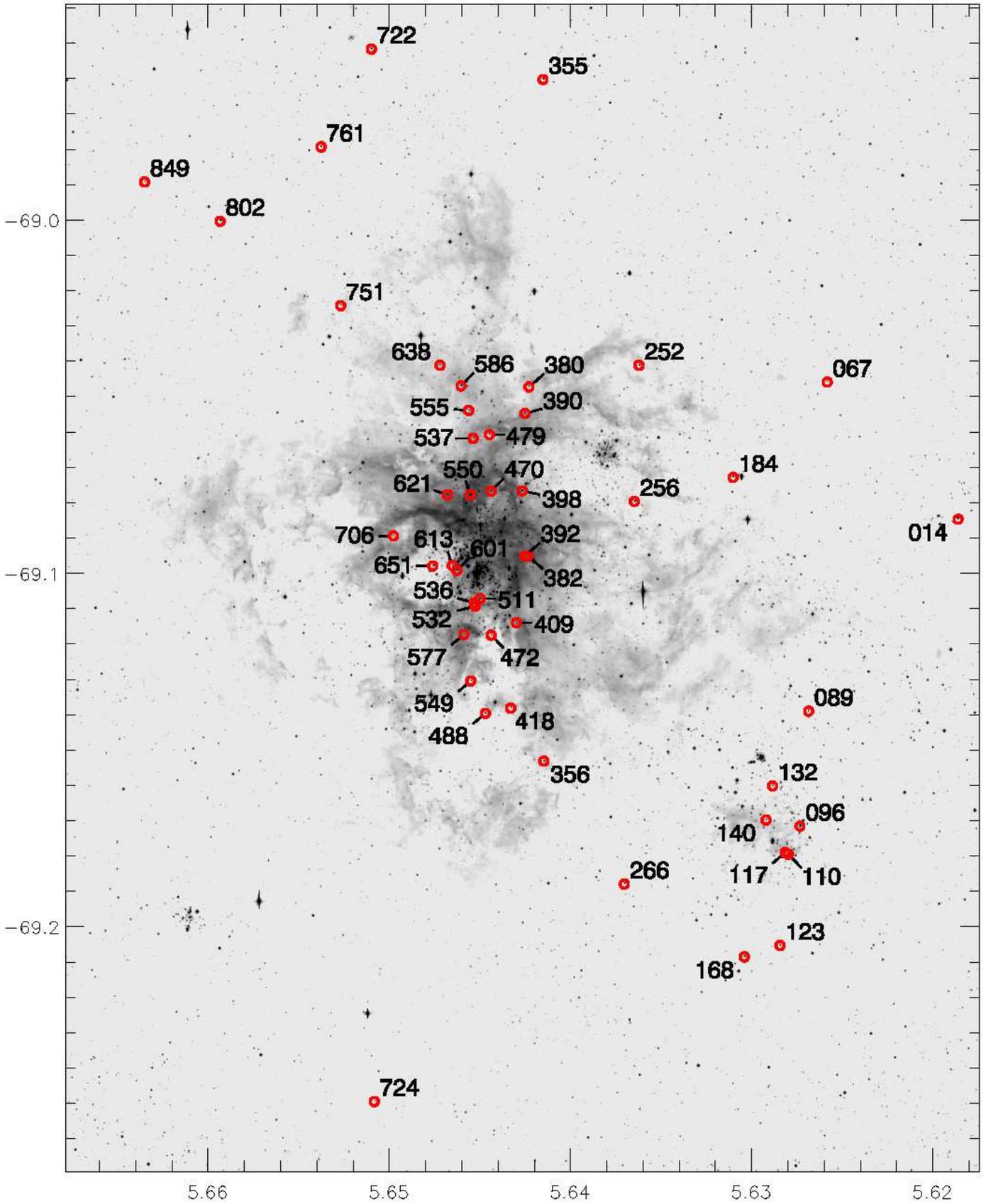}
  \caption{Spatial distribution of the Vz stars.}
  \label{fig: 16}
\end{figure*}

Moreover, their spatial distribution appears essentially inverse to that of the rapid rotators, exhibiting a strong concentration toward the ionizing clusters NGC~2070 and 2060 (Figure~16).  The average projected distance from R136 of 27 non-n/nn Vz stars within 81~pc is $27 \pm 3$~(m.e.)~pc, which is not statistically very different from the result for the rapid rotators given in Section~3.2 above, despite the distinctly contrary impression from inspection of Figures~7 and 16.  On further consideration, however, it is seen that the visual impression arises rather from the concentration of the Vz stars toward the NS {\it axis} of the elongated Tarantula nebulosity.  Their average separation in RA from R136 (i.e., from the NS axis) is $9 \pm  2$~(m.e.)~pc, which is indeed significantly different from the corresponding statistic for the rapid rotators ($30 \pm 5$~(m.e.)~pc). 

In addition, there are a number of Vz objects in an east-west band at the northern edge of the nebula, coinciding with the locations of the WN stars R144, R146, and R147 (Feast et al. 1960) as well as a compact \hb\ region that is a highly luminous {\it Spitzer} source (Walborn et al. 2013); this northern 30~Dor region of recent and current star formation has been newly recognized in this work.  The overall spatial morphology of the Vz objects in 30~Dor provides further evidence for the youth of most of them, i.e., a (near) zero-age main sequence (ZAMS) nature. 

In the discussion of the rapid rotators in Section~3.2, it was noted that 8 of them are also in the Vz class, which at first appears to contradict the different spatial distributions of the two categories with respect to the nebula and ionizing clusters.  However, by comparison of Figures~7 and 16, it can be seen that most of these stars lie at the boundaries of the two distributions, while a few belong to the northern Vz group.  The potential extreme, opposite runaway nature of VFTS~722 and 724 was also discussed in Section~3.2. 

It is noteworthy that two very underluminous O6.5~V((f))z stars, VFTS~089 ($M_V -4.30$) and 761 ($M_V -4.06$) have nitrogen-enhanced spectra.  How this might come about in very young objects raises interesting possibilities, such as chemically homogeneous evolution or binary mass transfer.  HD~12993 in the Galaxy is a similar case (Conti \& Leep 1974; Walborn 1976; Sota et al. 2011).  Further radial-velocity and quantitative analysis of these spectra may well provide important evolutionary insights (Rivero Gonz\'alez et al. 2012b).  

The high binary frequency of the O stars may provide an alternative origin for some Vz spectra: a composite of relatively early and late O-type normal dwarf spectra, in which the two components respectively dominate the \heb\ and \hea\ lines in the temperature-type classification ratios, while both contribute comparably to \heb\ $\lambda4686$, thus producing an apparent mid-O~Vz morphology.  Twenty-eight of the 48 Vz objects are definite or possible SB or visual multiples (Table~2).  Some Galactic systems of this nature are currently under analysis within GOSSS and the associated high-resolution programs OWN (Barb\'a et al. 2010) and IACOB (Sim\'on-D\'{\i}az et al. 2011).

Quantitative analyses of VFTS Vz spectra to investigate the hypothesis of smaller ages and lower luminosities have been carried out by Sab\'{\i}n-Sanjuli\'an et al. (2014).  Their results raise some further interesting questions about the class that will be discussed in the context of the absolute visual magnitudes and HRDs in Section~4 below.  The Of effect is strongly related to the stellar winds.  A key insight of the quantitative analysis is that, because the winds are weaker at lower metallicity, the Vz phenomenon may well be more frequent and have a greater duration in the LMC than in the Galaxy.

\subsection{O~V((fc)) and O~III(fc) Spectra}

\begin{figure*}
  \centering
  \includegraphics[angle=90,width=\textwidth]{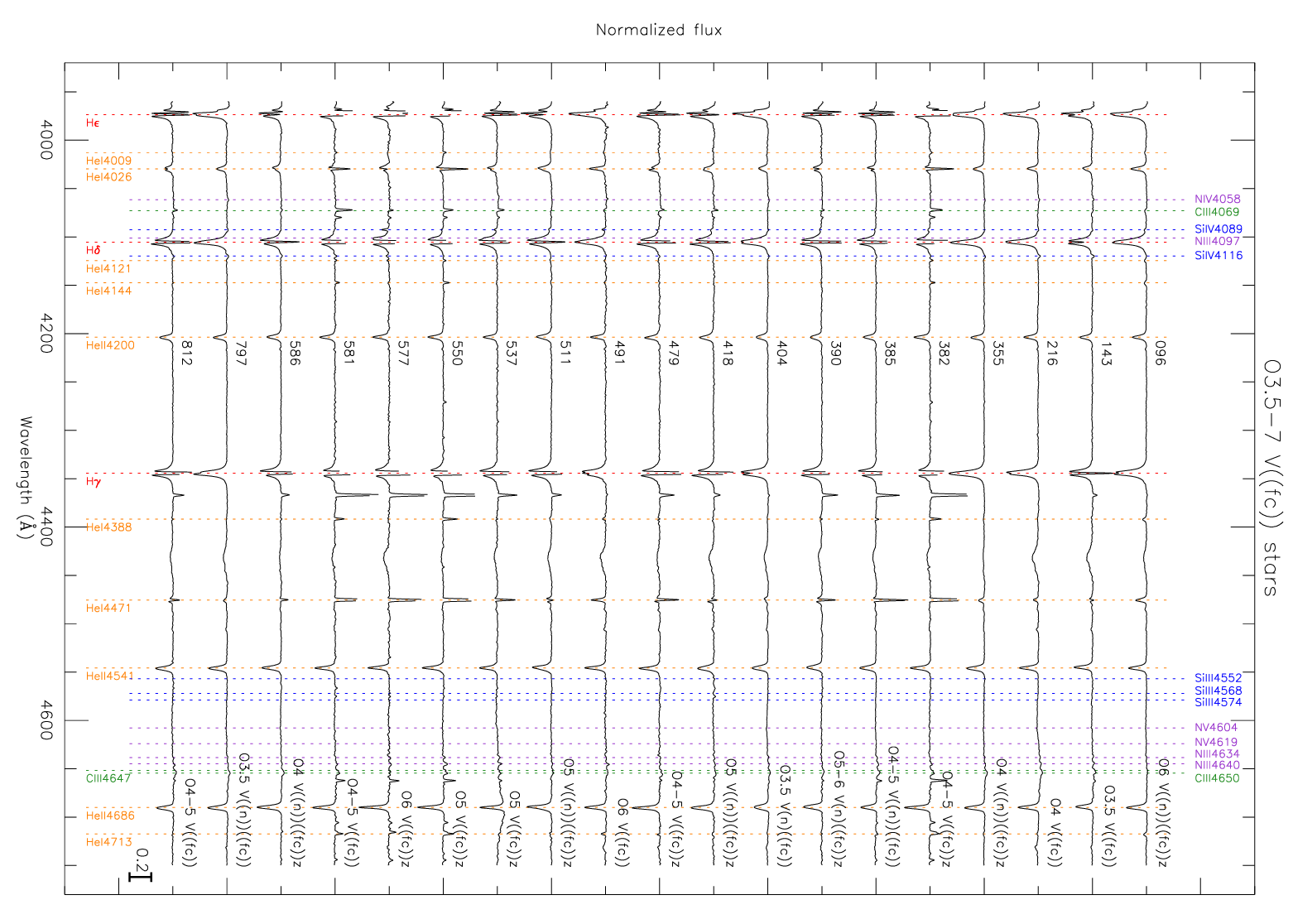}
  \caption{VFTS V((fc)) spectra.  VFTS~537 is contaminated by a nearby emission-line spectrum on the detector.}
  \label{fig: 17}
\end{figure*}

A new category of O-type spectra that emerged from GOSSS was described by Walborn et al. (2010b).  The defining characteristic is \cc\ $\lambda\lambda4647-4650-4652$ emission intensities comparable to those of the classical Of \nc\ $\lambda\lambda4634-4640-4642$ features.  The new qualifier ``fc'' was introduced for this category, with the correlated progression of 
\heb\ $\lambda4686$ from strong absorption, through neutralized, to emission indicated by double, single, and no parentheses, respectively and analogously to the notation for ``f'' spectra.  In the Galaxy, this phenomenon is peaked about spectral type O5 at all luminosity classes; in addition, there is a tendency for it to appear in certain clusters or associations but not others.  Another surprise in the VFTS sample is the presence of 19 class V spectra with this characteristic (Figure~17), although their spectral-type range is broader than in the Galactic counterparts, and the \cc\ emission tends to be {\it stronger} than the \nc.  Two Small Magellanic Cloud (SMC) objects with these characteristics were reported by Walborn et al. (2000); they were classified as OC partly on the basis of supporting evidence from the UV wind profiles, and of course the fc class did not yet exist.  However, relationships to these VFTS objects and the lower metallicities in both galaxies are not excluded.  

\begin{figure*}
  \centering
  \includegraphics[angle=90,width=\textwidth]{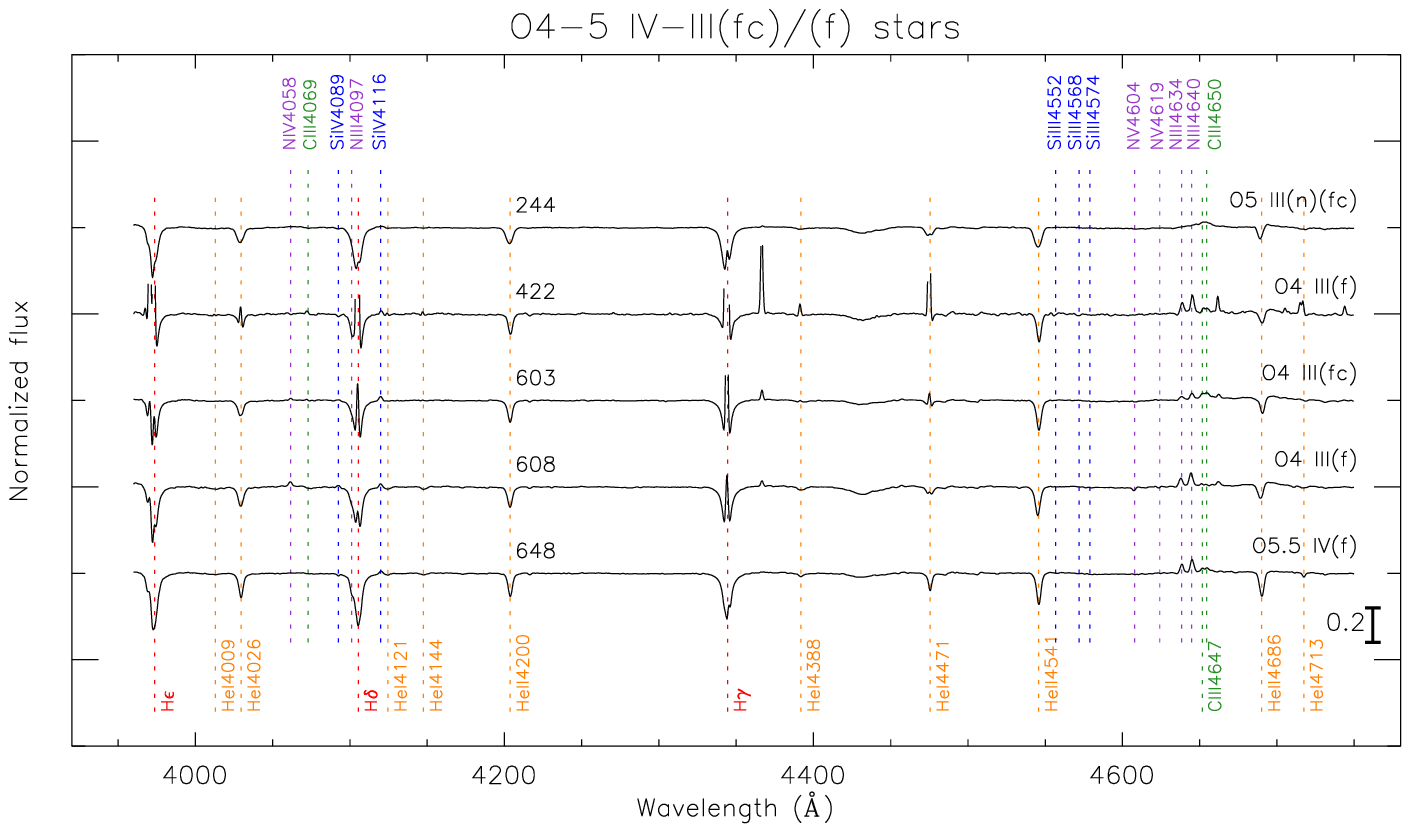}
  \caption{VFTS III(fc) and III(f) spectra.}
  \label{fig: 18}
\end{figure*}

Two early O~III(fc) spectra in the VFTS sample are shown in Figure~18, along with some O~III(f) comparisons.  It should also be noted that the primary component of the R139 = VFTS~527 SB2 system has a strong O6.5~Iafc spectrum (Taylor et al. 2011; Appendix~A here).

\begin{figure*}
  \centering
  \includegraphics[angle=90,width=\textwidth]{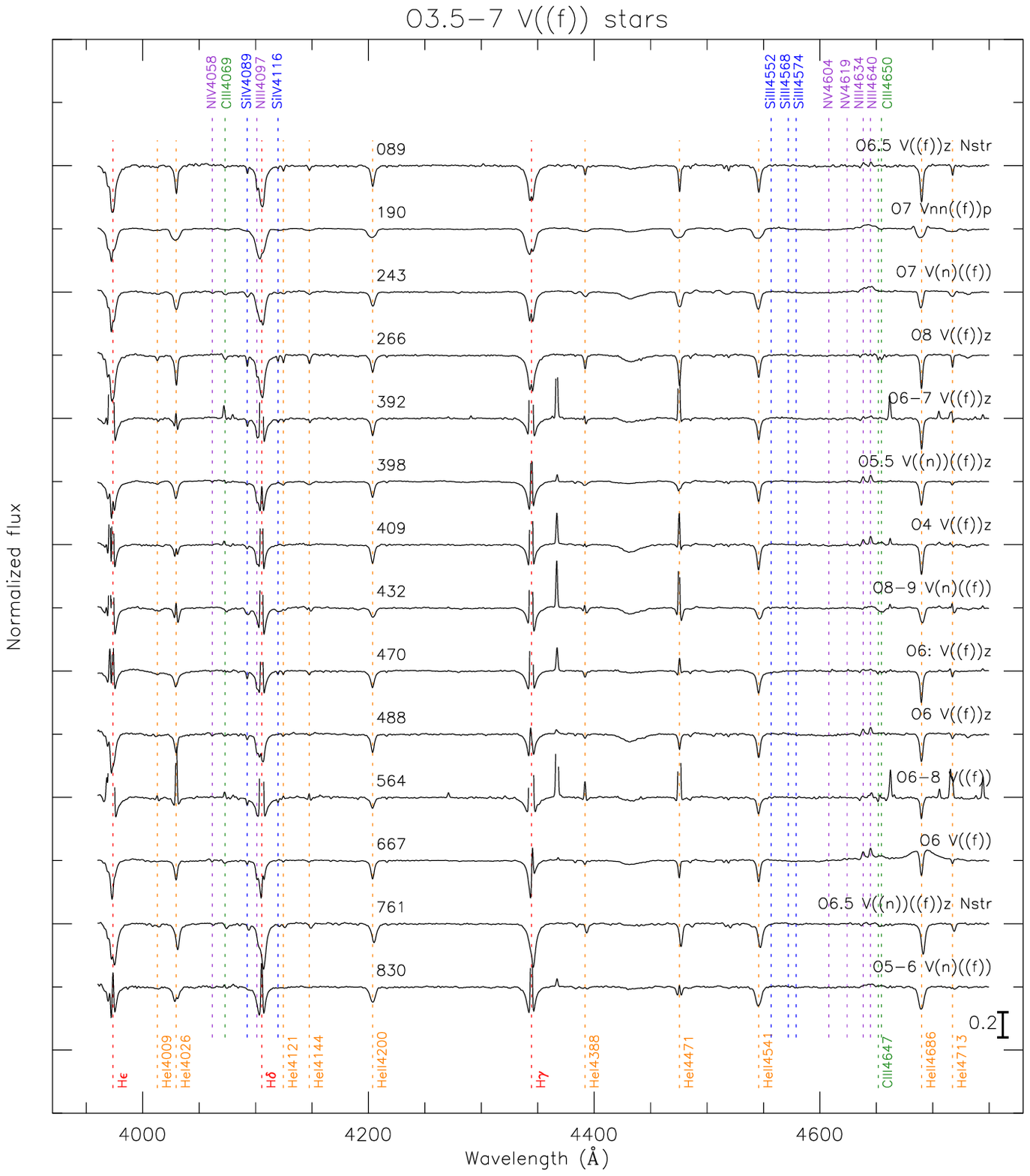}
  \caption{VFTS V((f)) spectra.  VFTS~667 is contaminated by a nearby emission-line spectrum on the detector.}
  \label{fig: 19}
\end{figure*}

\begin{figure*}
  \centering
  \includegraphics[angle=0,width=\textwidth]{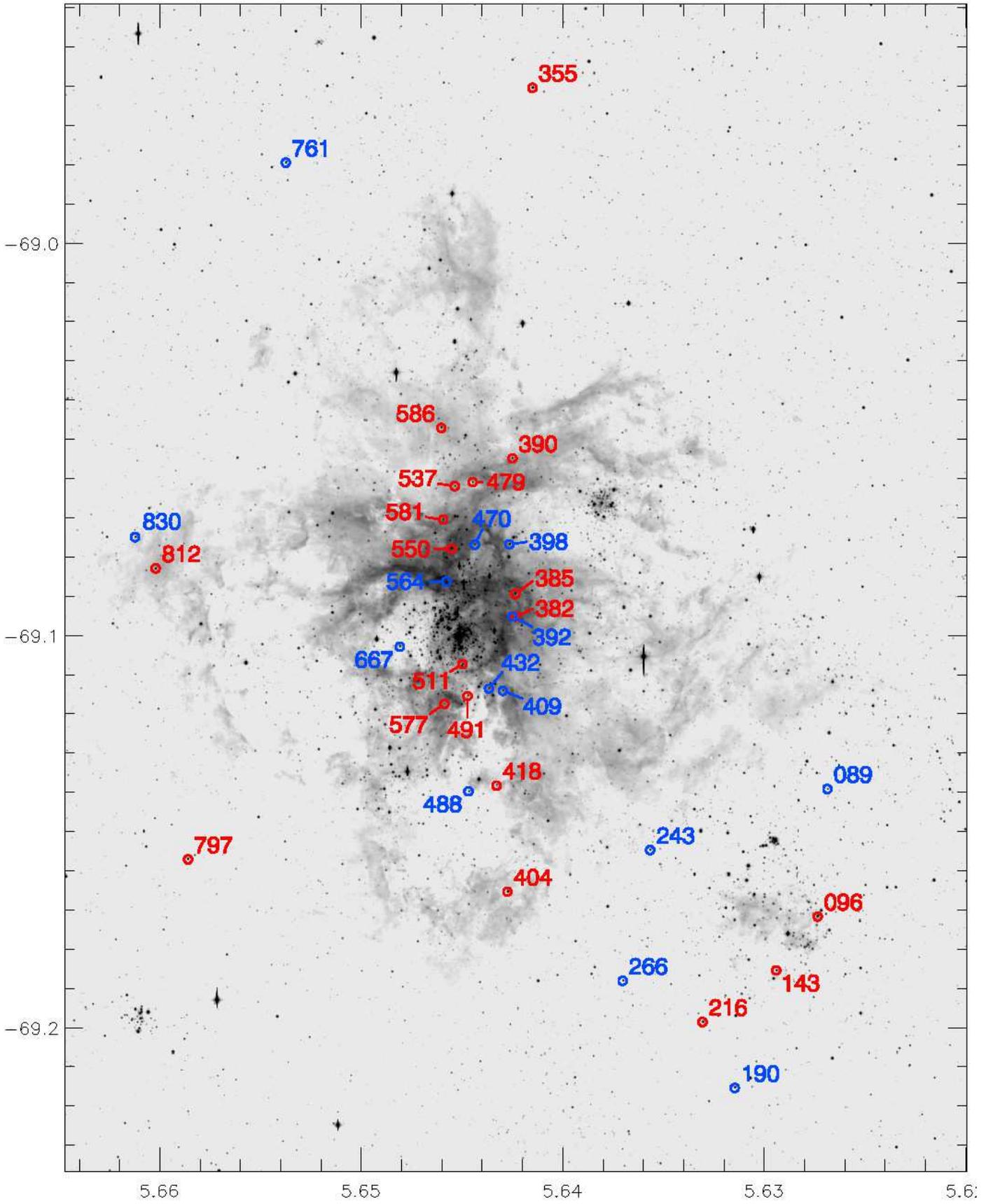}
  \caption{Spatial distributions of V((fc)) (red) and V((f)) (blue) stars.}
  \label{fig: 20}
\end{figure*}

\begin{figure*}
  \centering
  \includegraphics[angle=90,width=\textwidth]{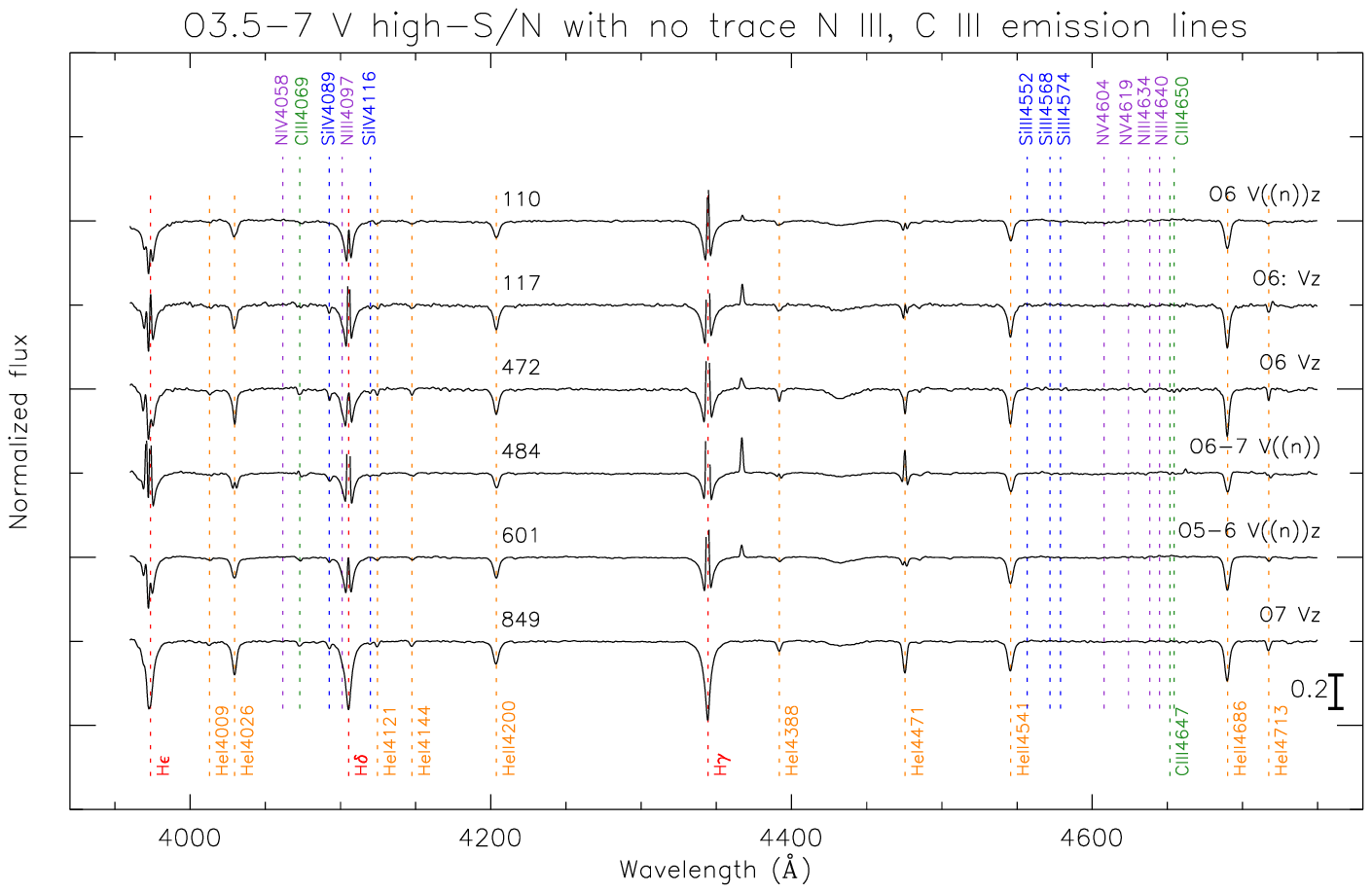}
  \caption{VFTS class V spectra with high S/N but no trace of \nc\ or \cc\ emission lines.}
  \label{fig: 21}
\end{figure*}

Fourteen normal V((f)) spectra are likewise present in our sample (Table~3).  Their spectra are shown in Figure~19, and the space distributions of these two luminosity class V subcategories are shown in Figure~20; they appear to be well mixed.  Moreover, we have found six O~V stars in our sample with very high-quality data but no trace of any emission lines (Table~3; Figure~21).  Clearly these distinct variants of O main-sequence spectral morphology in 30~Doradus present interesting challenges for physical modeling, with significant structural and/or evolutionary insights likely to follow.  The intricacies of N~III and C~III line formation in O-type spectra have recently been rediscussed by Rivero Gonz\'alez et al. (2011) and Martins \& Hillier (2012), respectively; it will be interesting to apply those results to detailed analysis of these spectra, to ascertain whether the model physics is now adequate to explain them or not.  As those papers emphasize, the effects of various parameters such as metallicity on these lines are far from straightforward.   

As listed in Table~2, 14 of the 19 V((fc)), 7 of the 14 V((f)), and 1 of the 6 non-emission stars are definite or possible SB.  There are also 11/19 V((fc)), 8/14 V((f)), and 5/6 V(no-emission) overlaps with the Vz, and some with other categories.  These relationships should be taken into account in further analyses; at a minimum, they are obviously not excluded, and some of them may turn out to be significant beyond that basic fact.  E.g., there appears to be a high fraction of SB among the V((fc)); and there is nearly complete correspondence between no emission and Vz that may be consistent with extreme youth and weak winds (Rivero Gonz\'alez et al. 2011). 

\subsection{Late-ON Spectra}

\begin{figure*}
  \centering
  \includegraphics[angle=0,width=\textwidth]{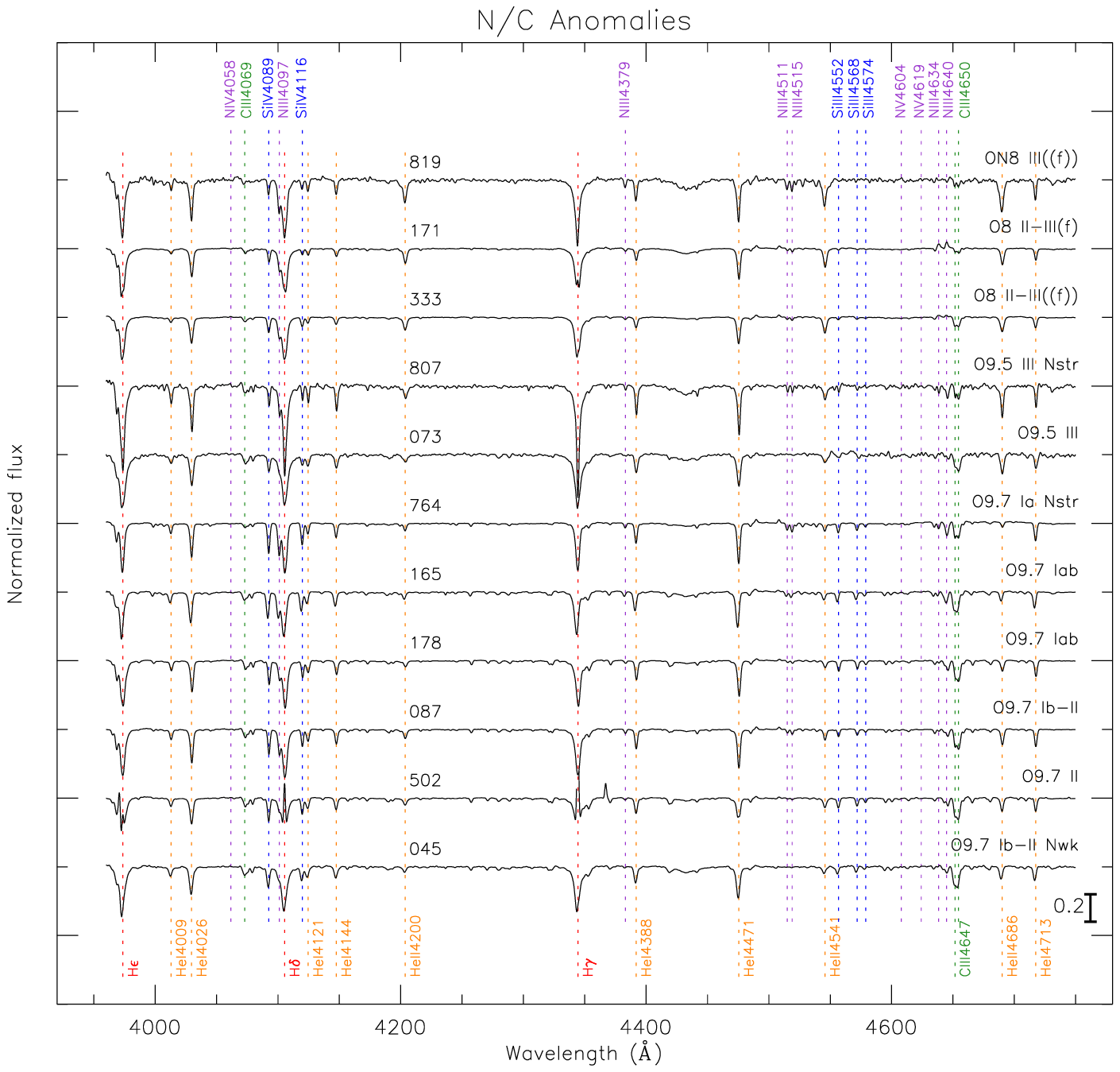}
  \caption{VFTS spectra of luminous late-O stars with nitrogen and carbon anomalies, together with morphologically normal comparison spectra of similar types.}
  \label{fig: 22}
\end{figure*}

A recent discussion of late-O, nitrogen-enhanced (ON) spectra was presented by Walborn et al. (2011); that paper specifically addressed rapid rotators, but it includes complete references to the general topic of N and C anomalies in O-type spectra, as well as possible evolutionary interpretations, that will not be replicated here.  The morphology of these anomalies in luminous late-O spectra is quite distinctive.  We have found only 3 such spectra among the 213 VFTS O-type AAA data (VFTS~764, 807, 819), a result no doubt related to the evolved nature of such objects versus the youth of the majority of this sample.  Note that all three are located near the southeastern edge of the observed field, far from the nebula. In fact, VFTS~764~=~Sk~$-69^{\circ}$~252 belongs to an older, dispersed association far south of 30~Dor; it is the southernmost star marked in Fig.~2.  Also note that despite its giant classification, VFTS~807 has a dwarf luminosity (Section~4 below).  Nevertheless, these spectra are displayed in Figure~22, along with several morphologically normal spectra and one N weak (VFTS~045) of similar types in this sample.  Although not associated with 30~Dor, these rare objects merit detailed physical analysis, since they are relevant to nuclear processing and mixing or mass transfer in OB stars or binary systems of the corresponding mass and age ranges (Rivero Gonz\'alez et al. 2012b).

\subsection{Other Interesting Spectra}

\begin{figure*}
  \centering
  \includegraphics[angle=0,width=\textwidth]{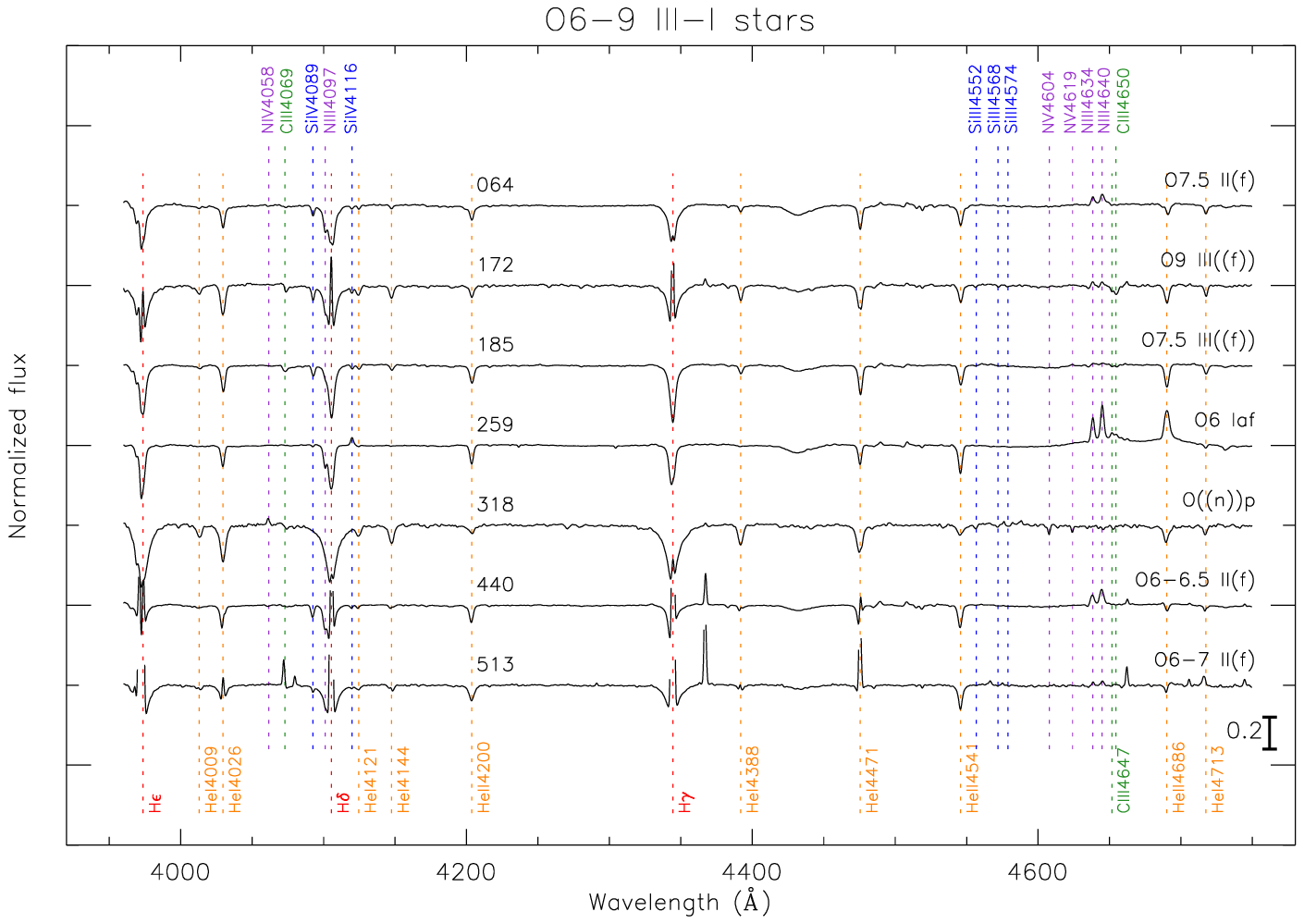}
  \caption{Other interesting VFTS spectra.  }
  \label{fig: 23}
\end{figure*}

Several additional spectra of special interest at types O6--O9 are displayed in Figure~23.  Most of them are not peculiar, but they are of relatively rare, luminous Of types so they are presented here for reference.  Other spectra listed in Table~2 but not discussed and displayed in Section~3 will be posted on the VFTS public web page, as will those in Appendix A.

However, the spectrum of VFTS~318 is highly peculiar, even unique to
date, so it is presented here even though it cannot be consistently
classified and appears in the BBB list of Appendix A for that reason.
The \heb/\hea\ line ratios would imply a type of O9.5~III, but the
absence of any Si~IV or other metal lines in the vicinity of H$\delta$,
together with the strong Balmer wings, produces the appearance of a B2~V spectrum there.  Most remarkably, though, the spectrum also contains a narrow emission line of \nd~$\lambda$4058 and narrow absorption lines of \ne~$\lambda\lambda$4604--4620, which if stellar would be characteristic of an O2 spectrum!  The object is located well away from any clusters, indicating that it is not very young, unless it were a runaway.  Moreover, it has a remarkably faint $M_V$ of only $-2.6$, corresponding to a normal B1-1.5 V star (Walborn 1972), so all of the O-type features must have an anomalous origin. The very high-ionization lines may originate in dilute material, perhaps compatible with their extremely narrow profiles; the possibility that they are excited by EUV or X-rays should also be considered.  Alternatively, they would have to originate in a subluminous, very hot star.  This object may be a result of advanced binary evolution; compatible with that possibility, radial-velocity variations indicate an SB.   

\section{Luminosities and HRDs}

As indicated in Table~2, absolute visual magnitudes and physical parameters of most AAA stars have been derived; they are presented as Hertzsprung-Russell Diagrams (HRDs) in this Section.  

The $M_V$ and luminosity values were calculated with the Bayesian photometry 
package CHORIZOS (Ma\'{\i}z Apell\'aniz 2004) and the following inputs:  (a) the $BVJHK$ photometry of Paper I; (b) effective temperatures derived from the spectral types of this paper and the calibration of Martins et al. (2005; to spectral type O3), shifted upward by 1000~K to account for the different metallicities of the Milky Way and the LMC (Rivero Gonz\'alez et al. 2012a; Doran et al. 2013; whence the values for O2); and (c) a distance modulus of 18.5 as adopted by VFTS. The LMC stellar grid of Ma\'{\i}z Apell\'aniz (2013a) was used, leaving three free parameters, which were determined for each star individually: (photometric) luminosity class, amount of extinction ($E(4405-5495)$), and type of extinction ($R_{5495}$).  A wide range of the $R$ values was found, the majority lying between 3 and 5 among the individual sightlines, with a few outliers reaching nearly 7.  The extinction laws have been derived from the present spectral types, new WFC3 $UBVI$ photometry, and Paper I $JHK$ photometry (Ma\'{\i}z Apell\'aniz 2013b; Ma\'{\i}z Apell\'aniz et al.  2014). In a future paper of the VFTS series, we shall analyze the spatial distributions of $E(4405-5495)$ and $R_{5495}$, and their relationships with that of the Diffuse Interstellar Bands (van Loon et al. 2013).

\subsection{Observational HRD}

\begin{figure*}
  \centering
  \includegraphics[angle=0,width=\textwidth]{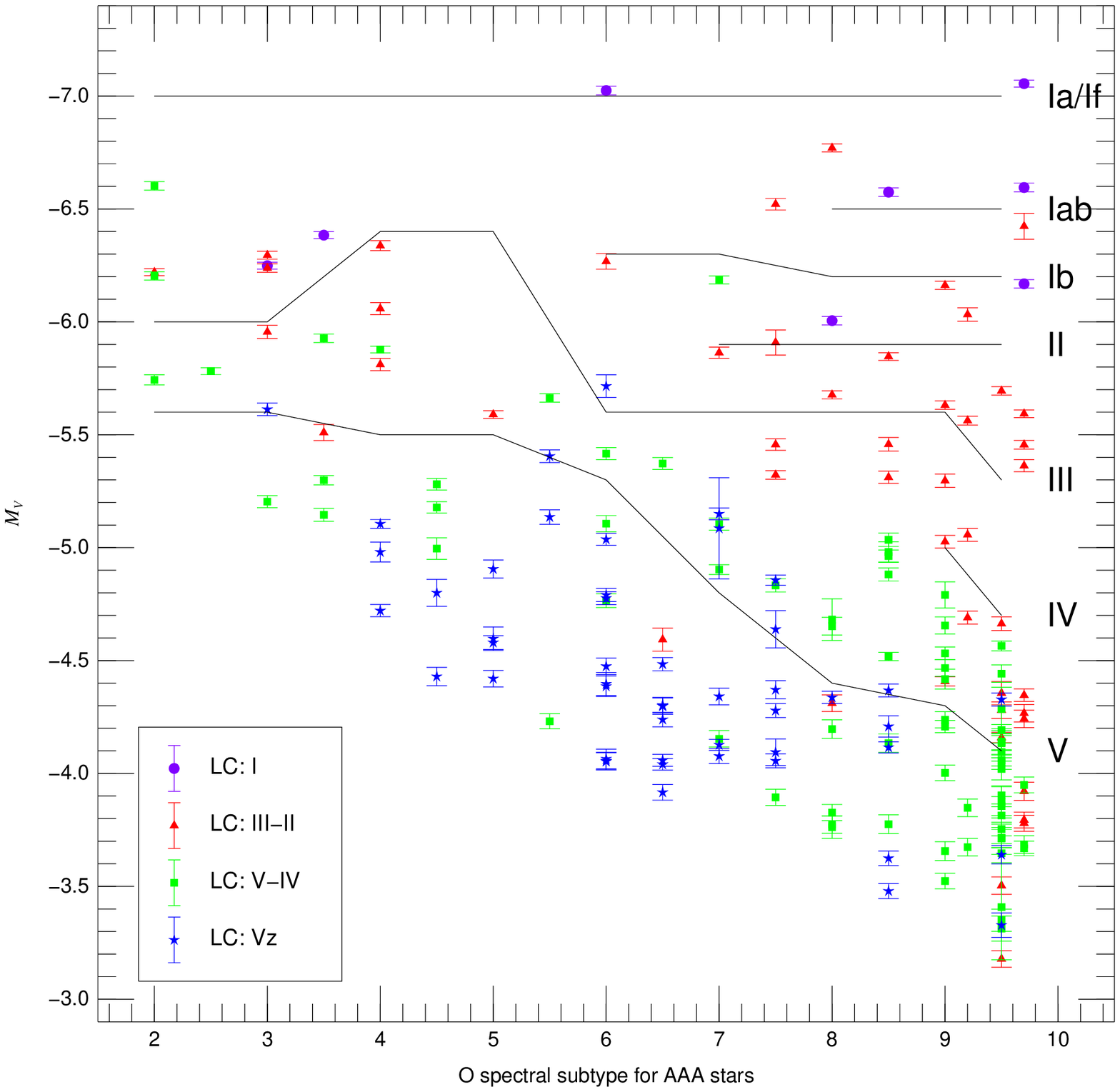}
  \caption{Observational HRD, with the absolute magnitude calibration of Walborn (1973) superimposed.}
  \label{fig: 24}
\end{figure*}

Figure~24 presents first the observational HRD (i.e., $M_V$ vs. spectral type) for all 180 AAA stars with derived parameters.  The $M_V$ calibration of Walborn (1973), supplemented with values for spectral types O2--O3 from Walborn et al. (2002a), is superimposed.  There is reasonable agreement between the calibration and the individual magnitudes, with some scatter and exceptions.  Individual magnitudes that are too bright are likely to be multiple systems; a case in point is VFTS~585 = Parker~1231 (spectral type O7~V(n) but $M_V -6.19$), with equal components separated by only 0\farcs09 (Walborn et al. 2002b); it is in the AAA list and this HRD because it was missed in the initial survey of the WFC3 images, and to serve as a reminder of what remains below that limit.  As is well known, the averaged luminosity class V calibration lies well above the ZAMS for O-type stars.  In particular, most of the Vz stars fall below the class V line--as do a fair number of normal class V types, especially at the earliest and latest spectral types.  This distribution is further discussed in the next Section.

Objects classified here as late-O giants but which fall among the dwarfs in the HRD are an important issue that is discussed in detail in the Appendix, in the context of a large number of BBB classifications with that property.  An interrelated combination of classification and physical factors is likely involved.  Some AAA objects in this category are seen in Fig.~24.  The 5 worst cases, falling below the class V calibration line, are VFTS~210, 235, 574, 631, and 843; 574 and 843 are among the rapid rotators with giant classifications discussed in Section~3.2.  There are also 7 cases just above the class V line, VFTS~012, 091, 399 (these 3 are also rapid rotators, completing that subset), 688, 753, 769, and 807. It is significant that VFTS~631 and 688 have remarks about weak \sid\ lines in Table~2; VFTS~569 and 620 have related remarks, but are not in the HRD because no parameters could be derived for them.  These 14 objects must be added to the related 39 BBB cases discussed in the Appendix.

Scatter between spectroscopic luminosity classifications of the OB stars and their derived absolute magnitudes is well known and has been often discussed, e.g., by Walborn (1973) who showed that the typical internal spread within associations is 0.5 mag.  Scatter within 30~Dor was discussed by Walborn \& Blades (1997).  Systematic effects due to uncertainties in association distances and memberships were considered by Walborn (2002).  In addition to the (physical) sources of random discrepancies noted in the previous paragraphs, differing reddening laws even within the same young region must be addressed, as we have done here.  A more recently recognized issue is the effect of initial rotational velocity on evolutionary tracks; some striking possible examples in the SMC were found by Walborn et al. (2000) and analyzed by Hillier et al. (2003).  Even more recently, the fundamental effects of binary evolution are being elucidated (Sana et al. 2012; de Mink et al. 2013).  Nevertheless, two-dimensional spectral classification remains an important method to chart the fundamental parameters of the majority of the stars, while discovering new categories and issues for further investigation.   

\subsection{Theoretical HRD}

\begin{figure*}
  \centering
  \includegraphics[angle=0,width=\textwidth]{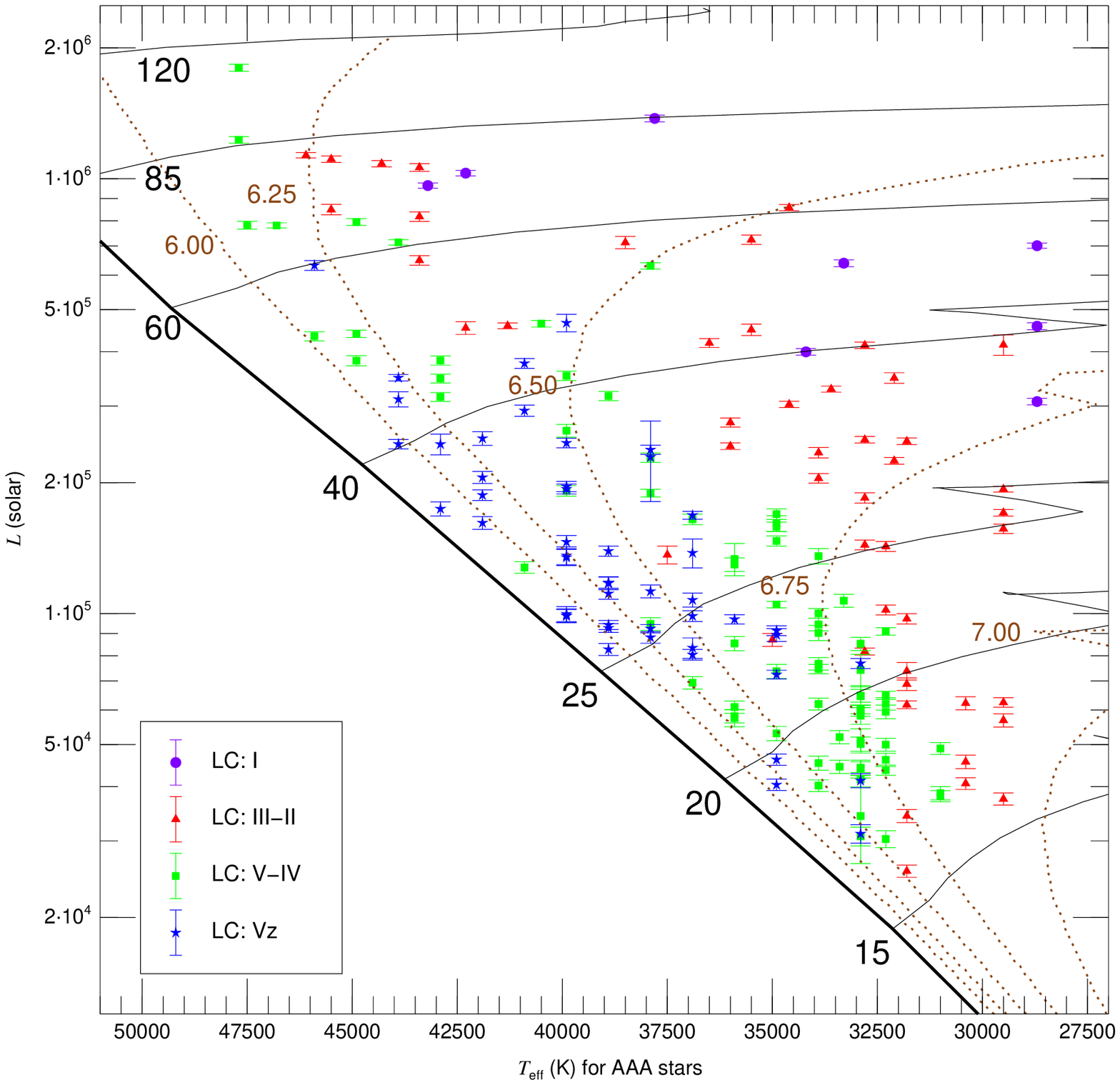}
  \caption{Theoretical HRD of the full field, with evolutionary tracks specified by initial mass in solar units, and isochrones by log age.}
  \label{fig: 25}
\end{figure*}

Figure~25 presents the theoretical (bolometric luminosity vs. effective temperature) HRD for the same stars as Fig.~24.  Here the ZAMS, nonrotating evolutionary tracks, and isochrones for an LMC metallicity of 0.4~$Z_{\sun}$ of Lejeune \& Schaerer (2001) are also plotted.  The isochrones are for ages of 1.0, 1.8, 3.2, 5.6, and 10.0~Myr.  As previously known (e.g., Walborn \& Blades 1997; Selman et al. 1999; Sabbi et al. 2012), a wide range of ages is present in the field of 30~Dor, even in the innermost region near R136. Nevertheless, a gap can be discerned between a sequence with ages 1--2~Myr and the remaining essentially continuous distribution older than about 3~Myr.  All the stars are younger than 10~Myr, consistent with the O-type selection.  (Trial HRDs were constructed only for stars within 80~pc of R136 in projection, but they appear essentially as sparser versions of the full-sample HRDs and so are not presented here.) 

In view of the gap around that age in at least the upper HRD, stars with ages $\leq2.5$~Myr have been denoted by a ``y'' (for ``young'') following the numeral in the HRD column of Table~2; this subgroup comprises all 32 analyzed stars with $T_{\rm eff} \geq 41,900$~K plus 23 cooler ones.  Their correlations with some of the special categories discussed above are considered here.  Most strikingly, 16/18 analyzed V((fc)) stars are in the young group.  This result suggests a strong hypothesis that the characteristic may arise from a combination of the small LMC N/C abundance ratio and stellar youth, i.e., minimal internal mixing of processed material as yet.  (It should be noted that the Galactic fc category discussed by Walborn et al. 2010b may have a different physical origin; some morphological distinctions between the Galactic and these 30~Dor spectra were pointed out in Section~3.5.)  Interestingly, the two V((f))z stars with strong nitrogen lines mentioned in Section 3.4 are also in the young group; clearly their evolutionary path or status must be different from that of the V((fc)) objects.

Of the analyzed Vn/nn/nnn objects, 7/17 are in the young group, including the Vnnn VFTS~285, providing independent evidence of a runaway nature for those that have peripheral locations, as does the latter.  Regarding the two Vnnz objects at the NS extremes of the field pointed out in Sections 3.2 and 3.4, VFTS~722 is in the young group while its opposite counterpart VFTS~724 just misses at about 3~Myr.  In 2.5~Myr they would reach their current locations on the sky with respect to R136 at tangential speeds of 53~km~sec$^{-1}$.

A further interesting feature of Fig.~25 is the distribution of the Vz stars, in comparison with that of their class V counterparts.  First, it is seen that the majority of them are in fact among the youngest objects; 27/45 are in the young group just discussed, while only 10 of them lie significantly beyond the 3.2~Myr isochrone; at least some of the latter may be composite spectra simulating the Vz morphology, or otherwise multiple.  However, a fair number of class V stars are also present in the younger domain--but predominantly at the highest and lowest temperatures (or earliest and latest O spectral types).  The intermediate temperatures at the smaller ages are dominated by the Vz class; indeed, the {\it only} class V-IV (green square) point below the 1.0~Myr isochrone is VFTS~830, which is flagged as ``z?'' in Table~2.  These results are consistent with the hypothesis of relatively small ages for that class, but with some complications that both expand knowledge and require explanation.    

This peculiar relative $T_{\rm eff}$/spectral-type distribution of the classes Vz and V spectra was first discovered and extensively discussed by Sab\'{\i}n-Sanjuli\'an et al. (2014) in their astrophysical analysis of the same sample.  In addition, they have shown that this $T_{\rm eff}$ distribution may be explained by \hea\ and \heb\ line-formation effects that tend to favor the Vz morphology at intermediate temperatures in 30~Dor; and that, to within the accuracy of the determinations, there is no separation in log~g between the Vz and V stars.  Finally, they offer the insight that the Vz phenomenon, which is related to reduced filling in of the \heb~$\lambda4686$ absorption by wind emission, is expected to last longer and consequently be more frequent at the lower LMC metallicity with respect to the Galaxy, since the wind strengths depend on the latter.  That may well be the reason for the surprisingly large number of Vz spectra in 30~Dor\footnote{The same effect may be an additional source of scatter in the present HRDs at higher luminosities, i.e., some objects may have luminosities underpredicted by their classifications because of weaker wind emission filling at \lam4686.  Such a phenomenon would of course go in the opposite sense to the ``subluminous giants'' discussed elsewhere in this paper, in which the \lam4686 absorption is weaker than expected for their actual luminosities.}.

Several factors in addition to the line-formation effects likely contribute to the relative dearth of Vz stars at the earliest and latest spectral types.  At the former, winds are stronger and evolution more rapid, so that the Vz phase may well be shorter than at later types.  It should also be noted that some early Vz spectra had to be dropped from the quantitative analysis, because of difficulties caused by the strong nebular contamination in their spectra.  Factors relevant to the paucity of class Vz at late O spectral types may be the rapid decline of the Of effect at those types, and possibly longer embedded phases in the younger population that would push them below the VFTS limit because of higher extinction.  Indeed, there is only one late-O star below the 1.8~Myr isochrone in Fig.~25, and it is a Vz (VFTS~638).  On the other hand, the relatively larger number of class V at late types must also be an effect of the larger range of ages they represent; indeed, most of them lie {\it above} the 3.2~Myr isochrone. 

%From the most basic equations of stellar structure, it can be inferred that for classes Vz and V stars of the same temperature (postulated), same gravity (observed), and lower luminosity for the former (generally observed), the radius and mass of the Vz will be smaller.  That is, the class V star has evolved from a more massive and hotter Vz, and the latter class is younger.  

It will be essential to analyze UV spectra of Vz and associated class V objects, to determine their mass-loss rates with higher accuracy than provided by differences in \heb~$\lambda4686$ (or H$\alpha$) absorption strengths alone.  Attention is drawn to the extreme SMC OC6~Vz object NGC~346-113 for which both the optical and UV spectra are shown by Walborn et al. (2000).  This star likely has the weakest wind known at that spectral type.  Unfortunately, no UV spectra are yet available for the Vz sample in 30~Doradus, although {\it HST} is quite capable of obtaining them.

%\subsection{Tarantula Nebula Observational and Theoretical HRDs}

%\begin{figure*}
%  \centering
%  \includegraphics[angle=0,width=\textwidth]{fighrd80obs.ps}
%  \caption{Observational HRD as in Fig.~24, but only for stars within projected distances from R136 less than 80~pc.}
%  \label{fig: 26}
%\end{figure*}

%\begin{figure*}
%  \centering
%  \includegraphics[angle=0,width=\textwidth]{fighrd80theo.ps}
%  \caption{Theoretical HRD as in Fig.~25, but only for stars within projected distances from R136 less than 80~pc.}
%  \label{fig: 27}
%\end{figure*}

%Figures~26 and 27 are analogous HRDs reduced to 117 stars within a projected distance of 80~pc from R136, to ascertain whether they might show improved discrimination of age sequences.  In general, that is not the case; overall they appear as more sparsely populated versions of Figs.~24 and 25, respectively, although a high fraction of the older stars have disappeared, as expected.  This result is not too surprising, since older stars and clusters are known to exist even very near to R136 (Selman et al. 1999; Sabbi et al. 2012).  It is pleasing albeit not obviously significant that half of the deviant Vz and most of the discrepant late-O ``giants'' discussed above have disappeared as well.

\section{X-Ray Sources}

With observations from the {\it Chandra X-Ray Observatory}, Townsley et al. (2006) presented a study of point sources within the 30~Doradus Nebula.  It is of interest to compare those results with the VFTS spectral classifications.
The strongest X-ray point sources in 30~Dor are the composite WN6~+~WC4 system R140N and the WN5h R136c, a bright member of the massive central core cluster.  Other WR stars in the field are also among the stronger sources, due either to their massive winds or to binary-system mechanisms, as will be further discussed elsewhere.  However, it is interesting that a very hard, likely nonthermal, and heavily extincted source apparently associated with the WN7h~+~OB system R135 could instead arise from the adjacent {\it Spitzer} source S3 discussed by Walborn et al. (2013). 

As listed in the Tables 2 and A.1 comments, {\it Chandra} sources have been identified with 12 VFTS AAA O-type classifications and 6 BBB.  The AAA sources include seven of types O2--O5 (VFTS~267, 468, 479, 506, 512, 601, 830), which may be reasonably hypothesized to have embedded stellar-wind shock origins, although five of them are also SB1 or composite.  VFTS~333, 513, and 664 have bright giant luminosity classes, signifying \heb~$\lambda4686$ significantly filled in by emission, which could however also be related to the X-rays.  VFTS~564 is classified as a possible Oe star.

However, the most intriguing X-ray source in the sample, VFTS~399, has the prosaic spectral type of O9~IIIn, but with a dwarf $M_V$ of only $-4.4$, i.e., it is in the group of subluminous ``giants'' discussed in Sections~3.2, 4.1, and A.2, with spectra displayed in Fig.~10.  It is located in isolation a tenth of a degree south of R136.  In one of three {\it Chandra} observations (Townsley 2012), it underwent a large flare lasting of order 10~ks, during which it became the third brightest source in 30~Dor, at about an order of magnitude above its quiescent level.  The other observation had a flux intermediate between the two extremes.  The source is very hard and likely nonthermal.  Immediately upon learning of this behavior, we found very broad H$\alpha$ emission in two VFTS observations and possible evidence for SB2 line profiles (Table~2).  There appears to be a trend of order hundreds of days in VISTA $K$ data, but no period could be discerned in OGLE $I$. Further investigation of these remarkable data is ongoing and will be reported elsewhere.                 
  
Four of the BBB sources are SB2 (VFTS~217, 445, 500, 527), while the other two (455, 579) are a large-amplitude SB1 and SB?, respectively.  VFTS~527~=~R139 is the massive O6.5~Iafc~+~O6~Iaf system discussed by Taylor et al. (2011).  Colliding winds provide a likely hypothesis for the origin of the X-rays from these systems.

\section{Summary and Outlook}

This work provides basic spectroscopic descriptive information to complement and support a wide array of analyses undertaken by the various specialists of the diverse VFTS Team.  Specifically, detailed spectral classifications have been presented for 352 O and B0 stars, many of which have been found to belong to several categories of special interest.  These include the earliest (O2--O3) spectral types, of which nine new members have been found and for an equal number of which refined types are given.  A salient result is the detection of a group of 18 very rapidly rotating O main-sequence stars, including the fastest known to date with \vsini\ $\sim600$~km~s$^{-1}$, for which radial-velocity and spatial distributions support a hypothesis of ejection from the 30~Dor clusters. Other noteworthy objects include new members of the also rapidly rotating but evolved Onfp class, and a surprisingly large number (48) of Vz objects, likely related to the lower metallicity and weaker winds in the LMC.  They have been hypothesized to be in a very early evolutionary state, which is substantiated by their locations both on the sky and in the HRDs, although further interesting aspects of their interpretation have emerged from parallel quantitative analysis as cited below.  Further intricacies and diversity of O-type main-sequence spectral morphology have been uncovered, including 19 members of the O V((fc)) class with C~III emission features equal to or stronger than the usual N~III, as well as six with very high S/N but no trace of either emission species; this complexity will no doubt yield significant physical diagnostics when modeled and understood.  Finally, several objects with morphological anomalies in their CNO spectra have been found, which are related to massive stellar evolution and mixing or transfer of processed material.              

Basic information about VFTS has been provided by Evans et al. 2011 (Paper~I), and further global analyses of these data from different perspectives have appeared or are in progress.  Multiplicity among the O-type sample based on radial velocities has been investigated by Sana et al. (2013; Paper~VIII) with the result that the spectroscopic binary fraction is at least 50\%.  The interstellar content of the spectra has been extracted by van Loon et al. (2013; Paper~IX), including the detection of high-velocity absorption-line components in \naa; while the nebular emission-line data have been presented by Torres-Flores et al. (2013), revealing several new kinematic structures in the nebula.  A census of the hot stars and their feedback, incorporating the present spectral classifications, has been compiled and discussed by Doran et al. (2013; Paper~XI).  The rotational velocities of the single O stars have been derived by Ram\'{\i}rez-Agudelo et al. (2013; Paper~XII).  The physical properties of the Vz class have been investigated by Sab\'{\i}n-Sanjuli\'an et al. (2014; Paper~XIII), providing important insights into its atmospheric and evolutionary significance. Detailed atmospheric and wind analyses of 62 VFTS O, Of/WN, and WN stars are presented by Bestenlehner et al. (2014; Paper~XVII).   Ultimately, exhaustive quantitative stellar spectral analysis of the full VFTS sample (complemented by {\it HST} spectroscopy in R136, PI P.~Crowther; proper motions, PI. D.~Lennon; photometry, Sabbi et al. 2013), as well as studies of extinction (Ma\'{\i}z Apell\'aniz et al. 2014; Paper~XVI) and nebular lines, will support theoretical analyses of massive stellar evolution and the nebular environment, by means of which this unprecedented dataset for the unique 30~Doradus region will advance understanding of both massive stars and their starburst habitats.       

\begin{acknowledgements}
We thank Leisa Townsley for providing the {\it Chandra} X-ray identifications together with parameter derivations prior to publication, and Selma de Mink for the {\it HST}/WFC3 image cutouts used to determine visual multiplicity among the target stars.  We also thank our VFTS colleagues Paul Crowther, Selma de Mink, Artemio Herrero, Norbert Langer, Joachim Puls, and Jorick Vink as well as an anonymous referee for useful comments on the manuscript.  STScI is operated by AURA, Inc. under NASA contract NAS~5-26555.  SSD acknowledges funding by the Spanish Government Ministerio de Econom\'{\i}a y Competitividad under grants AYA 2010-21697-C05-04, Consolider-Ingenio 2010 CSD2006-00070, and Severo Ochoa SEV-2011-0187, and by the Canary Islands Government under grant PID2010119.  JMA acknowledges support from the Spanish Government Ministerio de Ciencia e Innovaci\'on through grants AYA 2010-17631 and AYA 2010-15081, and from the Junta de Andaluc\'{\i}a through grant P08-TIC-4075.    
\end{acknowledgements}

\clearpage

\onecolumn
\begin{landscape}
\begin{center}
\setcounter{table}{+1}
{\small
\begin{longtable}{lllrrlrclll}
%\tablewidth{8in}
%\tablenum{2}
%\label{table:2}
%\tablecolumns{11}
%\flushleft{
\caption[]{\label{table:2} \leftline{AAA-Rated Spectral Classifications and Supporting Data for 213 VFTS O Stars}}\\
\hline
VFTS & SpT & $V$  & $B-V$  & $M_V$  & $T_\mathrm{eff}$  & $L/L_{\sun}$ & HRD  & Mult & AltID & Comment\\
\hline
\endfirsthead
\caption[]{\it{continued}}\\
\hline
VFTS & SpT & $V$  & $B-V$  & $M_V$  & $T_\mathrm{eff}$  & $L/L_{\sun}$ & HRD  & Mult & AltID & Comment\\
\hline
\endhead
\hline
\multicolumn{11}{r}{\it{continued on next page}}\\
\endfoot
\hline
\endlastfoot
012  &O9.5 IIIn               &15.83  &0.15  &$-$4.16  &31800   &6.17E4 &1 &\nodata &\nodata &runaway\\
014  &O8.5 Vz                 &\nodata &\nodata &\nodata &\nodata &\nodata &0   &SBs &\nodata &\nodata\\
016  &O2 III-If*              &13.55  &0.04  &$-$6.22  &46100 &1.13E6   &1y &\nodata   &\nodata &runaway; Evans et al. 2010\\
021  &O9.5 IV                 &15.57  &0.09  &$-$4.19  &32300   &6.49E4   &1   &SB? &\nodata &\nodata\\
045  &O9.7 Ib-II Nwk          &15.30  &0.38  &$-$5.46  &29500  &1.70E5   &1   &SBl        &ST1-04 &\nodata\\  
046  &O9.7 II((n))            &14.65  &0.13  &$-$5.36  &29500  &1.57E5   &1   &SB?        &ST1-05 &\nodata\\
064  &O7.5 II(f)              &14.62  &0.13  &$-$6.52  &35500  &7.24E5   &1   &SB1s       &ST1-12 &\nodata\\
065  &O8 V(n)                 &15.99  &0.06  &$-$3.77  &35900   &5.81E4   &1y &\nodata &\nodata &\nodata\\
067  &O9.5 Vz                 &16.83  &0.16  &$-$3.33  &32900   &3.12E4   &1   &SB?	   &\nodata  &\heb\ wings\\
072  &O2 V-III(n)((f*))       &13.70 &$-$0.14  &$-$5.74  &47500  &7.82E5   &1y   &NC	        &BI253                &runaway?\\
073  &O9.5 III                &16.14  &0.38  &$-$4.66  &31800   &9.73E4   &1   &SB1l       &ST1-19 &\nodata\\
074  &O9 Vn                   &16.53  &0.14  &$-$3.66  &33900   &4.53E4   &1 	&\nodata &ST1-20 &\nodata\\
076  &O9.2 III                &15.24  &0.19  &$-$5.06  &32300  &1.43E5   &1 &\nodata &ST1-21&\nodata \\
086  &O9 III((n))             &14.47  &0.12  &$-$6.16  &32800  &4.14E5   &1   &SB1l       &ST1-24 &\nodata\\
087  &O9.7 Ib-II              &13.58 &$-$0.14  &$-$5.59  &29500  &1.93E5   &1   &SBvs &\nodata &\nodata\\
089  &O6.5 V((f))z Nstr       &16.08  &0.20  &$-$4.30  &38900  &1.17E5   &1y         &\nodata &ST1-25 &\nodata\\
090  &O9.5 V                  &15.78  &0.19 	&\nodata &\nodata &\nodata &0 &SBl VM2? &ST1-26 &\nodata\\
091  &O9.5 IIIn               &15.98  &0.20  &$-$4.36  &31800   &7.37E4   &1 &\nodata &ST1-27 &\nodata\\
093  &O9.2 III-IV             &15.03  &0.10  &$-$4.69  &32300  &1.02E5   &1   &SBvs &\nodata &\nodata\\
094  &O3.5 Inf*p + sec?       &14.12  &0.11  &$-$6.38  &42300 &1.03E6   &1y   &SB2?        &ST1-28 &\nodata\\
096  &O6 V((n))((fc))z        &13.91  &0.00  &$-$5.72  &39900  &4.66E5   &1   &SBvs VM2   &ST1-29 &\nodata\\
098  &O9 III(n)               &15.07  &0.15  &$-$5.03  &32800  &1.44E5   &1 &SBl &ST1-30 &\nodata\\
103  &O8.5 III((f))           &16.20  &0.47  &\nodata  &\nodata &\nodata &0  &VM2?   &\nodata &near O9\\
110  &O6 V((n))z              &15.69  &0.07  &$-$5.04  &39900  &2.47E5   &1   &VM2?       &ST1-40               &no emission lines\\
117  &O6: Vz                  &16.64  &0.18  &$-$4.06  &39900   &9.89E4   &1y   &SB?	&\nodata &no emission lines\\
123  &O6.5 Vz                 &15.78 &0.10 &$-$4.04 &38900 &9.26E4 &1y &SB?  &\nodata &((fc))?\\
130  &O8.5 V((n))             &16.67  &0.16  &$-$4.52  &34900  &1.05E5   &1 &\nodata &\nodata &\nodata\\
132  &O9.5 Vz                 &16.09  &0.07  &$-$3.64  &32900   &4.13E4   &1 &\nodata	&ST1-51               &\heb\ broad\\
138  &O9 Vn                   &15.63 &$-$0.09 &$-$3.52 &33900 &4.02E4 &1 &SB2? &\nodata &runaway\\
140  &O8.5 Vz                 &16.05  &0.23  &$-$4.37  &34900   &9.13E4   &1   &SBl        &ST1-56 &\nodata\\
143  &O3.5 V((fc))            &15.36  &0.19  &$-$5.30  &44900  &4.40E5   &1y   &SB1l       &ST1-60 &\nodata\\
149  &O9.5 V                  &16.44  &0.16  &$-$3.65  &32900   &4.15E4   &2 &\nodata &ST1-61 &\nodata\\
154  &O8.5 V                  &14.94  &0.05  &$-$5.04  &34900  &1.69E5   &1   &SBs        &T88-7 &Walborn et al.\ 1999 O7.5~V((f))\\
160  &O9.5 III((n))           &14.17  &0.03  &$-$5.69  &31800  &2.49E5   &1   &SB1s &\nodata &\nodata\\
165  &O9.7 Iab                &13.72  &0.07  &$-$6.17  &28700  &3.07E5   &1 &\nodata &\nodata &runaway\\
168  &O8.5 Vz                 &15.46  &0.08  &$-$4.12  &34900   &7.24E4   &1   &SB?	 &\nodata &\heb\ broad\\
169  &O2.5 V(n)((f*))         &14.59  &0.03  &$-$5.78  &46800  &7.81E5   &1y   &SB?        &ST1-71               &\cd\ emission\\
171  &O8 II-III(f)            &14.06 &$-$0.05  &$-$5.68  &34600  &3.03E5   &1   &SBs        &ST1-72 &\nodata\\
172  &O9 III((f))             &\nodata &\nodata  &\nodata  &\nodata  &\nodata &0 &SBs &\nodata &rare use of f at O9\\
177  &O7n(f)p                 &14.63  &0.24  &-5.68  &37900 &3.91E5 &0 &\nodata &\nodata &\nodata\\
178  &O9.7 Iab                &12.91 &$-$0.05  &$-$6.60  &28700  &4.58E5   &1   &SBvs       &ST1-76 &\nodata\\
180  &O3 If*                  &13.54 &$-$0.08  &$-$6.25  &43200  &9.64E5   &1y   &SBvs       &ST1-78               &Crowther \& Walborn 2011\\
184  &O6.5 Vnz                &15.38 &$-$0.09  &$-$4.30  &38900  &1.18E5   &2y   &SBl &\nodata &\nodata\\
185  &O7.5 III((f))           &14.45  &0.07  &$-$5.32  &36000  &2.43E5   &1 	        &\nodata &ST1-80 &\nodata\\
190  &O7 Vnn((f))p            &14.67 &$-$0.04  &$-$4.90  &37900  &1.89E5   &1   &SB? &\nodata &runaway\\
191  &O9.5 V                  &15.74  &0.12  &$-$4.06  &32900   &6.02E4   &1   &SB1l       &ST1-84               &\heb\ wings\\
197  &O9 III                  &13.86 &$-$0.06  &$-$5.63  &32800  &2.51E5   &1   &SB1l       &ST1-87 &\nodata\\
208  &O6(n)fp                 &14.65 &0.33 &-5.94 &39900 &5.77E5	&0  &SBs  &ST1-93&\nodata\\
210  &O9.7 II-III((n))        &15.60  &0.00  &$-$3.78  &29500   &3.75E4   &2 	        &\nodata &ST1-94 &\nodata\\
216  &O4 V((fc))              &14.41  &0.25  &$-$5.88  &43900  &7.14E5   &1y   &SB?        &ST1-97 &\nodata\\
223  &O9.5 IV                 &14.77 &$-$0.05  &$-$4.56  &32300   &9.10E4   &2   &SBvs       &ST1-101 &\nodata\\
235  &O9.7 III                &15.48 &$-$0.06  &$-$3.79  &30400   &4.07E4   &2   &SB? &\nodata &\nodata\\
243  &O7 V(n)((f))            &15.26  &0.21  &$-$5.10  &37900  &2.28E5   &2   &SB1l &\nodata &\nodata\\
244  &O5 III(n)(fc)           &14.04 &$-$0.10  &$-$5.59  &41300  &4.59E5   &2   &SBvs &\nodata &\nodata\\
249  &O8 Vn                   &15.52 &$-$0.03  &$-$3.83  &35900   &6.09E4   &2y &\nodata &\nodata &\nodata\\
250  &O9.2 V((n))             &15.74  &0.02  &$-$3.85  &33400   &5.20E4   &2 &\nodata &\nodata &\nodata\\ 		   
251  &O9.5 IV                 &15.62 &$-$0.05  &$-$3.81  &32300   &4.61E4   &2   &SB?    &\nodata &\heb\ wings\\
252  &O8.5 Vz                 &15.46 &$-$0.13  &$-$3.62  &34900   &4.62E4   &2y &\nodata &\nodata &\nodata\\
256  &O7.5-8 V((n))z          &15.02 &$-$0.10  &$-$4.28  &36900   &9.87E4   &2   &SB1l &\nodata &\nodata\\
259  &O6 Iaf                  &13.65  &0.21  &$-$7.02  &37800 &1.38E6   &2   &SB1s &\nodata &\nodata\\
266  &O8 V((f))z              &15.38  &0.12  &$-$4.34  &35900   &9.70E4   &1 &\nodata &\nodata &\nodata \\
267  &O3 III-I(n)f*           &13.49 &$-$0.05  &$-$6.30  &44300 &1.08E6   &2y   &SB1vs &\nodata &\ne\ weak; X-ray source\\
277  &O9 V                    &15.04  &0.01  &$-$4.53  &33900  &1.00E5   &2   &SBl &\nodata &\nodata\\
280  &O9 V((n))               &15.40  &0.00  &$-$4.24  &33900   &7.67E4   &2 &\nodata &\nodata &\nodata\\
285  &O7.5 Vnnn               &15.63 &$-$0.06  &$-$3.89  &36900   &6.93E4   &2y 	&\nodata 	&\nodata &runaway\\
290  &O9.5 IV                 &15.67  &0.04  &$-$3.90  &32300   &5.00E4   &2   &SB?  &\nodata &\heb\ wings\\
303  &O9.5 IV                 &15.39  &0.00 &\nodata &\nodata  &\nodata &0  &VM2 &\nodata &\heb\ wings; \sid\ weak\\
306  &O8.5 II((f))            &14.08  &0.03  &$-$5.85  &33600  &3.28E5   &2   &SB?        &Mk80 &runaway\\
332  &O9.2 II-III             &14.07  &0.08  &$-$5.56  &32100  &2.24E5   &2   &SB1s       &Mk70 P32 &\nodata\\
333  &O8 II-III((f))          &12.49 &$-$0.06  &$-$6.77  &34600  &8.58E5   &2   &SBvs       &R133 P42 &X-ray source\\
355  &O4 V((n))((fc))z        &14.12 &$-$0.19  &$-$5.11  &43900  &3.48E5   &1y   &SB2 NC &\nodata &runaway\\
356  &O6: V(n)z               &15.87  &0.16  &$-$4.47  &39900  &1.46E5   &2y   &SB?        &P153                 &runaway\\
361  &O8.5 V                  &15.82  &0.02  &$-$4.96  &34900  &1.58E5   &2 &\nodata &P171 &\nodata\\
369  &O9.7 V                  &16.66  &0.15  &$-$3.69  &31000   &3.86E4   &2 	 &\nodata &P214       &\heb\ wings\\
380  &O6-7 Vz                 &16.18  &0.04  &$-$3.92  &38900   &8.27E4   &2y &\nodata &\nodata &\nodata\\
382  &O4-5 V((fc))z           &15.88  &0.14  &$-$4.80  &42900  &2.45E5   &2y &\nodata &\nodata &\nodata\\ 	 
385  &O4-5 V((n))((fc))       &14.65  &0.00  &$-$5.18  &42900  &3.47E5   &2y   &SBs	&P288 &\nodata\\
386  &O9 IV(n)                &14.75  &0.20 &\nodata &\nodata &\nodata &0   &SB1l VM2  &Mk58(W) P294 & Walborn et al.\ 2002b O7.5~V; inadvertently ``Mk58(e)'' in Paper~I\\
390  &O5-6 V(n)((fc))z        &15.49  &0.14 &\nodata &\nodata  &\nodata &0   &SB1l       &P316 &\nodata\\
392  &O6-7 V((f))z            &16.10  &0.13  &$-$4.48  &38900  &1.39E5   &2 &\nodata &\nodata &\nodata\\
398  &O5.5 V((n))((f))z       &14.40 &$-$0.03  &$-$5.14  &40900  &2.93E5   &2   &SBvs &Mk59 P341  &\nodata\\
399  &O9 IIIn                 &15.83  &0.08  &$-$4.41  &32800 &8.18E4 &1 &SB2? &\nodata &X-ray flare source; broad, variable H$\alpha$ emission\\
404  &O3.5 V(n)((fc))         &14.14  &0.02  &$-$5.93  &44900  &7.95E5   &2y   &SB1l &\nodata &\nodata\\
406  &O6 Vnn                  &14.30  &0.01  &$-$5.42  &39900  &3.52E5   &2 	&\nodata        &Mk55 P370            &contaminated; runaway\\
409  &O4 V((f))z              &15.75  &0.59  &$-$4.98  &43900  &3.11E5   &2y   &SB1l       &P404 &\nodata\\
410  &O7-8 V                  &16.03  &0.76  &\nodata 	&\nodata &\nodata &0  &VM3 &P409   &Knot3; z? Walborn et al.\ 2002b, 2 components O8.5~V, O9~V\\
415  &O9.5 V                  &15.48 &$-$0.10  &$-$3.41  &32900   &3.42E4   &2   &SB1l       &P466 &\nodata\\
418  &O5 V((n))((fc))z        &16.12  &0.21  &$-$4.42  &41900  &1.62E5   &2y   &SB?        &P473 &runaway\\
419  &O9: V(n)                &15.40  &0.09  &$-$4.79  &33900  &1.36E5   &2 &\nodata	        &P485 &\nodata\\
422  &O4 III(f)               &15.14  &0.24  &$-$5.81  &43400  &6.49E5   &2y   &SB1l  &\nodata &\nodata\\
429  &O7.5-8 V                &14.69 &$-$0.11  &$-$4.83  &36900  &1.65E5   &2   &SB2       &Mk57 P541            &z?\\
432  &O8-9 V(n)((f))          &15.65  &0.25  &$-$4.98  &34900  &1.61E5   &2   &SB2?        &P547 &\nodata\\
435  &O7-8 V                  &16.86  &0.53  &$-$4.64  &36900  &1.38E5   &2 	        &\nodata &P566   &z?\\
436  &O7-8 V                  &15.92 &$-$0.05  &$-$4.09  &36900   &8.34E4   &2   &SB?  &\nodata                  &z?\\
440  &O6-6.5 II(f)            &13.66  &0.02  &$-$6.27  &38500  &7.13E5   &2   &SB2?  &Mk47 P607 &\nodata\\
441  &O9.5 V                  &15.07 &$-$0.07  &$-$4.44  &32900   &8.52E4   &2   &SB2       &P613 &\nodata\\
466  &O9 III                  &15.60  &0.37  &$-$5.30  &32800  &1.84E5   &2 	  &\nodata  &P696 &\nodata\\
468  &O2 V((f*)) + OB         &14.59  &0.04  &\nodata &\nodata &\nodata &0   &VM4        &Mk36 P706            &\cd\ emission; X-ray source\\
470  &O6: V((f))z             &15.46 &$-$0.08  &$-$4.05  &39900   &9.87E4   &2y &\nodata &P716 &\nodata\\
472  &O6 Vz                   &16.39  &0.26  &$-$4.06  &39900   &1.00E5   &2y  &\nodata  &P712                 &no emission lines\\
479  &O4-5 V((fc))z           &15.90  &0.14  &$-$4.43  &42900  &1.74E5   &2y   &SB1l       &P747  &X-ray source\\
481  &O8.5 III                &14.16 &$-$0.04  &$-$5.46  &33900  &2.34E5   &2   &SB1l       &P9017 &\nodata\\
483  &O9 V                    &16.38  &0.12  &\nodata &\nodata &\nodata &0 &SB? &\nodata &\nodata\\ 	   
484  &O6-7 V((n))             &15.07  &0.14  &$-$5.37  &38900  &3.16E5   &2 	&\nodata 	&\nodata &no emission lines\\
488  &O6 V((f))z              &15.87  &0.26  &$-$4.79  &39900  &1.96E5   &2  	&\nodata  &P791 &\nodata\\
491  &O6 V((fc))              &15.62  &0.29  &$-$5.11  &39900  &2.63E5   &2   &SB?        &P803 &\nodata \\
493  &O9 V                    &16.74  &0.33  &$-$4.42  &33900   &9.03E4   &2 &\nodata &\nodata &\nodata\\
494  &O8 V(n)                 &16.73  &0.42  &$-$4.20  &35900   &8.53E4   &2 	&SB2?       &P822       &((f))?\\
498  &O9.5 V                  &16.45  &0.19  &$-$4.29  &32900   &7.42E4   &2 &\nodata &\nodata &\nodata\\
502  &O9.7 II                 &13.76  &0.07  &$-$6.42  &29500  &4.15E5   &2   &SBvs &Mk27W P850           &contaminated\\
503  &O9 III                  &14.87  &0.00  &\nodata &\nodata &\nodata &0   &SB1vs &Mk27E	             &contaminated\\
505  &O9.5 V-III              &16.24 &$-$0.02  &\nodata &\nodata &\nodata &0   &VM2? &\nodata &\nodata\\
506  &ON2 V((n))((f*))        &13.31  &0.02  &$-$6.60  &47700 &1.80E6   &2y   &SB1s       &Mk25 P871 &X-ray source\\
508  &O9.5 V                  &15.98  &0.17  &$-$4.13  &32900   &6.46E4   &2   &SB1l       &P872 &\nodata\\
510  &O8.5 V                  &16.28  &0.27  &\nodata &\nodata &\nodata &0  &SB11 &\nodata &\nodata\\
511  &O5 V((n))((fc))z        &15.28  &0.12  &$-$4.91  &41900  &2.53E5   &2y   &SB1s       &P884 &\nodata\\
512  &O2 V-III((f*))          &14.28  &0.20  &$-$6.20  &47700 &1.23E6   &2y   &SB1l       &P885 &X-ray source\\
513  &O6-7 II(f)              &16.20  &0.06  &$-$4.59  &37500  &1.36E5   &2 &\nodata	&\nodata    &near (fc); X-ray source\\
517  &O9.5 V-III((n))         &14.72 &$-$0.08  	&\nodata &\nodata &\nodata &0 &\nodata &Mk29 P909 &\nodata\\
518  &O3.5 III(f*)            &15.11  &0.27  &$-$5.51  &42300  &4.54E5   &2y   &SBs        &P901 &\nodata\\
521  &O9 V(n)                 &15.34  &0.17  &\nodata &\nodata &\nodata &0  &VM2        &P905 &\nodata\\
526  &O8.5 I((n))fp           &14.92  &0.54  &$-$6.57  &33300  &6.39E5   &2   &SB1l       &P925                 &near O9\\
530  &O9.5 III:nn             &16.28 &$-$0.02  &\nodata &\nodata &\nodata &0   &SBl  &P965 &\nodata\\
531  &O9.5 III:nn             &14.50 &$-$0.11  &\nodata &\nodata &\nodata &0   &SB1l &Mk22W P983           &runaway\\
532  &O3 V(n)((f*))z + OB     &14.76  &0.20  &$-$5.61  &45900  &6.32E5   &2y   &SB1l SB2?  &P974 &\nodata\\
536  &O6 Vz                   &16.21  &0.23  &$-$4.39  &39900  &1.35E5   &2y   &SB? &\nodata &\nodata\\
537  &O5 V((fc))z             &15.99  &0.07  &$-$4.58  &41900  &1.87E5   &2y 	 &\nodata       &P1022       &contaminated\\
546  &O8-9 III:((n))          &15.36  &0.04 &\nodata &\nodata &\nodata &0 &\nodata &P1052 &\nodata\\
549  &O6.5 Vz                 &16.51  &0.35  &$-$4.30  &38900  &1.17E5   &2y   &SB?        &P1063 &\nodata\\
550  &O5 V((fc))z             &15.25  &0.09  &$-$4.60  &41900  &2.06E5   &2y   &SB?        &P1077 &\nodata\\
554  &O9.7 V                  &16.61  &0.04  &\nodata &\nodata &\nodata &0 &\nodata &\nodata &\nodata\\
555  &O9.5 Vz                 &15.88  &0.08  &$-$4.33  &32900   &7.68E4   &2   &SB1l       &P1109 &\nodata\\
560  &O9.5 V                  &16.59  &0.05  &$-$3.31  &32900   &3.07E4   &2 	  &\nodata      &P1139                &\heb\ wings\\
564  &O6-8 V((f))             &16.02  &0.17  &$-$5.09  &37900  &2.29E5   &2 	&\nodata 	  &\nodata &((fc))? z? Oe?; X-ray source\\
566  &O3 III(f*)              &14.05  &0.06  &$-$5.96  &45500  &8.49E5   &2y &\nodata     &Mk23 P1163 &\nodata\\
569  &O9.2 III:               &16.09  &0.07  &\nodata &\nodata &\nodata &0  &\nodata	        &P1170                &p? discrepant \heb~\lam4686 vs. \sid\ luminosity criteria\\
574  &O9.5 IIIn               &15.89 &$-$0.12  &$-$3.18  &31800   &2.56E4   &1   &SB2? NC &\nodata &\nodata\\
577  &O6 V((fc))z             &16.64  &0.40  &$-$4.40  &39900  &1.36E5   &2y 	&\nodata &P1189 &\nodata\\
581  &O4-5 V((fc))            &16.07  &0.18  &$-$5.00  &42900  &3.15E5   &2y  	  &\nodata &P1218                &z?\\
582  &O9.5 V((n))             &16.83  &0.16 &\nodata &\nodata &\nodata &0 &\nodata &\nodata &\nodata\\ 
585  &O7 V(n)                 &13.65  &0.12  &$-$6.19  &37900  &6.30E5   &2   &SB1l       &P1231 &Walborn et al.\ 2002b VM2: O6~V SB2?, O7~V\\
586  &O4 V((n))((fc))z        &15.04 &$-$0.12  &$-$4.72  &43900  &2.45E5   &2y   &SB? &\nodata &\nodata\\
592  &O9.5 Vn                 &16.40  &0.14  &$-$3.72  &32900   &4.42E4   &2 &\nodata &\nodata &\nodata\\
596  &O7-8 V((n))             &15.23  &0.03 &\nodata &\nodata &\nodata &0  &SB1l VM2&\nodata     &z?\\
597  &O8-9 V(n)               &15.56  &0.03  &$-$4.13  &34900   &7.36E4   &2 &\nodata	&P1288 &\nodata\\
599  &O3 III(f*)              &13.80  &0.08  &$-$6.24  &45500 &1.11E6   &2y   &SB1s       &P1311 &\nodata\\
601  &O5-6 V((n))z            &14.69  &0.09  &$-$5.41  &40900  &3.76E5   &2 	 &\nodata       &Mk14N P1317    &no emission lines; X-ray source\\
603  &O4 III(fc)              &13.99  &0.04  &$-$6.34  &43400 &1.06E6   &2y   &SB1l       &Mk10 P1341 &\nodata\\
604  &O8.5 V                  &14.94  &0.04  &$-$4.88  &34900  &1.47E5   &2   &SB2? VM2?  &P1340&\nodata \\
608  &O4 III(f)               &14.22  &0.16  &$-$6.06  &43400  &8.18E5   &2y   &SB1l       &Mk14 P1350 &\nodata\\
609  &O9-9.5 V-III            &16.76 &$-$0.01  &\nodata &\nodata &\nodata &0   &SB?        &P1354 &\nodata\\
611  &O8 V(n)                 &16.16  &0.13  &$-$3.76  &35900   &5.75E4   &2y &\nodata &\nodata &\nodata\\
613  &O8.5 Vz                 &15.78  &0.16  &$-$4.21  &34900   &8.93E4   &2   &SBl VM2?   &P1369 &\nodata\\
615  &O9.5 IIInn              &15.89  &0.02  &\nodata &\nodata &\nodata &0  &VM3&\nodata &\nodata\\
619  &O7-8 V(n)               &15.98  &0.12  &$-$4.06  &36900   &8.04E4  &2   &SB1l       &P1401                &z?\\
620  &O9.7 III(n)             &16.61  &0.03  	&\nodata &\nodata &\nodata &0  &SB?        &P1416                &\sid\ weak\\
621  &O2 V((f*))z             &15.39  &0.27  &\nodata &\nodata &\nodata &0  &VM3        &P1429                &Knot2; ON? Walborn et al.\ 2002b O3-4 V\\
626  &O5-6n(f)p               &14.90  &0.26  &-5.62 &40900 &4.58E5 &0  &SB?        &P1423 &\nodata\\
627  &O9.7 V                  &15.38 &$-$0.15  &$-$3.67  &31000   &3.80E4   &1 &\nodata &P9034 &\heb\ wings\\
630  &O9.7 V-III              &16.21  &0.17  &\nodata &\nodata &\nodata &0 	&\nodata        &P1455 &\nodata\\
631  &O9.7 III(n)             &16.00  &0.14  &$-$3.92  &30400   &4.56E4   &2   &SBl        &P1459                &\sid\ weak\\
635  &O9.5 IV                 &15.52  &0.07  &$-$4.09  &32300   &5.94E4   &2   &SBvs       &P1468 &\nodata\\
638  &O8.5 Vz                 &15.63 &$-$0.14  &$-$3.48  &34900   &4.04E4   &2y 	&\nodata &\nodata &\heb\ broad\\
639  &O9.7 V                  &15.41 &$-$0.01  &$-$3.95  &31000   &4.89E4   &2   &SB? &\nodata &\nodata\\
645  &O9.5 V((n))             &16.29  &0.16  &$-$3.86  &32900   &5.01E4   &2   &SBl        &P1519 &\nodata\\
648  &O5.5 IV(f)              &14.16  &0.08  &$-$5.66  &40500  &4.64E5   &2   &SBvs       &Mk8 P1531            &\cc\ emission but not (fc)\\
649  &O9.5 V                  &16.07  &0.05  &$-$3.71  &32900   &4.40E4   &2   &SB2 &\nodata &\nodata\\
651  &O7 V(n)z                &14.70  &0.08  &$-$5.15  &37900  &2.38E5   &2   &SB1l       &Mk7 P1553 &\nodata\\
654  &O9 Vnn                  &15.71  &0.05  &$-$4.21  &33900   &7.46E4   &2   &SBl	&P1560 &\nodata\\
656  &O7.5 III(n)((f))p       &14.24  &0.06  &$-$5.46  &36000  &2.75E5   &2   &SB1l VM2?  &Mk6 P1563 &\nodata\\
657  &O7-8 II(f)              &15.45  &0.44  &$-$5.91  &35500  &4.49E5   &2   &SB1l       &P1573 &\nodata\\
660  &O9.5 Vnn                &15.92  &0.03  &\nodata &\nodata &\nodata &0 	        &\nodata &P1586 &\nodata\\
663  &O8.5 V                  &16.52  &0.23  &$-$3.78  &34900   &5.30E4   &2   &SB?	   &\nodata &z? runaway\\
664  &O7 II(f)                &14.25  &0.14  &$-$5.86  &36500  &4.19E5   &2   &SB?        &Mk4 P1607 &X-ray source\\
667  &O6 V((f))               &15.03  &0.07  &$-$4.78  &39900  &1.93E5   &2   &SB?        &P1614                &z? contaminated\\
669  &O8 Ib(f)                &14.18  &0.27  &$-$6.01  &34200  &4.00E5   &2   &SB1s       &P1619 &\nodata\\
677  &O9.5 V                  &16.68  &0.40  &\nodata &\nodata &\nodata &0  &VM3        &P1696 &\nodata\\
679  &O9.5 V                  &16.73  &0.25  &$-$3.87  &32900   &5.10E4   &2   &SB?        &P1698                &\heb\ wings\\
688  &O9.7 III                &15.60  &0.11  &$-$4.27  &30400   &6.22E4   &2   &SBl        &P1756                &\sid\ a bit weak\\
702  &O8 V(n)                 &16.31  &0.33  &$-$4.68  &35900  &1.34E5   &2   &SB1l       &P1829 &\nodata\\
704  &O9.2 V(n)               &16.76 &$-$0.07  &\nodata &\nodata &\nodata &0   &SB?&\nodata  &\nodata\\
706  &O6-7 Vnnz               &15.77  &0.14  &$-$4.24  &38900  &1.11E5   &2y 	&\nodata        &P1838 &\nodata\\
710  &O9.5 IV                 &16.14 &$-$0.02  &$-$3.35  &32300   &3.03E4   &2   &SB?        &P1849                &\heb\ wings\\
716  &O9.5 IV                 &15.91  &0.06  &$-$4.14  &32300   &6.20E4   &2   &SBs        &P1890 &\nodata\\
717  &O9 IV                   &15.69  &0.19  &$-$4.66  &33300  &1.07E5   &2   &SB? VM?    &P1892 &\nodata\\
722  &O7 Vnnz                 &15.04 &$-$0.13  &$-$4.08  &37900   &8.81E4   &1y   &SB? NC	    &\nodata &runaway\\
724  &O7 Vnnz                 &16.80  &0.43  &$-$4.34  &37900  &1.12E5   &1   &NC	&\nodata &nebular lines oversubtracted; runaway?\\
736  &O9.5 V                  &15.85  &0.05  &$-$4.14  &32900   &6.47E4   &2  & SB1l VM2?  &P1998                &\heb\ wings\\
737  &O9 V                    &15.70  &0.13  &$-$4.47  &33900   &9.44E4   &2 	 &\nodata &P2000                &z? \heb\ broad\\
743  &O9.5 V((n))             &15.04 &$-$0.17  &$-$4.05  &32900   &5.98E4   &1   &SB1l NC &\nodata &\nodata\\
746  &O6 Vnn                  &15.38  &0.12  &$-$4.77  &39900  &1.91E5   &2 &\nodata &\nodata &\nodata\\
750  &O9.5 IV                 &15.43 &$-$0.11  &$-$3.75  &32300   &4.36E4   &1   &SB1l NC &\nodata &\nodata\\
751  &O7-8 Vnnz               &16.32  &0.15  &$-$4.37  &36900  &1.07E5   &2   &NC&\nodata &\nodata\\
753  &O9.7 II-III             &16.46  &0.42  &$-$4.35  &29500   &6.25E4   &1   &NC    &\nodata &\heb\ wings; sharp lines; H$\gamma$~p\\
755  &O3 Vn((f*))             &15.04  &0.14  &$-$5.20  &45900  &4.34E5   &2y  	&\nodata &P2041                &\cd\ emission; runaway\\
761  &O6.5 V((n))((f))z Nstr  &15.35 &$-$0.13  &$-$4.06  &38900   &9.42E4   &1y   &NC	&\nodata &runaway\\
764  &O9.7 Ia Nstr            &12.26  &0.09  &$-$7.06  &28700  &7.01E5   &1   &SB1s NC    &Sk$-69^{\circ}$~252 &\nodata\\
768  &O8 Vn                   &16.10  &0.20  &$-$4.65  &35900  &1.30E5   &1   & SB2? NC &\nodata &\nodata\\
769  &O9.7 II-III             &15.83  &0.08  &$-$4.24  &29500   &5.68E4   &2   &SBl        &P2099 &\nodata\\
770  &O7 Vnn                  &15.79  &0.08  &$-$4.15  &37900   &9.46E4   &2y &\nodata &\nodata &\nodata\\
775  &O9.2 V                  &16.85  &0.29  &$-$3.67  &33400   &4.44E4   &1   &SB? NC        &\nodata &\heb\ wings\\
777  &O9.2 II                 &15.30  &0.38  &$-$6.03  &32100  &3.48E5   &1   &SB? NC &\nodata &\nodata\\
778  &O9.5 V                  &16.64  &0.14  &$-$4.02  &32900   &5.82E4   &2 &\nodata &\nodata &\nodata\\
782  &O8.5 III                &15.47  &0.36  &$-$5.31  &33900  &2.05E5   &1   &NC &\nodata &\nodata\\
797  &O3.5 V((n))((fc))       &14.68  &0.05  &$-$5.15  &44900  &3.82E5   &2y   &SB?&\nodata	&runaway\\
802  &O7.5 Vz                 &14.14 &$-$0.19  &$-$4.86  &36900  &1.68E5   &1   &SB2 NC   &BI258 &\nodata\\
807  &O9.5 III Nstr           &16.37  &0.56  &$-$4.28  &31800   &6.88E4   &1 &\nodata &\nodata &\nodata\\
812  &O4-5 V((fc))            &14.81  &0.05  &$-$5.28  &42900  &3.82E5   &2y   &SB1l       &P2246 &\nodata\\
819  &ON8 III((f))            &16.79  &0.40  &$-$4.31  &35000   &8.70E4   &1   &NC &\nodata &\nodata\\
830  &O5-6 V(n)((f))          &15.39 &$-$0.03  &$-$4.23  &40900  &1.28E5   &2y   &SBl        &P2270     &((fc))? z? X-ray source\\
843  &O9.5 IIIn               &15.88 &$-$0.05  &$-$3.50  &31800   &3.42E4   &1   &NC&\nodata &\nodata\\
849  &O7 Vz                   &15.14 &$-$0.08  &$-$4.13  &37900   &9.23E4   &1y   &NC	&\nodata	&no emission lines\\
892  &O9 V                    &15.69  &0.05  &$-$4.00  &33900   &6.19E4   &1   &NC &\nodata	&\heb\ broad\\
\end{longtable}
%}
\tablefoot{Entry code explanations: HRD column: 0,~not included in the HRDs (Section~4) because of no or uncertain luminosity class, visual multiplicity, no photometry or poor SED fit (33 stars);  1,2,~included in the HRDs (180 stars); 2,~within 80~pc projected of R136 (117 stars); y,~age $\leq2.5$~Myr per HRD (55 stars).  Multiplicity column: SB,~spectroscopic binary; 1,~single lined; 2,~double lined; l,~large amplitude ($>$20~km~s$^{-1}$); s,~small amplitude (10--20~km~s$^{-1}$); vs,~very small amplitude ($<$10~km~s$^{-1}$); SB?,~stellar absorption displaced from nebular emission lines but no radial-velocity variation measured; VMn,~visual multiple of n components within 1\farcs2 Medusa fiber, as determined in {\it HST}/WFC3 image; NC,~no WFC3 coverage.  140/213 stars have SB, VM, or both entries; moreover, multiples with separations between about 5000~AU and 1~AU (or less than the latter with unfavorable inclinations) would not, in general, be detected by either technique, except that some SB? may be within that gap.  On the other hand, some SB? may instead have high velocities relative to the gas. Alternate ID column sources: ST1,~Schild \& Testor 1992; BI,~Brunet et~al.\ 1975; T88,~Testor et~al.\ 1988; Mk,~Melnick 1985; P,~Parker 1993; R,~Feast et~al.\ 1960; Sk,~Sanduleak 1970.}
}
\end{center}
\end{landscape}

\clearpage

\begin{appendix}

\section{BBB Classifications}

The Appendix addresses 139 objects in the VFTS O--B0 sample with lower-rated (``BBB'') spectral classifications, which are largely excluded from the discussion in the main text as well as entirely from the HRDs.  They are listed in Table~A.1.  The reasons for the lower ratings are discussed next.  These spectra are not displayed in this paper (except for VFTS~318 in Fig.~23) but will be at the VFTS website. 

\subsection{Reasons for Lower-Rated Classifications}

There are several distinct reasons for the BBB classifications.  The most obvious are relatively low S/N of the data and/or severe nebular contamination at the \hea\ lines, which render the classifications uncertain and frequently preclude luminosity classes.  Of course, these cases are generally among the fainter stars in the sample.  Table~A.1 also contains 10 B0 spectra to define the O--B boundary; some of them are of fine quality.    

A second BBB category comprises the majority of the prominent SB2.  While the data quality for these is often excellent, they cannot be included in the HRDs without full orbital analyses, which in general require further observations that are in progress.  (A few SB2 that were recognized later are included in the AAA list and HRDs, as are of course any that remain undetected; in general, they may be expected to have relatively fainter and less contaminating secondaries.)  As already noted, the SB2 have been classified by JMA with MGB using the best data near quadratures; while some of these classifications are necessarily uncertain, others are fine.

The most vexing BBB category is discussed separately in more detail in the following section.  

\subsection{\heb~$\lambda4686$ vs. \sid\ Luminosity Criterion Discrepancies at Late-O Types}

Thirty-nine of the 139 BBB spectra fall into this category, as do 14 AAA types discussed in Section 4.1; other late-O BBB also have weak \sid\ but luminosity classes could not be derived.  While the classical MK procedure in general relied upon subjective averages over (sometimes mildly conflicting) multiple criteria, the tendency in this work (as well as in GOSSS; Sota et al. 2011) is to define a primary criterion for each subcategory, in the interests of clarity and reproducibility.  In the case of luminosity classification at late-O types, the \heb~$\lambda4686$/\hea~$\lambda4713$ luminosity criterion, which decreases with increasing luminosity due to emission filling in the \heb\ absorption line, has been preferred over the \sid/\hea\ ratios because of the susceptibility of the latter to metallicity effects.  Rather severe discrepancies between the two criteria are encountered in many of the 30~Dor spectra at these types (as well as less frequently in the Galaxy).

Thus, this approach has unfortunately not been entirely successful in the current sample.  While the different criteria agree for many late-O types in the AAA list, for a few as well as for numerous BBB they do not.  A contributing factor may well be that the late-O horizontal classification criterion \sic~$\lambda4552$/\heb~$\lambda4541$, the unit value of which defines type O9.7, is also sensitive to metallicity.  This dependence, in combination with the rapid decline of the luminosity-dependent \heb~$\lambda4686$ emission filling with advancing type in this range, such that \heb~$\lambda4686$/\hea~$\lambda4713$ has similar values near unity at O9.7~II and B0~IV, entails substantial interrelated uncertainties in both dimensions.  However, additional sources of \heb~$\lambda4686$ emission unrelated to luminosity in this young LMC sample cannot be excluded a priori.  It also appears that blended multiples may be another source of this problem; several cases in which spectral classifications from {\it HST} data yielded lower luminosity classes are noted in the Table~A.1 Comments (VFTS~141, 150, 153, 389). It is likely that quantitative spectral analysis and in some cases observations with higher spatial resolution will be required to fully elucidate these issues.

To quantify this problem further, the average absolute visual magnitudes of the 12 AAA and 24 BBB stars in this category with parameters are compared to the predictions from their luminosity classes here.  The observed $M_V$'s available for the BBB objects are listed in the Comments field of Table~A.1; as for the discrepant AAA, the observed values are fainter than the calibration predictions (Walborn 1973) in all but one case (VFTS~389, ironically one of those with a fainter {\it HST} luminosity class).  For 25 stars classified O9.5--O9.7~III, the average observed value is $-3.86 \pm 0.10$~(m.e.), compared with a calibration value of $-5.3$.  For six O9.5--O9.7~II-III and two O9~III, the observed value is $-4.34 \pm 0.12$, whereas the calibration values are $-5.6$.  There is one O9.5~II with an observed $-4.6$ but calibration $-5.9$; and one O9.5~IV with values of $-3.8$ vs. $-4.7$, respectively, but another (VFTS~389) with a brighter observed value of $-5.1$.  The observed values do increase toward brighter luminosity classes, but except for the two class IV stars, the observed minus predicted are remarkably consistent at 1.3--1.4~mag, as also with the O9.5--O9.7~V calibration value of $-4.1$ and what would have been predicted from the small \sid/\hea\ ratios.  On the other hand, for three B0~V stars (VFTS~347, 496, 540) the observed values are more consistent with the calibration, average $-3.5$ vs. $-3.6$ (Walborn 1972), respectively, so that some of the apparent late-O giants could also be slightly cooler objects as discussed in the previous paragraph.                

\subsection{Multiplicity}

Sana et al. (2013) have determined from detailed radial-velocity analysis that 35\% of the VFTS O stars are spectroscopic binaries with amplitudes greater than or equal to 20~km~s$^{-1}$, corresponding to an intrinsic (bias-corrected) binary fraction of 51\% for the orbital-period range considered 
($P < 10^{3.5}$~days).  They also detect an additional 11\% with smaller amplitudes, some of which may have other physical origins, but all of which are characterized as ``SB'' in Tables~2 and A.1 here for simplicity, with the amplitude ranges specified.  All of the SB designations here are from the work of Sana et al., except for the SB? (with displacements between stellar and nebular lines but no significant variations detected) which have been added here.  In addition, visual multiples have been so designated from inspection of {\it HST}/WFC3 images here (but note that there was no WFC3 coverage at the time for some objects at the peripheries of the VFTS field).  As specified in the notes to the tables, all of these categories sum to similar total multiplicity rates of 140/213 or 66\% for the AAA spectral types (Table~2), and 96/139 or 69\% for the BBB (Table~A.1).  As already noted, most of the SB2 are in the BBB list, while the AAA SB are predominantly SB1.  As also pointed out in the table notes, these percentages are lower limits because of the detectability gap between radial velocities and direct imaging, as well as unfavorable inclinations in the case of the former.  The implications for astrophysical and population studies of O stars in the Magellanic Clouds and beyond are clear and should always be borne in mind.   

\onecolumn
\begin{landscape}
\begin{center}
{\small
%\begin{longtable}{cp{2in}llp{2.5in}}
\begin{longtable}{cp{2in}lll}
%\tablewidth{6.5in}
%\tablenum{A1}
%\label{table:A1}
%\tablecolumns{5}
%\flushleft{
\caption[]{\label{table:A.1} \leftline{BBB-Rated Spectral Classifications and Supporting Data for 139 VFTS O--B0 Stars}}\\
\hline
VFTS & SpT & Mult & AltID & Comment/$M_V$\\
\hline
\endfirsthead
\caption[]{\it{continued}}\\
\hline
VFTS & SpT & Mult & AltID & Comment/$M_V$\\
\hline
\endhead
\hline
\multicolumn{5}{r}{\it{continued on next page}}\\
\endfoot
\hline
\endlastfoot
017  &B0 V                                     &SB1l  &\nodata &\nodata\\
035  &O9.5 IIIn                                &SB2? VM2? &\nodata &\nodata\\
042  &O9.5 III((n))                            &SB1l   &\nodata                     &\sid\ weak, $-4.7$\\
047  &O9 V + O9.5 V                            &SB2 &\nodata &\nodata\\
049  &O9.7 V: + B                              &SB2 &\nodata &\nodata\\
051  &OBpe                       &\nodata                            &ST1-06 &\nodata\\
055  &O8.5 V + O9.5 IV                         &SB2 VM2 &\nodata &\nodata\\
056  &O6.5 V + O6.5 V                          &SB2          &ST1-07        &H, He profiles triangular\\
058  &O8.5: V + B                              &SB2           &ST1-09 &\nodata\\
059  &O9.5 III:                                &SBl           &ST1-10        &\sid\ very weak, $-3.8$\\
061  &ON8.5 III: + O9.7: V:                    &SB2 &\nodata &\nodata\\
063  &O5 III(n)(fc) + sec                      &SB2           &ST1-11 &\nodata\\
066  &O9.5 III(n)                              &SBl       &\nodata                  &\sid\ very weak, $-4.5$\\
070  &O9.7 II                                  &SB?  &\nodata                       &\sid\ very weak\\
077  &O9.5: IIIn             &\nodata                                &ST1-22        &\sid\ very weak, $-3.4$\\
080  &O9.7 II-III((n))      &\nodata                                 &ST1-23        &\sid\ weak, $-4.2$\\
097  &B0 IV                                    &SB?           &ST1-31 &\nodata\\
102  &O9: Vnnne+         &\nodata                                    &ST1-32        &\feb\ emission; Dufton et al.\ 2011; runaway\\
104  &O9.7 II-III((n))    &\nodata     &\nodata                                             &\sid\ very weak\\
105  &O8: Vz + O9-B0                           &SB2 VM3       &ST1-37  &\nodata\\
109  &O9.7 II:n              &\nodata                                &ST1-38        &\heb~$\lambda$4686 emission wings, ``Onfp''? \sid\ very weak\\
113  &O9.7 II  or B0 IV ?                      &SBs           &ST1-39        &very sharp lines; \heb\ broad; \sid\ very weak \\
114  &O8.5 IV + sec                            &SB2           &ST1-41 &\nodata\\
116  &O9.7: V: + B0: V:                        &SB2 VM2?      &ST1-42 &\nodata\\
120  &O9.5 IV:                                 &SB2           &ST1-43 &\nodata\\
125  &Ope                                      &SB2?          &ST1-47        &or n with \heb\ emission; \hea~\lam4713 double\\
128  &O9.5 III:((n))           &\nodata                              &ST1-49        &\sid\ weak, $-3.3$\\
131  &O9.7 &\nodata &\nodata &\nodata\\
141  &O9.5 II-III((n))                         &SB?           &T88-6         &\sid\ very weak, $-4.3$; Walborn et al.\ 1999 B0.2~V\\
142  &Op         &\nodata                                        &ST1-58        &composite: \heb/\hea\ \lam4541/\lam4471 implies O6.5, but \lam4541/\lam4387 \& \lam4200/\lam4144 O8.5; false Vz\\
145  &O8fp                                     &SB1s          &T88-3         &contaminated by Brey~73-1A; Walborn et al.\ 1999 O7-8~II\\
148  &O9.7 II-III(n)                           &SBl NC    &\nodata                  &\sid\ very weak\\
150  &O9.5 III                                 &SB2          &T88-4         &``Onfp''? \sid\ very weak, $-4.6$; Walborn et~al.\ 1999 O9.5~V\\
151  &O6.5 II(f)p                              &SBvs VM5      &T88-2         &composite: weak \nd\ \& \cd\ emission, \ne\ absorption; \\
&&&                                                                          &Walborn et al.\ 1999, 3 components: O4~III(f)p, O8~III, B1-2:p(e) \\
153  &O9 III((n))                              &VM3           &T88-5         &\sid\ weak, $-5.0$; Walborn et al.\ 1999 O7~V(z!) N~strong\\
163  &O8.5 IV         &\nodata     &\nodata                                                 &``((n))'' but profiles triangular \& variable\\
173  &O9.7: II-III(n)p                         &SBl           &ST1-73        &``Onfp''; \sid\ weak\\
174  &O8 V + B0: V:                            &SB2           &ST1-75 &\nodata\\
176  &O6 V:((f)) + O9.5: V:                    &SB2           &ST1-77 &\nodata\\
183  &B0 IV                                    &SB?    &\nodata &\nodata\\
187  &O9 IV: + B0: V:                          &SB2           &ST1-81 &\nodata\\
188  &O9.7: III:                               &SB?    &\nodata                     &\sid\ very weak, $-4.0$\\
192  &O9.7 II or B0 IV ?        &\nodata                             &ST1-85        &\sid\ weak\\
201  &O9.7 V + sec                             &SB2      &\nodata &\nodata\\
205  &O9.7 II((n)) or B0 IV((n)) ?    &\nodata                       &ST1-91        &\sid, \cc\ very weak\\
207  &O9.7 II((n))         &\nodata                                  &ST1-92        &\sid, \cc\ very weak\\
217  &O4 V((fc)): + O5 V((fc)):                &SB2           &ST1-98        &X-ray source\\
226  &O9.7 III        &\nodata      &\nodata                                                &runaway; \sid\ weak, $-3.2$\\
231  &O9.7 IV:(n) + sec                        &SB2           &ST1-103 &\nodata\\
241  &B0 IV    &\nodata        &\nodata                                                     &\sid\ weak\\
242  &B0 IV                                    &SB?   &\nodata                      &runaway, \sid\ weak\\
253  &O9.5 II     &\nodata    &\nodata                      &\sid\ very weak, $-4.6$; H strong\\
304  &O9.7 III                                 &VM2?  &\nodata                  &\sid\ weak; shallow WFC3 image appears extended, but deep symmetrical\\
310  &O9.7 V:      &\nodata     &\nodata                                                    &contaminated by R135\\
313  &B0 IV      &\nodata &\nodata                                                          &\sid\ weak\\
314  &O9.7 IV:(n) + sec:                       &SB2  &\nodata  &\nodata\\
316  &O9.7 V:      &\nodata    &\nodata                                                    &contaminated\\
318  &O((n))p                                  &SBl    &\nodata                     &composite: He lines imply O9.5~III; H$\delta$ B2~V from strong wings and no \sid, \cc;\\ 
&&&                                                                              &weak, narrow \nd\ emission, \ne\ absorption present; $-2.6$\\
327  &O8.5 V(n) + sec                          &SB2           &P15 &\nodata\\
328  &O9.5 III(n)                              &SB?   &\nodata                      &\sid\ weak, $-3.3$; runaway\\
329  &O9.7 II-III(n)                           &SBl           &P19           &\sid\ weak\\
339  &O9.5 IV(n)         &\nodata                                    &P83           &\sid\ weak, $-3.8$\\
345  &O9.7 III(n)                              &SBl           &P103          &\sid\ very weak, $-4.1$\\
346  &O9.7 III         &\nodata                                      &P113      &\sid\ very weak, $-3.6$\\
347  &B0 V          &\nodata                                         &P116 &$-2.8$\\
350  &O8 V                                     &SB2           &P124          &circular nebula, Rubio et al.\ 1998\\
352  &O4.5 V(n)((fc)):z: + O5.5 V(n)((fc)):z:  &SB2 &\nodata &\nodata\\
360  &O9.7      &\nodata                                             &P169          &\sid\ strong; no \hea~\lam4713\\
370  &O9.7 III                                 &SB?           &P222          &\sid\ weak, $-3.4$; runaway\\
371  &O9.5 V(n) + sec                          &SB2           &P240 &\nodata\\
373  &O9.5n        &\nodata                                          &P246 &\nodata\\
389  &O9.5 IV       &\nodata                                         &Mk58(E) P304  &\sid\ weak, $-5.1$; Walborn et al.\ 2002b O9.5~V; inadvertently ``Mk58(w)'' in Paper~I\\
393  &O9.5(n)      &\nodata                                          &P324 &\nodata\\
400  &O9.7            &\nodata                                       &P348 &\nodata\\
405  &O9.5:n &\nodata &\nodata &\nodata\\
412  &O9.7 &\nodata &\nodata &\nodata\\
443  &O7: V(n): + O7: V(n):                    &SB2 VM2       &P615 &\nodata\\
444  &O9.7      &SB? &\nodata           &\nodata\\
445  &O3-4 V:((fc)): + O4-7 V:((fc)):          &SB2           &P621          &X-ray source\\
446  &Onn((f))   &\nodata &\nodata &\nodata\\
450  &O9.7 III: + O7::                         &SB2           &Mk50 P643 &I.~Howarth et al., in prep.\\ 
451  &O(n)    &\nodata        &\nodata                                                     &contaminated; bright circumstellar nebulosity\\
455  &O5: V:n                                  &SB1l          &P661          &X-ray source\\
456  &Onn     &\nodata    &\nodata                                                          &((fc))?\\
460  &O7.5 V + O7.5 V                          &SB2           &P674 &\nodata\\
464  &O9.5:                                    &VM2           &P702 &\nodata\\
465  &On          &\nodata                                           &P700    &\nodata\\
475  &O9.7 III                                 &SBl           &P722          &\sid\ weak, $-4.5$\\
476  &O((n))  &\nodata &\nodata &\nodata\\
477  &O((n))    &\nodata &\nodata                                                           &contaminated? \hea~\lam4713 Oe?\\
487  &O6.5: IV:((f)): + O6.5: IV:((f)):        &SB2 VM2   &\nodata &\nodata\\
492  &O8 V + O9.5: V                           &SB2           &Mk21 P830 &\nodata\\
495  &O9.7 II-IIIn          &\nodata                                 &P9019         &\sid\ very weak\\
496  &B0 Vn                                    &SB?     &\nodata &$-3.8$\\
497  &O3.5 V((f))z + OB          &\nodata                            &R140d         &contaminated\\
500  &O6.5 IV((fc)) + O6.5 V((fc))             &SB2   &\nodata                      &X-ray source\\
514  &O9.7 III                                 &SBl           &P9021         &\sid\ weak, $-3.2$\\
515  &O8-9p                                    &VM2     &\nodata                 &(subtract 513!) ``Onfp''? variable? \sid\ \& \cc\ strong, \heb~\lam4686 weak imply high luminosity\\
519  &O3-4 ((f)) + OB + WN                     &SB1s          &R140c         &contaminated\\
522  &O6 II-Iab(fc) + O5.5 V((fc)):            &SB2 &\nodata &\nodata\\
525  &B0 Ia                                    &SB            &Mk38 P930 &\nodata\\
527  &O6.5 Iafc + O6 Iaf                       &SB2           &R139 P952     &Taylor et al.\ 2011; X-ray source\\
528  &O9.7(n)         &\nodata                                       &P956          &``Onfp''? \sid\ very weak\\
529  &O9.5(n)                                  &SB?           &P955 &\nodata\\
538  &ON9 Ia: + O7.5: I:(f):                   &SB2 VM2       &Mk22 P1024   &\nodata\\
539  &O9.5(n)       &\nodata                                         &P1012 &\nodata\\
540  &B0 V   &\nodata &\nodata &$-3.8$\\
543  &O9 IV + O9.7: V                          &SB2           &P1031 &\nodata\\
552  &O8.5 III: + B                            &SB2 &\nodata &\nodata\\
557  &On                                       &SB2  &\nodata &\nodata\\
559  &O9.7(n)           &\nodata                                     &P1133 &\nodata\\
561  &O9:(n)                                   &SB1l          &P1145 &\nodata\\
563  &O9.7 III: + B0: V:                       &SB2 VM2       &P1154 &\nodata\\
565  &O9.5:                                    &SB? &\nodata &\nodata\\
570  &O9.5ne+                                  &SB2    &\nodata                    &\feb\ emission\\
571  &O9.5 II-III(n)                           &SB2?     &\nodata                     &\sid\ weak\\
579  &O9:((n))                                 &SB? VM2       &P1201         &Knot~1; Walborn et al.\ 2002b O9.5~V; X-ray source\\
583  &O8 V + O8.5 V                            &SB2 &\nodata &\nodata\\
587  &O9.7:                                    &SB? &\nodata &\nodata\\
588  &O9.5                                     &SB1l      &\nodata                  &``Onfp''? \sid\ weak\\
594  &O9.7 &\nodata &\nodata &\nodata\\
607  &O9.7 III                                 &SB?      &\nodata                   &\sid\ weak, $-3.6$\\
622  &O9.7 III                                 &SB?    &\nodata                     &\sid\ weak\\
634  &O V                                      &VM2           &P1473         &z?\\
642  &O5 Vz: + O8 Vz:                          &SB2 &\nodata &\nodata\\
647  &O8: V:      &\nodata &\nodata &\nodata\\
652  &B2 Ip + O9 III:                          &SB2           &Mk5 P1552  &I.~Howarth et al., in prep.\\
661  &O6.5 V(n) + O9.7: V:                     &SB2           &P1594  &\nodata\\
703  &O7: V: + O8: V:                          &SB2           &P1828   &\nodata\\
711  &O9.7 III             &\nodata                                  &P1850         &\sid\ very weak, $-4.2$\\
728  &O9.7 II-III((n))                         &SBl           &P1966         &\sid\ weak\\
733  &O9.7p                                    &SBl           &P1988         &``Onfp''; \heb\ wings; \sid\ weak; \nb~\lam3995 too strong\\
771  &O9.7 III:(n)                             &SB2           &P2104 &\nodata\\
774  &O7.5 IVp + O8.5: V:                      &SB2 SB3?    &\nodata                &He strong?\\
787  &O9.7 III                                 &SB            &P2157         &\sid\ weak, $-3.5$\\
806  &O5.5 V((fc)):z + O7 Vz:                  &SB2 NC        &BI259  &\nodata\\
810  &O9.7 V + B1: V:                          &SB2           &P2242  &\nodata\\
859  &O9.5 IV + sec                            &SB2         &\nodata     &\nodata\\ 
887  &O9.5 II-IIIn                             &SB2 NC      &\nodata          &\sid\ very weak, $-4.4$\\
\end{longtable}
%}
\tablefoot{See Table~2 for explanation of notations in the Multiplicity and
Alternate~ID columns.  96/139 stars have multiplicity flags.  The SB2
have been classified by JMA with MGB.  There are 10 B0 stars, of which
one is a supergiant.  The ``\sid\ weak'' spectra are further
discussed in the text.  Because of the very steep spectral-type
dependence of the luminosity criteria at the O--B0 boundary,
\heb~$\lambda$4686/\hea~$\lambda$4713 of or near unity implies
class II at O9.7 but V--IV at B0; also, metallicity affects the
sensitive \sic~$\lambda$4552/\heb~$\lambda$4541 spectral-type
criterion, a value of or near unity defining type O9.7.}
}
\end{center}
\end{landscape}

\end{appendix}

\end{document}